\documentclass[acmsmall]{acmart}

\usepackage{subfig}
\usepackage{multirow}
\usepackage{physics}
\usepackage{colortbl}
\usepackage[export]{adjustbox}

\newcommand{\rev}[1]{{\color{black}#1}} 

\AtBeginDocument{%
  \providecommand\BibTeX{{%
    \normalfont B\kern-0.5em{\scshape i\kern-0.25em b}\kern-0.8em\TeX}}}

\definecolor{mygreen}{RGB}{68, 179, 25}
\definecolor{mydarkred}{RGB}{215, 48, 39}
\definecolor{myorange}{RGB}{252, 141, 89}
\definecolor{myyellow}{RGB}{254, 224, 144}
\definecolor{mydarkestgrey}{RGB}{51, 51, 51}
\definecolor{mydarkergrey}{RGB}{102, 102, 102}
\definecolor{mygrey}{RGB}{153, 153, 153}
\definecolor{mylightgrey}{RGB}{204, 204, 204}
\definecolor{mylightestblue}{RGB}{224, 243, 248}
\definecolor{mylightblue}{RGB}{145, 191, 219}
\definecolor{myblue}{RGB}{69, 117, 180}
\setcopyright{acmlicensed}
\acmJournal{PACMHCI}
\acmYear{2024} \acmVolume{8} \acmNumber{CSCW1} \acmArticle{21} \acmMonth{4}\acmDOI{10.1145/3637298}
\begin{document}
\vspace*{-2cm}

\noindent
\begin{minipage}{\linewidth}
\centering 
\textit{This is the author’s version of the article that has been accepted in CSCW 2023 and will be published in PACMHCI}
\vspace*{0.5cm}
\end{minipage}



\title[Belief Miner]{\textsc{Belief Miner}: A Methodology for Discovering Causal Beliefs and Causal Illusions from General Populations}

\author{Shahreen Salim}
\orcid{0009-0004-9214-450X}
\email{ssalimaunti@cs.stonybrook.edu}
\affiliation{%
  \institution{Stony Brook University}
  \streetaddress{Engineering Drive}
  \city{Stony Brook}
  \state{New York}
  \country{USA}
  \postcode{11790}
}

\author{Md Naimul Hoque}
\orcid{0000-0003-0878-501X}
\affiliation{%
  \institution{University of Maryland}
  \streetaddress{4130 Campus Dr}
  \city{College Park}
  \state{Maryland}
  \country{USA}}
\email{nhoque@umd.edu}

\author{Klaus Mueller}
\orcid{0000-0002-0996-8590}
\affiliation{%
 \institution{Stony Brook University}
  \streetaddress{Engineering Drive}
  \city{Stony Brook}
  \state{New York}
  \country{USA}
  \postcode{11790}
}






\renewcommand{\shortauthors}{Salim, et al.}

\begin{abstract}
Causal belief is a cognitive practice that humans apply everyday to  reason about cause and effect relations between factors, phenomena, or events. Like optical illusions, humans are prone to drawing causal relations between events that are only coincidental (i.e., causal illusions). Researchers in domains such as cognitive psychology and healthcare often use logistically expensive experiments to understand causal beliefs and illusions. In this paper, we propose \textbf{\textsc{Belief Miner}}, a crowdsourcing method for evaluating people’s causal beliefs and illusions. Our method uses the (dis)similarities between the causal relations collected from the crowds and experts to surface the causal beliefs and illusions. \rev{Through an iterative design process, we developed a web-based interface for collecting causal relations from a target population. We then conducted a crowdsourced experiment
with 101 workers on Amazon Mechanical Turk and Prolific using this interface and analyzed the collected data with Belief Miner. We discovered a variety of causal beliefs and potential illusions, and we report the design implications for future research.} 
 
\end{abstract}


\received{January 2023}
\received[revised]{July 2023}
\received[accepted]{November 2023}


\maketitle
\section{Introduction}

In the view of psychology, a belief is ``the mental acceptance or conviction in the truth or actuality of some idea''~\cite{schwitzgebel2011belief}. Accordingly, a \textit{causal} belief is a belief about one or more factors that are thought to cause or contribute to the development of a certain phenomenon, such as an illness or the outcome of an intervention. A causal belief is not restricted to a single relation; it can embrace entire causal mechanisms. As Blanzieri~\cite{blanzieri2012role} writes, causal beliefs are halfway between actual knowledge about a physically objective reality and a socially-constructed reality. They are different from causality and causal inference which are strictly derived from hard data via well-defined statistical principles~\cite{blanzieri2012role}. 

An interesting phenomenon is that of \textit{causal illusion} which occurs when people develop the belief that there is a causal connection between two events that are in fact just coincidental \cite{matute2015illusions}. Causal illusions are the underpinnings of pseudoscience and superstitious thinking and they can have disastrous consequences in many critical areas such as health, finances, and well-being~\cite{freckelton2012death}. A mild form of a causal illusion is the good-luck charm that people carry to special events or in general, but more serious are bogus medicines that can inhibit people from taking up scientifically more credible treatments to restore or preserve their health. 



The first step towards combating a causal illusion is to detect it. Researchers typically use \textit{contingency judgement tasks} for this purpose~\cite{matute2015illusions}. In these experiments, participants are exposed to a number
of trials (around 50) in which a given cause is present or absent, followed by
the presence or absence of a potential outcome. At the end of the experiment, participants are asked to judge the causal relations between the cause and the effect (or outcome). While effective, these experiments have several limitations: 1) they lack a mechanism to expose complex causal structures (with many links and chains) beyond a single cause and effect; 2) they are unable to elicit complex cognitive conflicts such as how people fare with competing causes to the same effect; and 3) they are highly controlled laboratory studies and logistically expensive to conduct. We believe that crowd-sourcing these activities can provide a powerful alternative tool to address these challenges.

 We propose \textbf{\textsc{Belief Miner}}, a crowdsourcing methodology for discovering causal beliefs and illusions from the general population. While prior works have showcased the capability of crowdsourcing in identifying causal relations~\cite{berenberg2018efficient, yen2021narratives+, yen2023crowdidea}, they often approach the topic from a predominantly data-driven perspective, focusing either on
creating large networks or training datasets for Causal ML. Prior works were not designed to model contingency judgment tasks and identify causal illusions using crowdsourcing. ``Belief Miner'' aims to fill this gap.

Our methodology draws on the rich literature in psychology, causal illusion, and crowdsourcing (Section \ref{sec:background}, \ref{sec:related_work}, and \ref{sec:design_overview}). Informed by the literature, our method uses crowdsourcing to collect a dataset of causal relations from a general population on a topic of interest and contrasts these with causal relations obtained from domain experts to understand causal beliefs and detect illusions via several metrics and mechanisms we propose. Thus, Belief Miner also offers a nuanced and formalized post-hoc analysis~\cite{berenberg2018efficient} in causal crowdsourcing, which current studies and systems lack~\cite{berenberg2018efficient, yen2021narratives+, yen2023crowdidea}.

 To validate our method, we designed an interactive web-based tool that would allow crowd workers to interactively create small causal networks (and alter their created network if needed) from a set of variables. We then conducted a formative study with 94 crowd workers on Amazon Mechanical Turk (AMT). We asked participants to create causal relations between randomly chosen variables relevant to climate change. We selected the theme of climate change due to its propensity for controversial views and potential to expose causal illusions. For instance, in a study conducted in 2019~\cite{leiserowitz2019climate}, a slight majority (59\%) of Americans believed climate change is human-caused, and nearly a third (30\%) believed that natural variability is the primary cause. This belief and confusion can be harmful to necessary policy-making to counter climate change~\cite{fleming2021causalmisconceptions}. 
 
 As mentioned, in addition to the crowdsourced data, we also collected causal relations from a group of domain experts (e.g., climate scientists). Then, with both datasets in place, we employed our method on them to discover causal illusions. While the results were generally positive, we identified two issues: 1) there was a moderate possibility of selection bias due to the order in which the variables were presented to the participants; and 2) the completion time was longer than expected which pointed to possible usability issues.
 
 Based on the findings of this formative study, we revised the design of the interface and the experimental protocol. Using the revised design, we conducted another study with 101 crowd workers from AMT and Prolific. We observed a stronger alignment between the causal beliefs of the crowd and the experts and a reduced completion time of the crowdworkers compared to the formative study.  Our findings also reveal various discrepancies between the causal relations created by crowd workers and experts. We observe 1) that a significant number of workers overestimated the impact of certain attributes, 
 and 2) that participants with flawed causal beliefs (i.e., illusions) assigned lower confidence scores to their networks in both studies, suggesting that our mechanism effectively counters illusion and has the potential to increase awareness among individuals. 

 In summary, our contributions are as follows: 1) \textit{Belief Miner}, a methodology that includes a web-based interactive system and evaluation method for discovering causal beliefs and illusions; and 2) Two crowdsourcing studies on Amazon Mechanical Turk and Prolific with 94 and 101 crowd workers. The collected data from the experiment shows an application of our methodology in the domain of climate change.

\section{Background: Causal Belief and Causal Illusion}
\label{sec:background}


In this section, we will provide the background that has guided our research and development. We will build on a principal metric, the $\Delta p$ index \cite{allan1980note}, which researchers use to design the level of contingency in an experiment.


In the simplest case, there is one potential cause $C$ and one observed outcome $O$.  Given these two variables there are then four possible configurations: 1) both $C$ and $O$ occur, 2) $C$ occurs but $O$ may not, 3) $C$ may not occur but $O$ still does, and 4) neither $C$ nor $O$ occur. We can capture these four configurations into the contingency table shown in Table~\ref{fig:contingemcy-table}. $\Delta p$ is defined as follows \cite{allan1980note, matute2015illusions}:


\begin{table*}[t]
    \centering
    \footnotesize
    \begin{tabular}{ccc}
    \toprule
        & \textbf{Outcome present} & \textbf{Outcome not present} \\
    
    \midrule
    
    \textbf{Cause present} & $O|C$ & $\neg O|C$ \\
    
    \textbf{Cause not present} & $O|\neg C$ & $\neg O|\neg C$ \\
    
    \bottomrule
    \end{tabular}
    \caption{\textbf{Contingency table components}. In this table, $C$ is the cause and $O$ is the outcome.
    }
    \label{fig:contingemcy-table}
\end{table*}


\begin{table*}[t]
    \centering
    \footnotesize
    \begin{tabular}{p{1.3cm}p{1.2cm}p{1.4cm}p{1.8cm}p{1.2cm}p{1.4cm}p{1.8cm}}
    \toprule
    \multicolumn{4}{c}{\textbf{Trial Matrix 1}} & \multicolumn{3}{c}{\textbf{Trial Matrix 2}} \\
    \midrule
        & \textbf{Outcome present} & \textbf{Outcome not present} & \textbf{Probabilities} & \textbf{Outcome present} & \textbf{Outcome not present} & \textbf{Probabilities} \\    
    \midrule
    
    \textbf{Cause present} & 80 & 20 & $P(O|C)=0.8$ & 20 & 80 & $P(O|C) = 0.2$ \\
    
    \textbf{Cause not present} & 20 & 80 & $P(O|\neg C) = 0.2$ & 80 & 20 & $P(O|\neg C) = 0.8$ \\

    & & & $\Delta p = 0.6$ & & & $\Delta p = -0.6$ \\

    \midrule
    \multicolumn{4}{c}{\textbf{Trial Matrix 3}} & \multicolumn{3}{c}{\textbf{Trial Matrix 4}} \\
    \midrule
    & \textbf{Outcome present} & \textbf{Outcome not present} & \textbf{Probabilities} & \textbf{Outcome present} & \textbf{Outcome not present} & \textbf{Probabilities} \\
    \midrule
    \textbf{Cause present} & 80 & 20 & $P(O|C)=0.8$ & 20 & 80 & $P(O|C) = 0.2$ \\
    
    \textbf{Cause not present} & 80 & 20 & $P(O|\neg C) = 0.8$ & 20 & 80 & $P(O|\neg C) = 0.2$ \\

    & & & $\Delta p = 0.0$ & & & $\Delta p = 0.0$ \\
    
    \bottomrule
    \end{tabular}
    \caption{\textbf{Trial matrices emerging from different contingency table configurations}. Each quadrant is an example of one of the four types of trial matrices. The two matrices on the bottom represent null contingencies. All numbers are in \%.}
    \label{fig:trialmat-all}
\end{table*}


\begin{align}
    \Delta p &= P(O|C)-P(O|\neg C) \nonumber \\
    P(O|C) &= \norm{O|C}/(\norm{O|C}+\norm{\neg O|C}) \nonumber\\
    P(O|\neg C) &= \norm{O|\neg C}/(\norm{O|\neg C}+\norm{\neg O|\neg C}) \nonumber
\end{align}

A true causal relationship exists when $\Delta p$  is non-zero. When $\Delta p$ is positive then $C$ is said to promote $O$, while when $\Delta p$ is negative $C$ is said to inhibit $O$. 

The contingency table gives rise to four distinct cases of trial matrices shown in Table~\ref{fig:trialmat-all}. In the first case, $\Delta p>0$, $C$ might be an evidence-based medicine to treat a cold and $O$ is the disappearance of the cold. This is shown in trial matrix 1 where row 1 is the treatment that divides the response of a cohort of patients who took the medicine, while row 2 is the control that divides the response of a cohort of patients who did not take the medicine and instead took a placebo.

In the second case, $\Delta p<0$, $C$ might be an effective medicine to control a person’s cholesterol level, and $\neg O$ is the cholesterol level that remains in check and will not rise. This case is illustrated in trial matrix 2 which is trial matrix 1 transposed. 

Trial matrix 3 is one for which $\Delta p=0$, which occurs when $P(O|C) = P(O|\neg C)$ –- the null contingency.  In this case, the treatment has no effect on the outcome. A patient may take the cold medicine or not, but the cold will always disappear on its own. A classic example of this scenario is alternative (homeopathic) medicine which typically lacks strong scientific evidence for its effectiveness \cite{national2015nhmrc, singh2008trick}. 


The symmetric case of trial matrix 4 is analogous. A person with no risk of high cholesterol might take, inspired by effective product marketing, a scientifically unproven medicine and feel confirmed in that choice when the level stays normal. 

\subsection{When a Causal Illusion Weakens Trust in a Proven Cause}
\label{subsec:illusion-weaken-proven}

Having formed a null contingency about a certain phenomenon can lead people to discount a true causal relationship of a scientifically acknowledged cause for the same outcome. As Matute et al. \cite{matute2015illusions} write: “The availability of more than one potential cause can result in a competition between both causes so that if one is considered to be a strong candidate, the other will be seen as a weak one.” This was verified via several experiments where it was found that participants who had established a prior belief about the effectiveness of an unproven \textit{bogus} medicine – a causal illusion – weakened the belief in the effectiveness of a proven medicine. In contrast, participants who did not have a chance to develop the illusion had sustained trust in the proven medicine \cite{vadillo2013fighting}. 


 Experiments have shown that the strength of a causal illusion can be effectively controlled by the frequency at which the cause is present, even when its effectiveness to drive the outcome remains the same. This has important implications on a person's belief in a proven medicine. The more frequently the competing (alternative) medicine has been administered and a confirmatory (yet specious) outcome has been experienced, either now or in the past, the smaller the belief in the proven medicine \cite{yarritu2015dark}.

 An effective way to convince people that they have fallen victim to the null contingency is to ask them not to take the treatment when they hope for an outcome to occur \cite{blanco2012mediating}. But this proves difficult when the cost of the treatment is low and the outcome is ubiquitous and persuasive. In fact, the ubiquity of both treatment and outcome are perfect conditions for a causal illusion to emerge \cite{vadillo2011contrasting, yarritu2014illusion}. It fosters trust in the belief that there must be a causal relationship between the two since there are many opportunities for coincidences \cite{blanco2013interactive}. 

While the examples given so far are relatively benign, there are more serious scenarios where these cognitive mechanisms can be harmful \cite{freckelton2012death}. A person with a natural medicine mindset, when receiving, say, a cancer diagnosis might resort to acupuncture, herbal treatments, fruit juice therapy, and spiritual consultations instead of seeking more conventional evidence-based interventions, such as surgery or radiation, and chemotherapy. This is what has been reported to have happened to Steve Jobs, the founder of Apple Computer and an exceptionally tech-savvy and forward-thinking individual \cite{cartwright2011alternative}. It vividly shows that gaining immunity from null contingencies is hard.

\subsection{Gauging Causal Illusions: From the Lab into the Wild}
\label{subsec:gauging_causal_illusions}


Causal illusion has been of interest to researchers for a long time. It can reveal many cognitive and behavioral practices, as described in Section \ref{subsec:illusion-weaken-proven}. To study different research questions, researchers typically tweak the trial matrices (Table~\ref{fig:trialmat-all}) to create desired scenarios and then conduct control experiments based on that.

These experiments typically involve hypothetical scenarios set in the medical domain. Participants are asked to impersonate doctors who are assigned a set of fictitious computer-modeled patients (i.e., the total number of trials).  In some scenarios, the participants take on an active role in prescribing medical treatments for some illnesses, while in others they simply observe the patients follow a certain treatment regime. As the experiment proceeds, participants see records of patients who either have or have not taken a certain medicine and then have or have not recovered from the disease (based on the trial matrices). At the end of the experiment, the participants are asked whether the treatment was effective or not (for example see~\cite{matute2015illusions}). 

The experiments are meticulously designed to expose the triggers and impacts of causal illusions. They are highly controlled laboratory studies and focus on very simple and elementary causal relationships. Our crowd-based belief miner takes these studies out of the research laboratory into the wild where an abundance of data awaits, reflecting complex causal chains with many links and paths. There are also many unexpected cause and outcome variables that might be discovered when the crowd-sourced data are studied in depth. It is a tool by which complex cognitive conflicts can be efficiently extracted and exposed for real-life phenomena of possibly high complexity. While this approach cannot directly replicate the usage of trial matrices shown in Table \ref{fig:trialmat-all}, it can produce the conclusions that are typically derived from the experiments that use trial matrices. Furthermore, it is easy to set up for practitioners. 

Another interesting aspect of our methodology is that it asks participants to engage in a critical assessment of their beliefs, by ways of reviewing the small causal network they construct. As we have already hinted at in the introduction, we found evidence of the implications of a theory proposed by Walsh and Sloman \cite{walsh2004revising}, who experimented with the concept of contradiction as a way to get people to revise their beliefs in a causal illusion. They suggest that a person might reduce their belief in the effectiveness of a certain treatment upon discovering that some elements of it have an outcome that is opposite of what they expected. 


\section{Related Work}
\label{sec:related_work}

Our goal in this paper is to discover causal beliefs and illusions using crowdsourcing. In this section, we discuss crowdsourcing, HCI, and CSCW concepts relevant to achieve that.

\subsection{Crowdsourcing and Quality Control in Crowdsourcing}

Bigham et al.~\cite{bigham2015human} identified three broad areas for collective intelligence and HCI. They are 1) \textit{directed crowdsourcing}, where a single person or a group guides a large set of people to accomplish a task (e.g., labeling a large dataset~\cite{DBLP:conf/cscw/KairamH16, DBLP:conf/chi/ChangAK17}, CommunityCrit~\cite{mahyar2018communitycrit}); 2) \textit{collaborative crowdsourcing}, where a group of people works together to accomplish a task (e.g., Wikipedia, Project Sidewalk~\cite{saha2019project}, ConceptScape~\cite{liu2018conceptscape}); and 3) finally, \textit{passive crowdsourcing}, where people do not coordinate and are not consciously aware of participating in a crowdsourced system, however, one can still mine their behavior to infer collective intelligence (e.g., mining search history). 

In this paper, we mainly focus on directed crowdsourcing (referred to as crowdsourcing from here on) since we provided explicit direction to our crowd workers. In a crowdsourcing experiment, a crowd worker typically completes a small part of the overall task (i.e., micro-tasks)~\cite{DBLP:journals/pacmhci/ChungSKHKL19, kittur2008crowdsourcing, kittur2011crowdforge}. These micro-tasks are eventually aggregated for inferring collective intelligence. In our case, a micro-task refers to creating a causal network between a small number of attributes.

Over the years, several methods and tools have been proposed for measuring and increasing the efficacy of the design of the micro-tasks based crowdsourcing~\cite{kittur2013future, kittur2012crowdweaver, rzeszotarski2012crowdscape, bernstein2010soylent,kittur2011crowdforge, noronha2011platemate}. Of particular interest is \textit{Quality Control} in crowdsourcing~\cite{kittur2013future}, which ensures the validity of the collected response from crowd workers. This is important since crowd workers tend to spend minimum effort and sometimes try to game the systems~\cite{kittur2013future}. The currently established methods for quality controls typically fall into two broad categories: task design and post-hoc analysis~\cite{kittur2013future}. Examples for task design include fault-tolerant subtasks~\cite{bernstein2010soylent,kittur2011crowdforge, noronha2011platemate, krishna2016embracingerrors}, attention check~\cite{agley2022quality}, peer review filters~\cite{dow2012shepherding, kittur2011crowdforge, bernstein2010soylent, horton2010labor}, intelligent task assignment~\cite{kim2018hitorwait} and optimizing instructions~\cite{kittur2008crowdsourcing, downs2010your, lasecki2015sequence,qiu2020conversationalmicrotask}. Examples for post-hoc analysis include comparison with gold standards~\cite{downs2010your,callison2009fast, berenberg2018efficient}, validation study~\cite{saha2019project}, agreement between crowd workers~\cite{callison2009fast, ipeirotis2010quality}, and behavior analysis~\cite{rzeszotarski2012crowdscape,robert2015crowddiversity, gadiraju2015malicious, chiang2021freetime,yin2018runningout}. 
%

Our work can be seen as a post-hoc analysis model for causal crowdsourcing. We use causal relations collected from experts (i.e., gold standards), a popular post-hoc method, to find illusions in the crowd-generated causal networks. Although relevant work in causal crowdsourcing \cite{berenberg2018efficient} employed comparison with gold standards on a smaller scale, our work both scales up the approach and adds a nuanced dimension to the post-hoc analysis model tailored for causal crowdsourcing. This makes our contribution one of the first to delve into such depth and granularity in this domain. Finally, we believe our findings will guide future task designs for causal crowdsourcing. Appropriate task designs can make people self-aware and help them avoid falling victim to causal illusions. We lay down this future direction in our discussion (Section \ref{sec:discussion}).

\subsection{Causality and Crowdsourcing}
\label{sec:lit_causality}

While causality is a core concept across several scientific domains,  designing crowdsourced experiments for collecting causal relations is a relatively new research area. Caselli et al. contributed to this area by focusing on annotating causal relations in narrative texts, specifically news data, through crowdsourcing experiments~\cite{caselli-inel-2018-crowdsourcing}. Their work analyzed parameters affecting annotation quality and compared crowdsourced and expert annotations, emphasizing the generation of structured data based on narrative strategies. The most relevant work in this space, however, is Iterative Pathway Refinement~\cite{berenberg2018efficient}, a network search strategy where workers modify a short linear pathway between attributes. The authors showed that their method is more efficient than a single line-based micro task, provides better contexts to crowd workers, and the union of the pathways can create a large network. Yen et al.~\cite{yen2021narratives+} extended this line of work by proposing CausalIDEA, an interactive interface where users can create small networks as well as provide textual explanations for creating specific causal relations. The causal diagrams and textual narratives add a new dimension to understanding causal beliefs. The authors investigated how a user's causal perception is affected by seeing the causal networks created by others. Furthermore, Yen et al. introduced CrowdIDEA~\cite{yen2023crowdidea}, a  tool that integrates crowd intelligence and data analytics to support causal reasoning, featuring a three-panel setup: enabling access to crowd's causal beliefs, data analytics, and the ability to draw causal diagrams. Their study also demonstrated that seeing the crowd's causal beliefs significantly improved the accuracy of causal relationships in the final diagrams of the participants and reduced reliance on data analytics, showcasing the tool's potential to enhance causal reasoning processes.

These prior works provide evidence that people can identify causal relations between pairs of events and collaboratively create causal networks that are information-rich. While inspiring, they consider causal crowdsourcing from an algorithmic or data science perspective, focusing either on creating large networks or training datasets for Causal ML~\cite{kusner2017counterfactual}. In contrast, we consider causal crowdsourcing as a tool to surface how causal illusions persist in a domain, which prior works have largely overlooked~\cite{berenberg2018efficient, yen2021narratives+}.

While we do not focus on data science or machine learning, our work has implications for Causal ML. For example, failing to detect causal illusions and erroneous causal relations can introduce biases into datasets, leading to skewed decision-making based on inferred causal relations~\cite{wang2015visual, hoque2021outcome, DBLP:journals/tvcg/GhaiM23}.  Broader ML  literature has also demonstrated that ML models trained with datasets containing spurious relations or errors may perform poorly and yield incorrect inferences ~\cite{yapo2018ethical, 10.1145/3512930, Panch2019-yv, kusner2017counterfactual}.

In summary, the absence of mechanisms to detect causal illusions is a significant obstacle to the practical application of causal crowdsourcing in different domains, including social science and policy-making (domain of this work), causal ML, finances, and health. We aim to bridge this gap.

\section{Method}\label{sec:design_overview}
%
%

In this section, we describe the evaluation method we have devised to help expose the potential causal illusions and the complex cognitive conflicts mentioned in Section \ref{sec:background}.


%

\subsection{Data}
\subsubsection{Causal Belief Data}


Causal beliefs are conceptualized through the presence of a \textit{cause} and an \textit{effect} and their \textit{relationship} in a certain phenomenon. Following the Structural Causal Model (SCM)~\cite{pearl2018book}, we define a \textit{causal relationship} through a directed link/edge between the cause and effect (\textit{cause }$\rightarrow$ \textit{effect}). Therefore, gathering causal relations from a domain of interest is synonymous with identifying the causal edges among a collection of relevant attributes that form an interconnected directed causal network~\cite{berenberg2018efficient}. 

Let us define a causal network $N = (V, E)$ with nodes $V$ and links $E$. Here, $V$ is a set of causal attributes and $E\subseteq\{{u\rightarrow v | u,v \in V\ and\ u \neq v}\}$, where $u\rightarrow v$ is a unidirectional causal relationship between attributes $u$ and $v$. 


\subsubsection{Ground Truth}
\label{subsubsec:ground-truth}
Comparison with gold standards or ground truth is common in crowdsourcing~\cite{downs2010your,callison2009fast, berenberg2018efficient}. In our case, ground truth refers to scientifically verified causal relations. However, establishing ground truth in a complex domain such as climate change where many interconnected factors may exist, can be a demanding task. We propose the following collaborative method with two \textit{phases} to establish ground truth for all possible causal links. 

\textbf{Phase1: Independent Ground Truth Generation.} First, a fixed number of experts are instructed to create their version of the causal networks independently using all the attributes. We propose a minimum of three experts. We require experts to provide scientifically verified references to the created relations.

\textbf{Phase2: Collaborative Meeting.} After the network creation phase, the union of their networks is considered for the discussion phase. Here, the experts work collaboratively to assign \textit{credibility scores} for all possible causal links in the unified network. Credibility scores denote the level of validity for a specific causal link. We chose this notion since there can be ``levels'' of correctness for a specific causal relation instead of them being just right or wrong. This can also be described as the strength of a causal relationship. The notion is motivated by mediation analysis in causality~\cite{mackinnon2007mediation}. According to mediation analysis~\cite{mackinnon2007mediation}, in a causal network, an attribute may have a direct effect on another attribute as well as indirect effects through other attributes (mediating attributes). Thus, the number of mediating attributes between two attributes indicates the strength of the causal relationship. 
    
    Based on this observation, we propose four levels for the credibility scores: all links that did not appear in any experts' causal network can be assigned the lowest possible credibility score (0), and the links that appear in all of them can be assigned the highest credibility score (3). Finally, the rest of the links are assigned a score of 1 or 2 after expert deliberation, depending on the number of mediating variables present in between. 
    



\subsection{Metrics}

We offer two angles of evaluation or analysis of the collected causal belief data: 1) an aggregated quantitative and qualitative overview of the causal beliefs; and 2) causal illusion detection. The metrics required are defined below: 

    \subsubsection{Aggregated Evaluation:} Aggregated statistics such as the distribution of total votes are common in crowdsourcing~\cite{kittur2013future}, including causal crowdsourcing~\cite{berenberg2018efficient,yen2021narratives+}. They are useful for obtaining an overview of the data and determining outliers. In addition to the total votes, we also examine the distribution for the \textit{Average Network Credibility Score (ANC)}. We calculate the average network credibility (ANC) score using the following equation:
\begin{align}\label{eq1}
    ANC_N &= \frac{\sum_{e \in E}{cs_{e}}}{|E|} \nonumber
\end{align}

    Here, $N$ is a small causal network created by an individual, and $cs_e$ is the credibility score (from Section \ref{subsubsec:ground-truth}) of link $e$ present in network $N$. Thus, $ANC_N$ can be calculated for each network created by separate individuals, and their distribution will indicate the credibility for the networks created by people.

\subsubsection{Causal Illusion Detection} 
    
    
In Section \ref{sec:background}, we define \textit{Causal Illusion} as the incorrect assumption of cause and effect in a certain phenomenon. Thus, causal illusion inherently denotes a discrepancy between people's causal beliefs and the ground truth.

    

We define two types of causal illusions that utilize ground truth data and the causal belief data collected from the people/crowd. We represent the crowd data using the \textit{crowd score (cr)} (a score representing the crowd's inclination toward a specific causal link) and the ground truth data using the \textit{credibility score (cs)}. There are several ways the \textit{crowd score (cr)} can be calculated, such as using the normalized total votes assigned to a causal link by the crowd.

Let us suppose we have a causal link $l$ with a crowd score of $cr_l$ and a credibility score (from ground truth) of $cs_l$. There can be two potential cases of causal illusion;
\begin{itemize}
    \item Where the crowd had a stronger inclination toward a causal link with a comparatively lower credibility score, i.e., $cr_l > cs_l$. We define this state as being \textit{potentially misinformed}. While we do not investigate the reasons behind this, they can indicate the consumption of less credible sources, shallow reading practices, and denial of climate change \cite{climatechangenotbelief}. 
    \item Where the crowd had a weaker inclination toward a causal link with a comparatively higher credibility score, i.e., $cr_l<cs_l$. We define this state as the state of being \textit{potentially uninformed/oblivious}. The cause behind obliviousness can be a lack of knowledge regarding that specific topic, i.e., people genuinely do not know about it enough to vote for it.  
\end{itemize}

\section{Initial Interface}
\label{sec:survey}

Our overarching goal in designing the interface was to foster critical thinking about a topic of interest (e.g., climate change) while ensuring ease of use in creating causal networks. We felt the necessity to develop our own data collection tool since there is no open-source tool available for causal belief collection. In pursuit of this objective, we employed an iterative design process with specific design goals to continually refine and improve the interface based on evolving needs.

\begin{figure}
    \centering
    \includegraphics[width=0.95\textwidth]{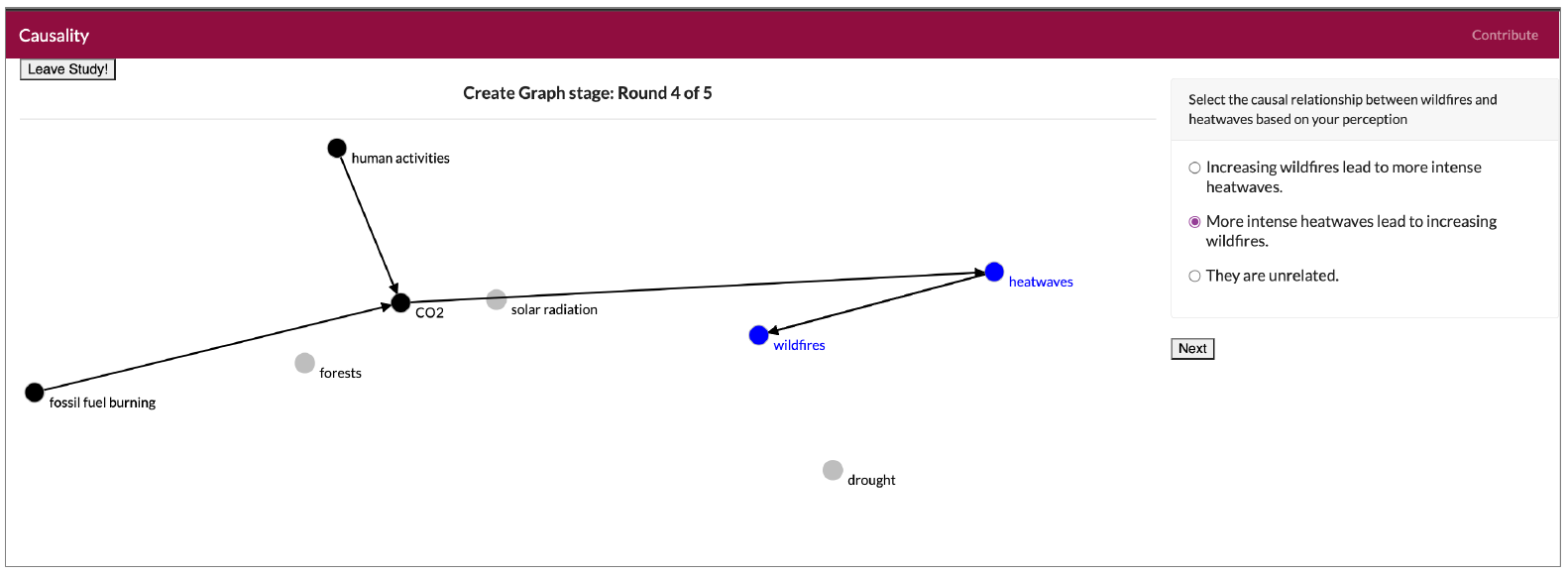}
    \caption{\textbf{Overview of the initial collection interface}. This example shows the steps participants followed to create the causal networks. In the center, we see the causal network created by the participant. The edge connecting the two nodes marked in blue is the newly created edge by the participant. We provide options to select the direction for the newly created causal edge on the right.}
    \Description{Figure \ref{fig:collection_interface_initial} shows a screenshot of the collection interface. It shows a case where a crowd worker is in the process of making a causal network.}
    \label{fig:collection_interface_initial}
\end{figure}

\subsection{Design Goals}
\label{subsec:design_goals}


In this section, we provide the initial design goals for the interface. Some part of the interface loosely follows the methodology proposed in prior works~\cite{berenberg2018efficient, yen2021narratives+}: people create small causal networks to demonstrate their beliefs, and then the small networks are aggregated into a large causal network. However, we non-trivially enhanced the methodology to meet the following design goals:


\paragraph{DG1. Interactively Create and Modify Causal Networks.}
Interactive visual interfaces can enhance people's understanding and decision-making of complex systems \cite{dietvorst2018overcoming}. We decided to utilize an interactive visual interface based on a Directed Acyclic Graph (DAG) to represent causal networks, enabling users to freely choose attributes and their causal relations. The interface should provide modification controls, such as changing link directions, to allow users to refine networks. 

\paragraph{DG2. Use of Natural Language to Narrate Causality}
Natural language texts are used in conjunction with visual representations to provide explanations and enhance comprehension of causal networks~\cite{yen2021narratives+}. This inspired us to use natural language as an explanation along with the interactive visual representations \cite{Choudhry2021OnceUA}. The narrative component should clarify potential confusion arising from graphical representations.

\paragraph{DG3. Quick Completion Time}
One important design goal in crowdsourcing studies is to ensure quick completion time to enhance the efficiency of data collection~\cite{kittur2013future}. By minimizing the time required for participants to create, modify, and evaluate causal networks, we want to harness the crowd's collective intelligence more effectively and gather a larger volume of diverse causal networks for analysis and insights. 


\subsection{Processing Raw Data and Generating Causal Attributes}
\label{subsec:causal_attribute} 

Before collecting causal beliefs, the set of single/multi-word attributes relevant to the domain must be identified. Experiments with techniques purposed to automatically extract attributes specific to ``climate change'' (our demonstration domain) from relevant text documents were only mildly successful. To identify the relevant attributes for our domain (climate change), we manually extracted them from reputable climate-related sources~\footnote{
https://www.climaterealityproject.org/blog/key-terms-you-need-understand-climate-change 

https://www.climaterealityproject.org/blog/10-indicators-that-show-climate-change   

https://opr.ca.gov/facts/common-denier-arguments.html}. We also determined specific words to represent upward and downward trends, ensuring natural comprehension, such as ``fewer (human activities)'', ``less (methane)'', and ``decreasing (solar radiation)''.  This process yielded a table of 17 attributes with trend terms (34 attributes in total, combining trends with attributes). We then validated these attributes via an informal discussion session with three climate science experts from our university. All experts hold PhDs in relevant fields and have been conducting climate science research for at least ten years. All agreed that the attributes are of importance to any climate science expert, and understanding people's perceptions about them is crucial. 

In addition to extracting the attributes, we also use Word2Vec~\cite{mikolov2013efficient}, a neural word embedding model trained on the English Wikipedia corpus containing many climate-related documents, to compute attribute coordinates based on semantic distances. We used these coordinates to lay out the variables in the 2D space of the initial interface (Section \ref{subsec:interface_modules-1}). The Word Mover Distance (WMD) was employed for multi-word attributes, calculating the minimum distance words need to travel between documents. We also explored other text embedding models, including BERT~\cite{devlin2018bert}, RoBERTa~\cite{liu2019roberta}, and GloVe~\cite{pennington2014glove}. After a discussion with the research team, we found Word2Vec's embeddings to be the most suitable option.

We note that we did not utilize the Word2Vec coordinates in our final and redesigned interface because we adopted a different layout approach (Section \ref{subsec:interface_modules}).

\subsection{Collection Interface Modules}\label{subsec:interface_modules-1}

The collection interface is a web-based interface implemented using Python as the back-end language and D3 for visualization \cite{bostock2011d3}. We used MongoDB as a database for our collected results. The input to the interface is a dataset containing nodes of a causal network ($V$) where the edge list ($E$) is unknown. Thus, this interface is our tool to infer $E$ from the crowd. The detail of the input dataset is provided in Section~\ref{subsec:causal_attribute}. We describe the visual component of the several interface modules in the following sections. The modules are independent and can be implemented according to the intended sequential workflow. Snapshots of each module are provided as supplemental material. We further describe our workflow specific to the experimental setup later in Section~\ref{sec:experiment1}.

\subsubsection{Instructions and Overview Module}
\label{instruction}
This module provides an overview of the interface and study tasks as a step-by-step guide, with necessary explanations and instructions for each step. It also presents necessary pictures of each page that people will encounter in their workflow.  


\subsubsection{Demographics Survey Module}
\label{demographics1}
This module collects participants' demographics and their perception of domain-specific (e.g., climate change) knowledge and awareness. We currently support multiple-choice and Likert scale questions in this module.





\subsubsection{Causal Network Creation Module}
\label{creation1}

The main module in the tool allows participants to create causal relations between pairs of attributes to build a small causal network \textbf{(DG1)}. The module follows \textbf{DG1} and \textbf{DG2} principles, visualizing causal links as node connections and describing them in natural language.  The attributes are presented as circular nodes in the interface, positioned based on their word-vector space.  The order of the attributes' appearances is pre-determined. Therefore, all individuals will see the attributes in the same order. We did this to observe the difference in people's perceptions given the same set of choices. This makes people's perception a random variable in our experiment instead of the order of the attribute's appearance.  

Participants perform two micro-tasks per causal link: choosing trends for attributes (e.g., increasing and decreasing for the attribute CO2) and selecting the causal relationship between them. This process is repeated in multiple rounds to create the network (Figure~\ref{fig:collection_interface_initial}). Following~\cite{berenberg2018efficient}, we decided to keep a narrative flow in the causal network. Therefore, after the first round, a participant needs to create a causal link between any previously chosen attribute and one new attribute. Participants need to make three different causal networks. 
Except for the first network, participants are presented with a mix of previously used and new attributes when creating a causal network. This ensures reduced learning requirements, which contributes to shorter completion times \textbf{(DG3)}.



\subsubsection{Alteration Module}
\label{alternation}
Following \textbf{DG1}, we developed this module to allow people to alter the network they have created by modifying each network link. The available options are (1) changing the originally selected link direction or (2) deleting the link entirely. This module always appears after the \textit{Creation Module}.



\subsubsection{Interpretation and Evaluation Module}
\label{interpretation}
In this final step, each individual is asked to evaluate their created (and possibly altered) network. Following \textbf{DG2}, we provided a narration of the network. To generate this narration, we combine a graph traversal algorithm with a reasonably simple text template to translate the network into a textual narration. 
The module then provides people with the opportunity to evaluate their created network. The evaluation is collected as a confidence level on a 5-point Likert scale. This module always appears after the \textit{Alteration Module}. 

\subsubsection{Usability Rating Module}
\label{evaluation_interface}
The purpose of this module is to provide participants with the opportunity to evaluate the complete data collection interface. Following the System Usability Scale (SUS) \cite{sus}, we present each individual with five usability-related statements. We also provide them with two knowledge-related statements to measure their evaluation of the interface from the perspective of gaining knowledge. A primary goal of the knowledge/learning-related statements was to gauge active thinking's effects by creating a causal network on people's original perceptions. The positive and negative usability statements are presented in alternating order. We mention the seven statements as supplemental materials.

\section{Formative Study}
\label{sec:experiment1}


We recruited $98$ crowd workers from Amazon Mechanical Turk (AMT) to collect causal perceptions on climate change. They used the initial collection interface to create small causal networks. We include all study materials in the supplement.

\begin{figure}
    \centering
    \includegraphics[width=0.9\textwidth]{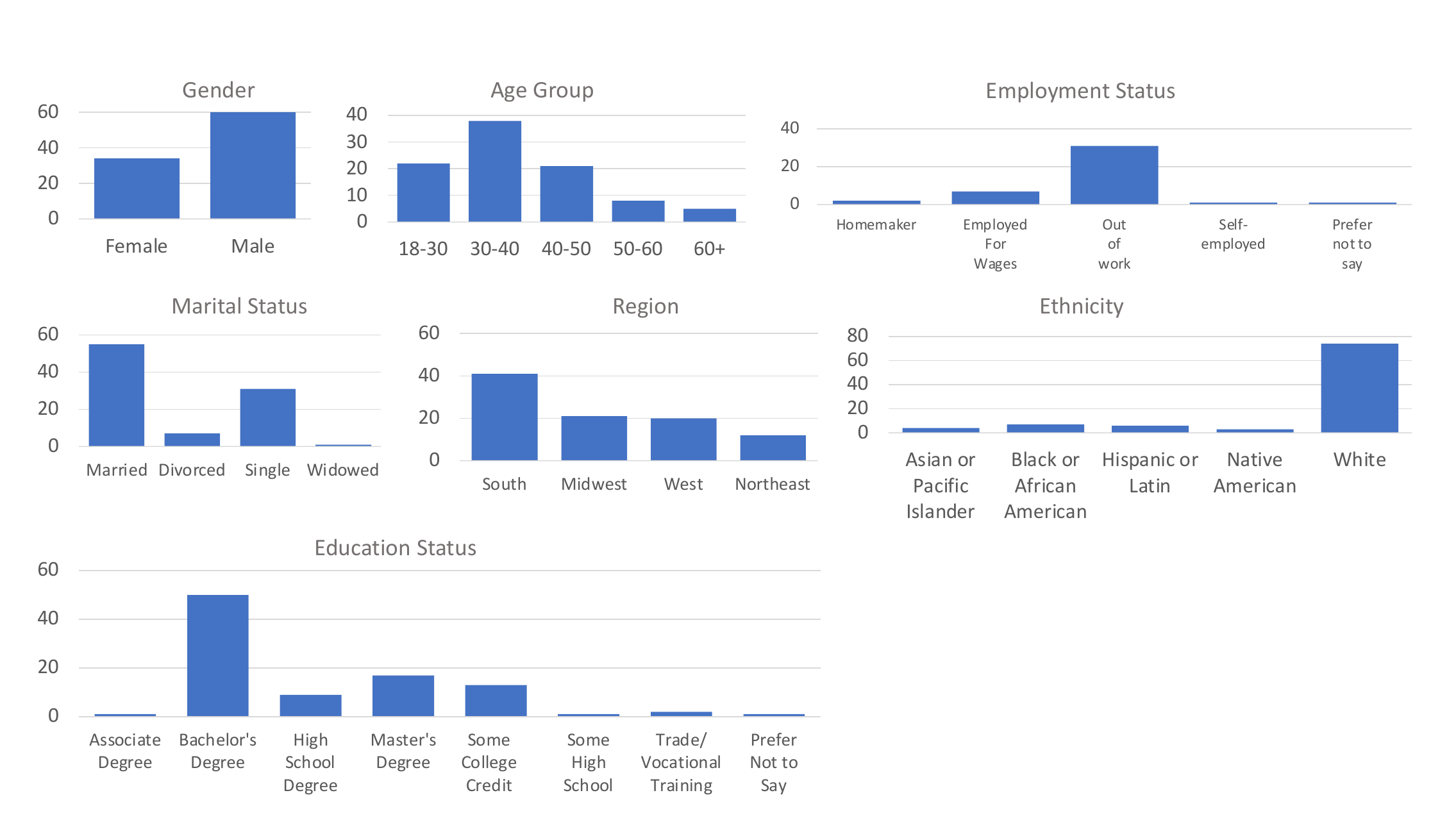}
    \caption{\textbf{Demographics of the crowd workers in the formative study.} Y-axes represent counts for each category.}
    \Description{Figure \ref{fig:demographics} shows a collection of bar charts, each containing statistics of various demographic variables. The variables in order from the top-left are: Gender, Age Group, Employment Status, Marital Status, Region, Ethnicity, and Education Status.}
    \label{fig:demographics}
\end{figure}

\paragraph{Crowd Workers' Demographics and Expertise Level Regarding Climate Change}
We needed to discard the work of 4 workers due to incompleteness, which led us to have 94 valid workers. We present various aspects of the crowd workers' demographics in Figure \ref{fig:demographics}. The majority of the crowd workers happened to be male, white, and within the age group of 20-40. More than half of the crowd workers finished their bachelor's degrees, and more than two-thirds are employed for wages. Geographically, we only collected results from the United States. A significant portion of the crowd workers is from the Southern region of the United States.  

We also examine the self-reported knowledge and agreement levels of crowd workers regarding climate change-related attributes and statements. Around 49\% the of participants consider themselves knowledgeable about climate change attributes. Additionally, there is a strong agreement among participants with climate change-related statements (around 90\%), indicating a belief in climate change. Further details can be found in Appendix \ref{formative-concern}.
\paragraph{Protocol}
We used the ``external HIT'' function on AMT, where the interface hosted on our server was accessible to the workers. Therefore, our interface did not require the crowd workers to log in and provide personal information beforehand.  We only accepted the results when the crowd workers completed every task in the workflow. The successful crowd workers were paid \$2.75 each upon completion. We initially estimated the work would take around 15 minutes. However, the average time taken by the workers was around 30 minutes, according to the AMT website. We provide the step-wise protocol below:
 \begin{enumerate}
      \item \textbf{Read the instructions and pass the test}: In the ``Instructions and Overview Module'' (Section~\ref{instruction}), the crowd workers received a step-by-step guide and explanations of each interface module. They had the option to revisit previous pages and restart if needed. A test required them to demonstrate their understanding, and they could retry it to improve their comprehension.
     \item \textbf{Complete the demographics survey}:  In the ``Demographics Survey Module'' (Section~\ref{demographics1}), the crowd workers answered 11 questions about their ethnicity, gender, marital status, geographical location (state), education, employment status, age group, knowledge, concern, and agreement towards climate change. They had the option not to provide their information for each demographic question. 
     \item\textbf{Create a causal network}: The crowd workers used the ``Causal Network Creation Module'' (Section~\ref{creation1}) to build a small causal network. They created five causal links by (i) selecting two attributes along with their trends (e.g., ``CO2'' with an ``increasing'' or ``decreasing'' trend) and (ii) choosing the causal relationship between the selected attributes (e.g., \textit{increasing emissions} leads to \textit{increasing CO2}). This process involved two micro-tasks for each link creation. After the first two links, the workers are asked to select a new node that has not been selected before and an already selected node in order to create a new causal relation, adding to the emerging small network. 
     \item\textbf{Alter causal network}: In the ``Alteration Module'' (Section~\ref{alternation}), the crowd workers could modify their created causal network. They selected a link, left-clicked on it, and chose from available options for alteration, including \textit{deletion}, \textit{changing the direction}, or \textit{no modification}.
     
     \item\textbf{Evaluate causal network}:  Crowd workers used the ``Interpretation and Evaluation Module'' (Section~\ref{interpretation}) to review and evaluate their created causal network. They could view it in a node-link diagram or Directed Acyclic Graph (DAG) format and read it in natural language text. Additionally, they provided a confidence level for each network on a scale of 1 to 5.

     \item \textbf{Repeat Task 3-5 two more times}: Crowd workers repeated the process of creating small networks, altering them, and evaluating them two more times. The interface provided necessary prompts and repetition for this task.
     
     \item \textbf{Evaluate the interface}: After creating three causal networks, the crowd workers used the ``Usability Rating Module'' (Section~\ref{evaluation_interface}) to rate the interface. They rated seven usability and learning statements on a 5-point Likert scale to provide feedback.

     \item \textbf{Verification and compensation}: At the end of their participation, each crowd worker received a unique code. This code was used for result validation, filtering incomplete data, and providing compensation for their contribution.
 \end{enumerate}

\section{Findings from the Formative Study}
\label{sec:evaluation1}

We collected three types of data in this experiment: demographic information (94 users), small causal networks (282 networks), and subjective feedback (94 users).  The causal networks were saved as vectors containing causal links or (source node, target node) pairs, along with confidence scores tied to the worker identifier. 




\subsection{Establishing Credibility Scores}

\begin{figure}
    \centering
    \includegraphics[width=\textwidth]{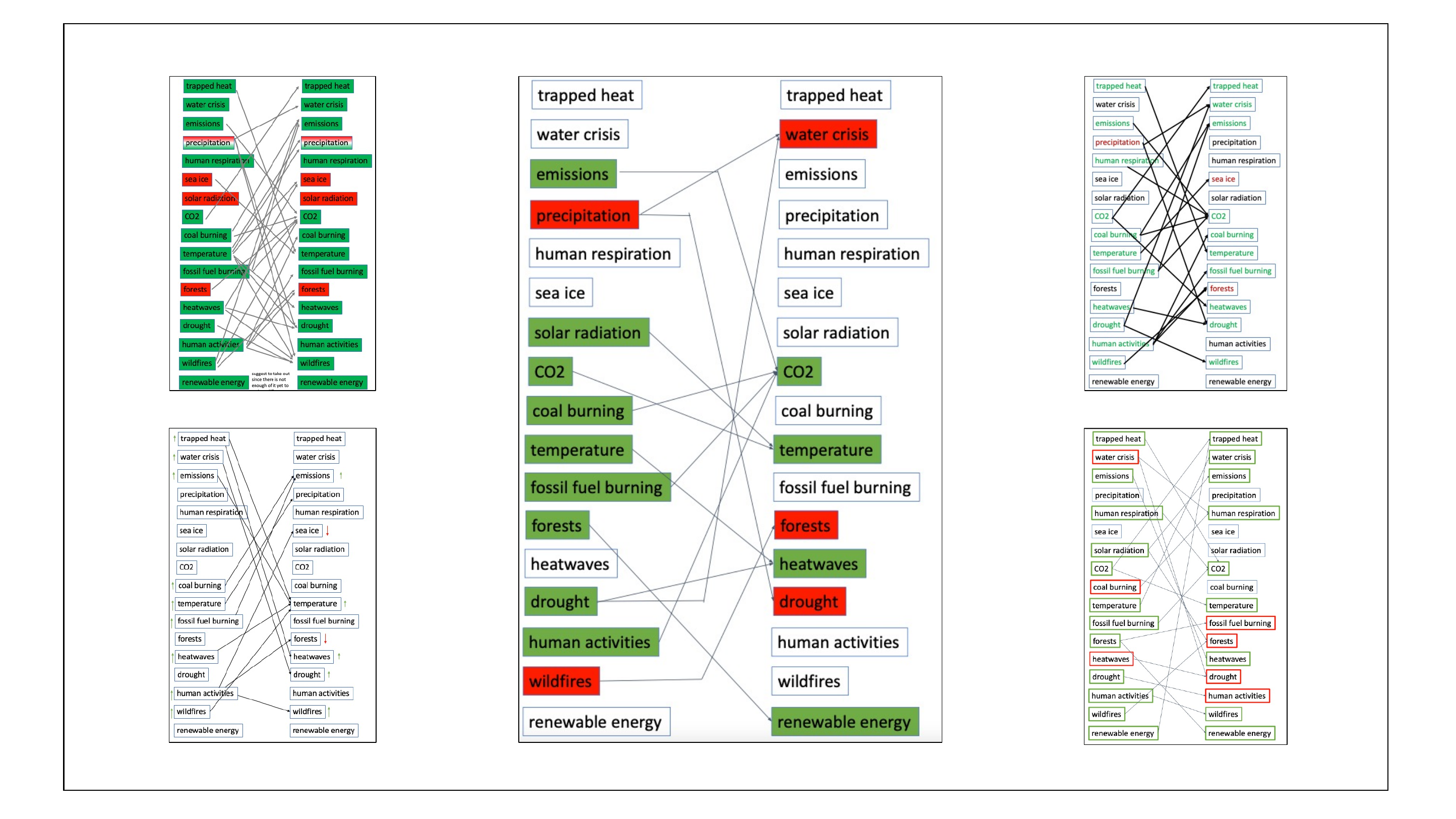}
    \caption{\textbf{Snippets of the ground truth networks created by the experts}. Each expert was provided with the same template containing all causal attributes. Both the left and the right columns contain the same attributes. The experts colored \textcolor{mygreen}{green} and \textcolor{red}{red} for expressing the upward and downward trend respectively and connected any two attributes of choice from the left column to the right column using an arrow ($\rightarrow$) to create a causal link.}
    \Description{Figure \ref{fig:gt} illustrates the examples of ground truth networks created by experts.}
    \label{fig:gt}
\end{figure}

We assigned credibility scores to the causal links based on the methodology proposed in Section \ref{sec:design_overview}. Five experts, including two authors of this paper, participated in this stage. The other three experts are climate science experts from our university. They hold PhDs in relevant fields and have been conducting climate science research for at least ten years. Experts independently created their version of the causal networks using the 34 attributes. We provided them with climate change-related literature that is publicly accessible and is more likely to be used by the general population~\footnote{
https://climate.nasa.gov/\\
https://www.epa.gov/\\
https://www.noaa.gov/\\
https://www.climatecentral.org/
 }. Figure \ref{fig:gt} shows the causal networks created by the experts. The ground truth establishment procedure yielded a dataset of all possible causal links and their credibility scores.
Based on our proposed evaluation metrics in Section \ref{sec:design_overview}, we conduct an extensive analysis of the causal belief data collected during the experiment. The results are presented next. 


\begin{figure}
    \centering
    \includegraphics[width=0.8\textwidth]{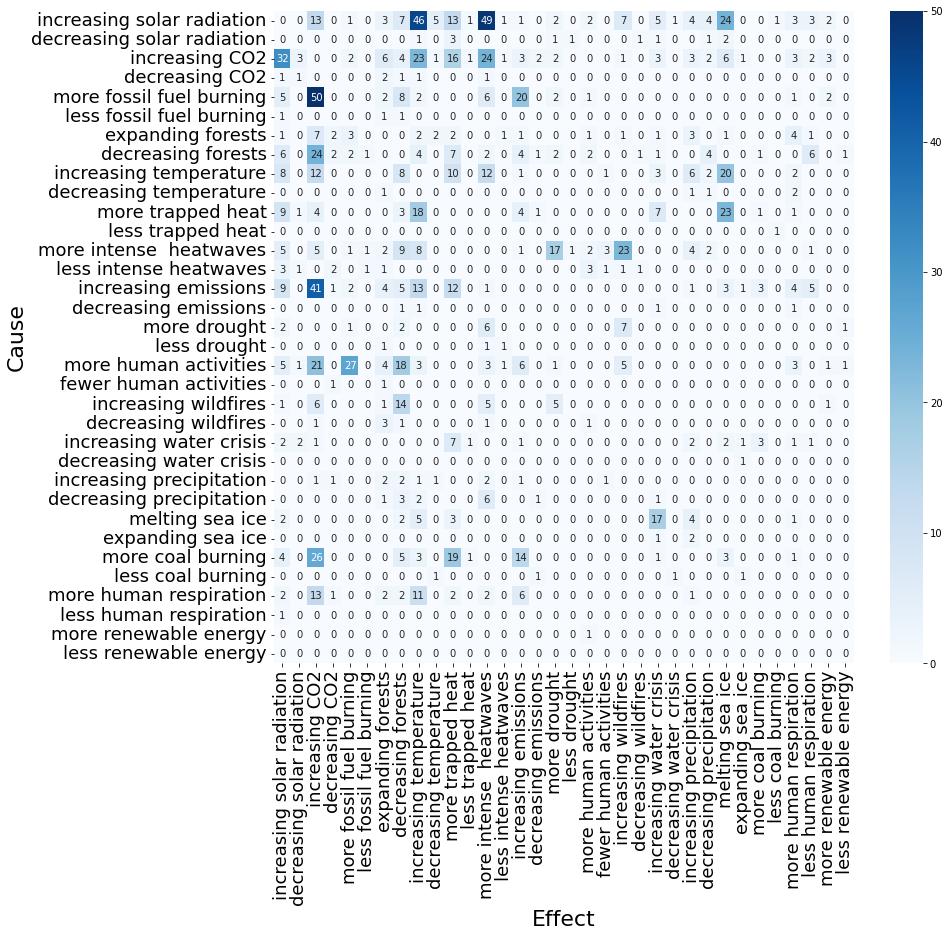}
    \caption{\textbf{The Adjacency Matrix Heatmap Representation of the collected in the formative study.} The cell values represent the total number of votes for that specific causal relation.}
    \Description{Figure \ref{fig:adjacency} demonstrates an adjacency matrix heatmap representation, where both rows and columns denote causal attributes and each cell shows the total number of votes of specific causal relations with numbers and color intensity.}
    \label{fig:adjacency}
\end{figure}

\subsection{Aggregated Evaluation}\label{subsec:aggregated}


\subsubsection{Combined Network and Total Votes Per Link}
We combined all 282 small causal networks by counting the votes for each causal link. Following \cite{yen2021narratives+}, we present the results as an adjacency matrix heat-map in Figure \ref{fig:adjacency}. The top-3 most voted links were, ``more fossil fuel burning $\rightarrow$ increasing CO2'', ``increasing solar radiation $\rightarrow$ more intense heatwaves'', and ``increasing solar radiation $\rightarrow$ increasing temperature'' with 50, 49, and 46 votes, respectively. The combined network is sparse, and most relations have zero votes. The Pearson Correlation Coefficient $(r)$ between the total votes and the credibility scores for the causal links is 0.4 with $p<10^{-43}$, indicating a moderate consensus between experts and crowd.


\subsubsection{Average Network Credibility Scores (ANC) and Average Confidence Score (AC)}

\begin{figure}
     \centering
     \subfloat[]{
        \includegraphics[width=0.45\textwidth]{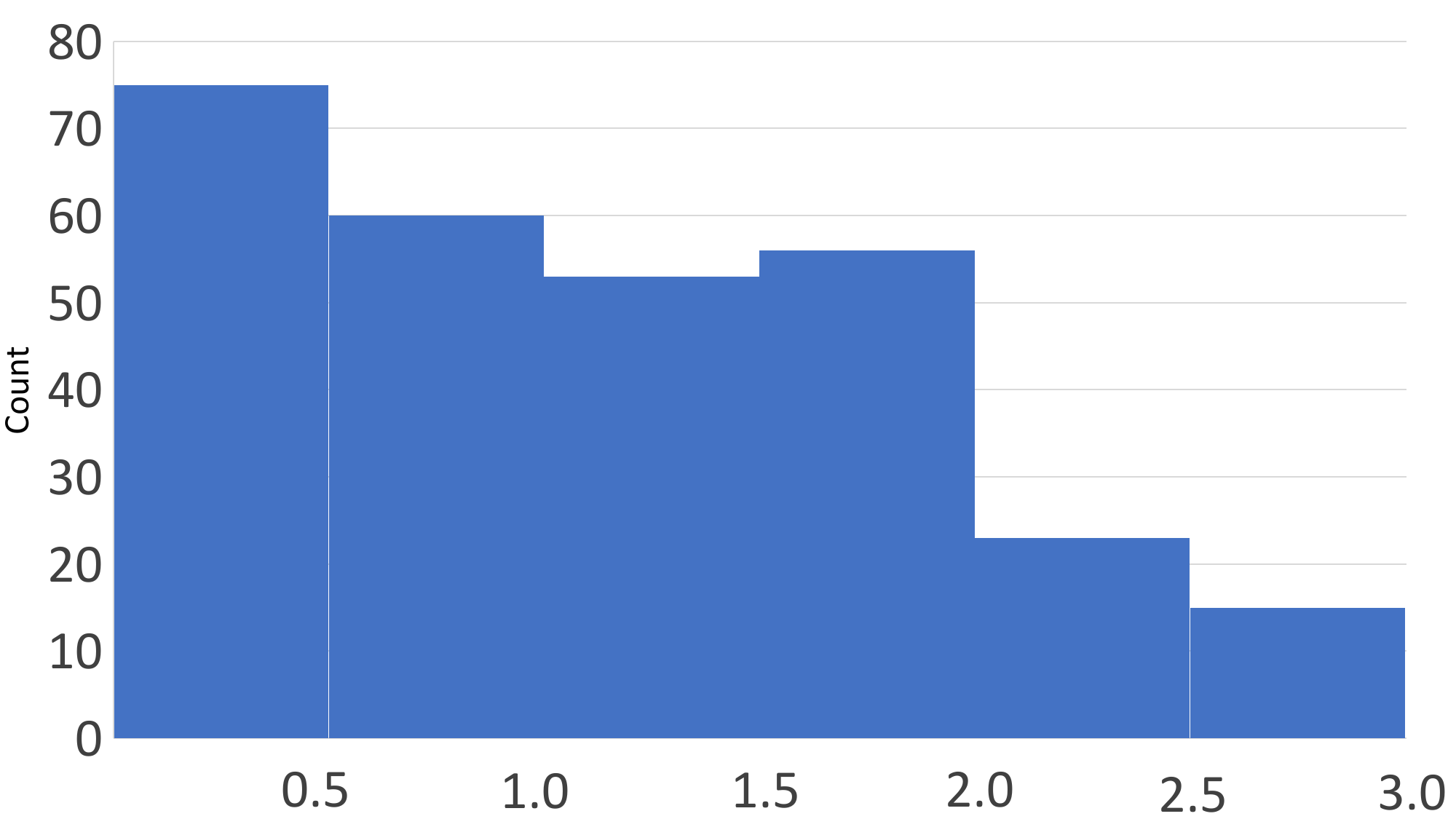}
        \label{fig:credibility_network}
     }
    \subfloat[]{
         \includegraphics[width=0.45\textwidth]{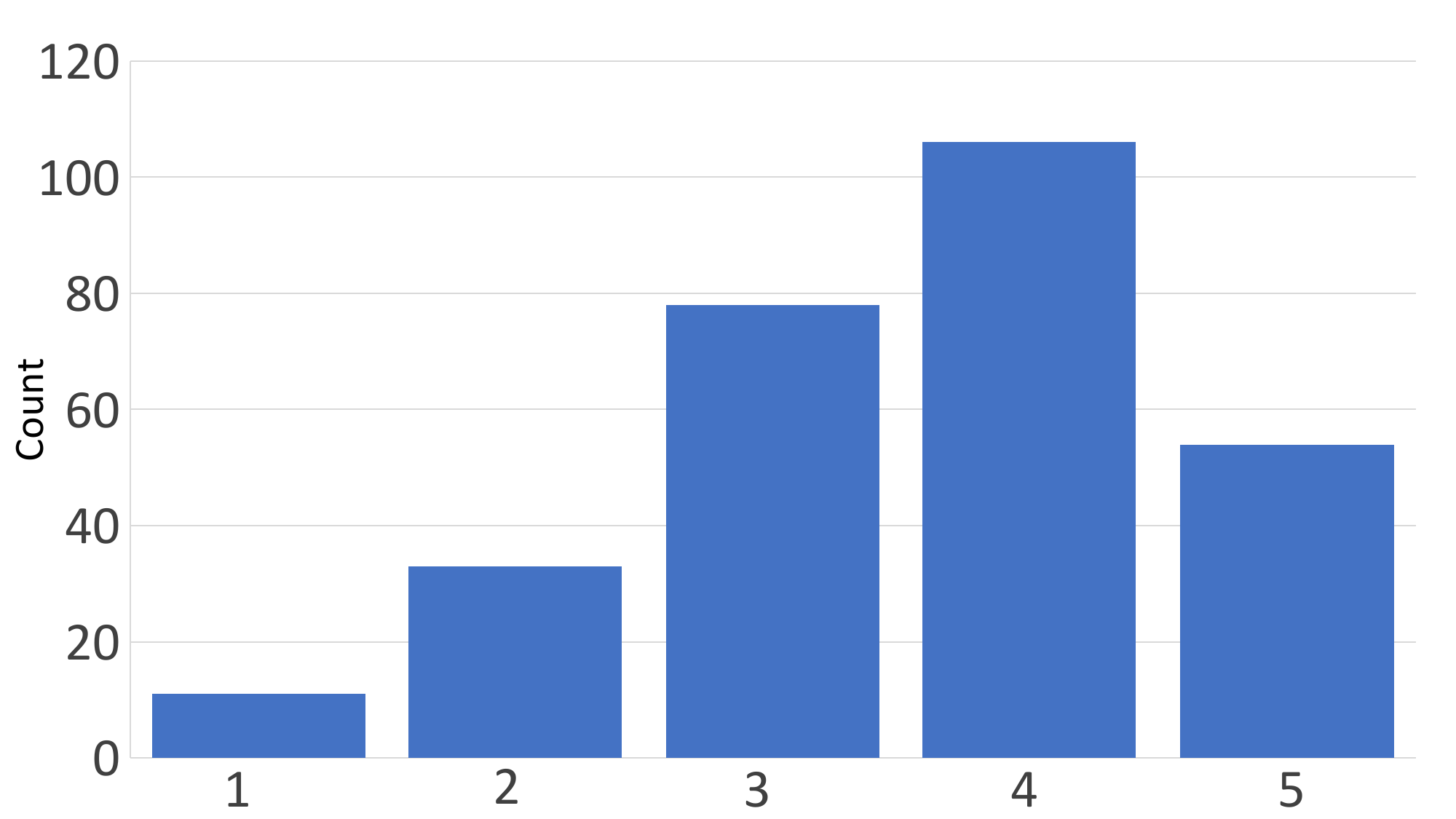}
         \label{fig:confidence}
    }
    \caption{\textbf{The average network credibility scores and the crowd's evaluations/confidence on the causal networks collected in the formative study.} (a) Distribution of Average Network Credibility Scores (0= incorrect link, 3= correct link). (b) Distribution of the crowd's provided confidence scores (1= not confident at all, 5= completely confident).}
    \Description{Figure \ref{fig:evaluation} contains two histograms of Average Network Credibility Scores and Confidence scores on the created causal networks. }
    \label{fig:evaluation}
\end{figure}

Figure \ref{fig:credibility_network} shows the distribution of ANC scores. The majority of created networks were less credible $(ANC<2)$, suggesting a shallow understanding of climate change topics and occasional worker reluctance. Additionally, Figure \ref{fig:confidence} displays the distribution of Average Confidence Scores. Less than 15\% of cases showed low confidence, while nearly 57\% indicated high confidence. These scores reflect the level of certainty associated with the crowd workers' network creations.


The \textit{Aggregated Evaluation} of the data we collected highlights the sparsity of the combined network, the tendency of less credible networks, and the presence of low confidence of the crowd workers in their own created network. These observations motivated us to dive deeper into the data and find the rationales behind them.  
A closer look at the most popular causal relations revealed that widely acknowledged scientific facts were reflected in the highly voted relations. However, there were also spurious relations, such as ``increasing solar radiation → more intense heatwaves'' and ``increasing solar radiation → increasing temperature''. The identified causal relations frequently lacked essential aspects of the climate change issue and resembled arguments used by climate change deniers who attribute it solely to natural factors like the sun \cite{nasa_sun}. These findings also align with the results of \cite{leiserowitz2019climate}, which indicate that people tend to view climate change as an effect of natural variability. It is essential to note that these unfocused relations received lower confidence scores than the more focused ones mentioned above (about 10\% lower).

\begin{table}[]
\resizebox{\textwidth}{!}{%
\begin{tabular}{lcclcc}
\hline
\textbf{Bogus Cause} &
  \multicolumn{1}{l}{\textbf{\begin{tabular}[c]{@{}l@{}}More intense \\ heat waves\end{tabular}}} &
  \multicolumn{1}{l}{\textbf{\begin{tabular}[c]{@{}l@{}}Less intense\\ heat waves\end{tabular}}} &
  \textbf{True Cause} &
  \multicolumn{1}{l}{\textbf{\begin{tabular}[c]{@{}l@{}}More intense \\ heat waves\end{tabular}}} &
  \multicolumn{1}{l}{\textbf{\begin{tabular}[c]{@{}l@{}}Less intense \\ heat waves\end{tabular}}} \\ \hline
\begin{tabular}[c]{@{}l@{}}Increasing\\ solar radiation\end{tabular} &
  (True) &
  (False) &
  \begin{tabular}[c]{@{}l@{}}More \\ fossil fuel burning\end{tabular} &
  True &
  False \\
\begin{tabular}[c]{@{}l@{}}Decreasing\\ solar radiation\end{tabular} &
  True &
  False &
  \begin{tabular}[c]{@{}l@{}}Less \\ fossil fuel burning\end{tabular} &
  False &
  True \\ \hline
\textbf{Bogus Cause} &
  \multicolumn{1}{l}{\textbf{\begin{tabular}[c]{@{}l@{}}Increasing\\ temperature\end{tabular}}} &
  \multicolumn{1}{l}{\textbf{\begin{tabular}[c]{@{}l@{}}Decreasing\\ temperature\end{tabular}}} &
  \textbf{True Cause} &
  \multicolumn{1}{l}{\textbf{\begin{tabular}[c]{@{}l@{}}Increasing\\ temperature\end{tabular}}} &
  \multicolumn{1}{l}{\textbf{\begin{tabular}[c]{@{}l@{}}Decreasing\\ temperature\end{tabular}}} \\ \hline
\begin{tabular}[c]{@{}l@{}}Increasing\\ solar radiation\end{tabular} &
  (True) &
  (False) &
  \begin{tabular}[c]{@{}l@{}}More \\ fossil fuel burning\end{tabular} &
  True &
  False \\
\begin{tabular}[c]{@{}l@{}}Decreasing\\ solar radiation\end{tabular} &
  True &
  False &
  \begin{tabular}[c]{@{}l@{}}Less \\ fossil fuel burning\end{tabular} &
  False &
  True
\end{tabular}%
}
\caption{\textbf{The trial matrices for the two (synonymous) outcome variables heat wave and temperature.} In each matrix the left matrix captures the bogus case and the right is the true cause.}
\label{tab:trial-matrix-5-6}
\end{table}
\subsection{Formulating the Trial Matrices}
\label{subsec:formulating-trial-matrices}
For the outcome variable ``increasing temperature'' (synonymous with ``more intense heat waves'' in climate science \cite{heatwaves}), we identified two competing causes: the simple bogus cause ``increasing solar radiation'' and the true cause ``more fossil fuel burning.'' The trial matrices for both outcome variables, as shown in Table \ref{tab:trial-matrix-5-6}, can be constructed following a similar approach as in Table \ref{fig:trialmat-all}.

The left trial matrix represents the classic bogus cause of ``increasing solar radiation.'' We have added the parentheses to the former since this is a truly imaginary cause as the solar radiation is not really increasing. It indicates that regardless of whether solar radiation increases or not, there will be more intense heat waves. The second column for the opposite outcome, ``less intense heat waves,'' is set to false for both conditions since this outcome is not observed in real life or simulations. The trial matrix on the right represents the true cause of ``more fossil fuel burning.'' It is scientifically established that ``more fossil fuel burning'' causes ``more intense heat waves,'' while ``less fossil fuel burning'' eliminates the outcome. The second column, ``less intense heat waves,'' reflects the inverse relationship, as expected in a genuine causal relationship.

It is important to note that our crowd-sourced tool does not provide values for each cell in these trial matrices. Our focus is not on conducting formal experiments with complete matrices, but rather on designing experiments to assess the degree of causal illusion in a general population. We compare the magnitudes of the upper left cells in the trial matrices for the bogus and true causes to gauge the level of causal illusion.  Most principled work on causal illusions also largely focuses on these types of results. Additionally, we identify critical links in the causal chain and assess potential knowledge gaps, which are discussed in Section \ref{subsubsec:illusion_quant}.

\subsection{Causal Illusion Detection}\label{potential_causal_illusion_1}
The low ANC scores in Section  \ref{subsec:aggregated} indicate a lack of credibility among the crowd-created networks. We explore this further by analyzing the concept of \textit{Causal Illusion}.

\subsubsection{Causal Illusions}

\begin{figure}
    \centering
    \includegraphics[width=\textwidth]{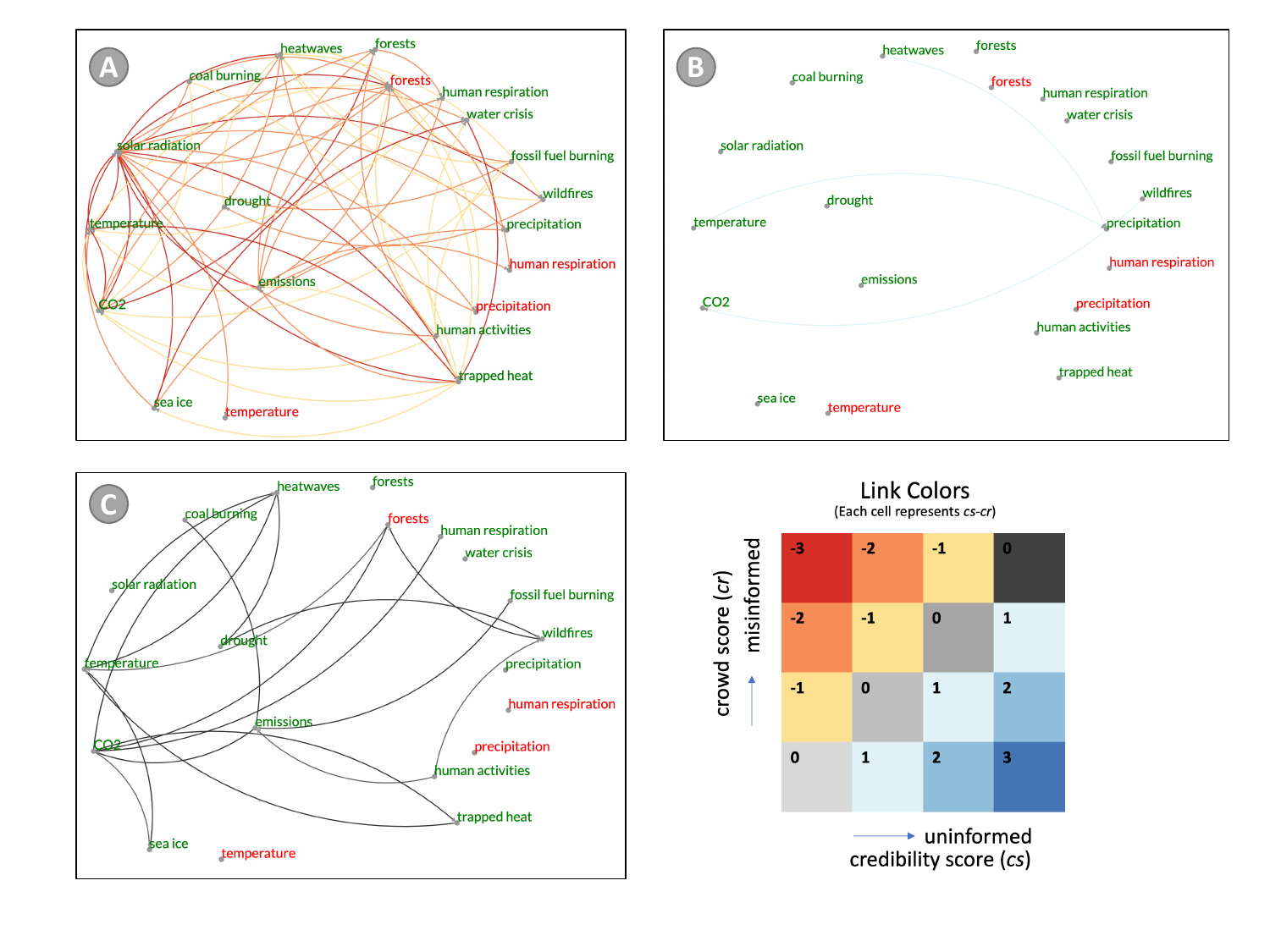}
    \caption{\textbf{The discrepancy networks generated from combined crowd network and ground truth network in the formative study.} Each link color represents the discrepancy between the crowd and ground truth for that specific causal relation. The link colors denote the degree of discrepancy or illusion and the type (\textit{being misinformed} or \textit{being oblivious}). (A) shows the cases of potentially misinformed links, (B) shows the cases of potentially oblivious links, and (C) shows the cases where the crowd correctly predicted the credibility scores. The legend table (row represents crowd score, column represents credibility score) on the right shows the discrepancy/illusion score $(cs-cr)$ (in each cell)  and the corresponding color. credibility score increases from left to right ($\rightarrow$), whereas the crowd score increases from bottom to top ($\uparrow
$). Only the significant links (total vote $\geq$ 4) and attributes are visible. }
\Description{Figure \ref{fig:discrepancy} shows three different discrepancy networks and one legend table describing the link colors.}
    \label{fig:discrepancy}
\end{figure}

In Figure \ref{fig:discrepancy}, we present the \textit{Discrepancy Network} that reveals various levels of discrepancies between the crowd and ground truth networks. The \textit{discrepancy/illusion score} is determined by subtracting the \textit{credibility score} $(cs)$ from the \textit{crowd score} $(cr)$ for each link. A non-zero discrepancy score indicates the presence of a causal illusion. The \textit{crowd score (cr)} is calculated by normalizing the link's total votes, considering the long-tailed distribution of causal links. To assign \textit{crowd scores (cs)} to each link, we utilize a modified equal depth/frequency binning technique. Both \textit{crowd scores (cs)} and \textit{credibility scores (cs)} range from 0 to 3. 
 
Each link color represents either the state of having a causal illusion (being \textit{misinformed} or \textit{oblivious}) or correct (no causal illusion). The colors for these links are coded by the legend on the right. The upper left colors label the state of being misinformed, the lower right colors label the state of being oblivious, and the grey diagonal labels scores where there was agreement. For the latter, we use shading to indicate credibility.

The dominance of \textit{misinformed} links is evident, outnumbering \textit{oblivious} and \textit{correct} links. However, analyzing different discrepancy/illusion scores (Table \ref{tab:stats-discrepancy}) reveals that correct links constitute almost half of all links (139 out of 281). Notably, when considering significant/visible links, correct links still account for over a quarter (18 out of 61). The lowest grey level shading represents links with minimal votes and is not displayed in the discrepancy network.

\begin{table}[t]
\begin{tabular}{cllll}
\toprule
\multicolumn{1}{l}{\textbf{Type}}& \textbf{Link Color}      &\textbf{Discrepancy Score (\textit{cs-cr})} & \textbf{Count (All)} & \textbf{Count (Visible)} \\ \midrule
\multirow{3}{*}{Misinformed} &\color{mydarkred}{Red}&-3                        & 16          & 16              \\
&\color{myorange}{Orange}& -2                        & 28          & 28              \\
&\color{myyellow}{Yellow}&-1                        & 77          & 21              \\ \midrule
\multirow{4}{*}{Correct}&\color{mydarkestgrey}{Grey(Darkest)}&0 (cs = 3)                & 14          & 14              \\
& \color{mydarkergrey}{Grey(Darker)}&0 (cs = 2)                & 4           & 4               \\
&\color{mygrey}{Grey}&0 (cs = 1)                & 11          & 0               \\
&\color{mylightgrey}{Grey(Light)}&0 (cs = 0)                & 110         & 0               \\ \midrule
\multirow{3}{*}{Oblivious}&\color{mylightestblue}{Blue(Very Light)}&1                         & 16          & 3               \\
&\color{mylightblue}{Blue(Light)}& 2                         & 4           & 0               \\
&\color{myblue}{Blue}&3                         & 1           & 0\\ \midrule
\multicolumn{1}{l}{}& &Total                     & 281         & 86                          \\ \bottomrule
\end{tabular}
\caption{\textbf{Statistics of various discrepancy/illusion scores in the formative study.}}
\Description{Table \ref{tab:stats-discrepancy} contains the statistics of various discrepancy scores.}
\label{tab:stats-discrepancy}
\end{table}

Consistently higher ratios of \textit{misinformed} links emphasize the crowd's vulnerability to misinformation compared to being \textit{oblivious/uninformed}. This vulnerability is particularly evident in the prevalence of misconceptions related to \textcolor{mygreen}{increasing solar radiation} among climate-change deniers (Figure \ref{fig:discrepancy}). Furthermore, other interesting and noteworthy examples, along with more details and analyses, are provided in Appendix \ref{formative-misconception-cases}. The occurrence of \textit{uninformed} crowd judgments is negligible for significant causal links, underscoring the importance of debunking misinformation and promoting accurate information on complex topics like climate change.

\subsubsection{Causal Illusion Quantification.}\label{subsubsec:illusion_quant} 
\begin{figure}
    \centering
    \includegraphics[width=0.8\textwidth]{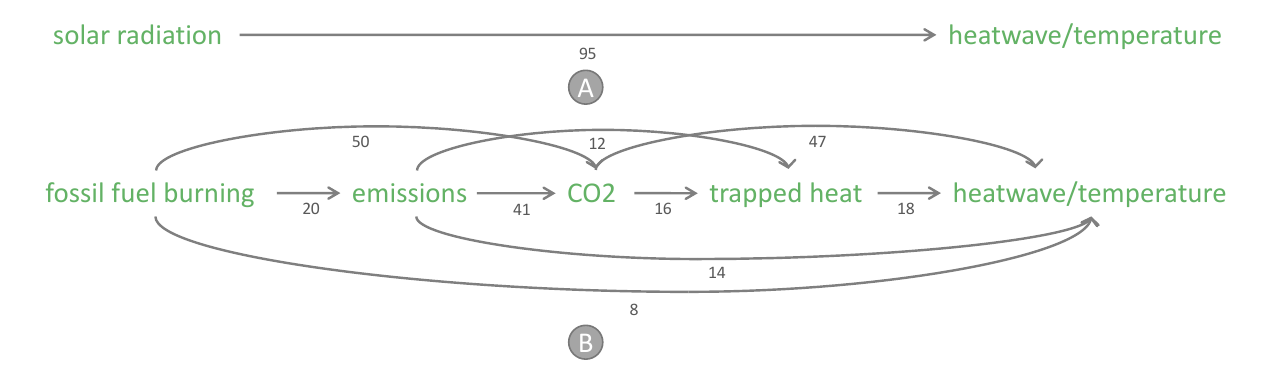}
    \caption{\textbf{Extracted causal chains in the formative study for the bogus cause and the true cause.} Panel A is the simple link that connects the bogus cause with the outcome. This is the link selected by the fraction of the crowd who fell victim to the causal illusion. Panel B are several pathways of varying complexity by which the crowd has linked the true cause with the outcome. The numbers below each link are the number of votes the link received in the compound causal network. The green color indicates rising values.}
    \label{fig:fourth}
\end{figure}

Using the votes obtained for each causal link, we quantified the presence of causal illusion by comparing the bogus cause to the true cause. In Figure \ref{fig:fourth}, we presented the causal links associated with two trial matrices in Table \ref{tab:trial-matrix-5-6}, demonstrating the contrasting results. Panel A depicted the simple bogus relation \textcolor{mygreen}{increased solar radiation} $\rightarrow$ \textcolor{mygreen}{increased temperature}/\textcolor{mygreen}{more intense heat waves} which received a combined vote of 95. In contrast, Panel B showed the more complex true relation starting with \textcolor{mygreen}{increased fossil fuel burning} and ends with \textcolor{mygreen}{increased temperature}/\textcolor{mygreen}{more intense heat waves}. Various traversals were observed in which the wisdom of the crowd navigated this chain. The most accurate 4-hop path (\textcolor{mygreen}{increased fossil fuel burning} $\rightarrow$ \textcolor{mygreen}{increasing emissions} $\rightarrow$ \textcolor{mygreen}{increasing CO2} $\rightarrow$ \textcolor{mygreen}{more trapped heat} $\rightarrow$ \textcolor{mygreen}{increased temperature}/\textcolor{mygreen}{more intense heat waves}) received a support of 16 votes, assuming the weakest link as the defining one, or 23.75 votes on average. Additionally, 3-hop and 2-hop paths also emerged, each offering varying degrees of accuracy. Evaluating the degree of illusion, we found that the ratio of votes for the bogus cause to the true cause ranged from 95/23.75=4 to 95/14=5.94, indicating a stronger belief in the causal illusion. On the other hand, if we compare the dominant true causal knowledge with the bogus cause we get a ratio of about 2.

\subsection{Design Issues}


\subsubsection{Selection Bias}
One important issue in the initial interface was the potential for selection bias due to the pre-determined order and the selective and gradual appearance of attributes in the interface. The order in which attributes are presented can influence participants' perceptions and choices. Further, it is possible that the positioning of certain attributes based on Word2Vec coordinates influenced participants to perceive and associate them more favorably compared to others. This may have contributed to the sparse nature of the combined network (Figure~\ref{fig:adjacency}).

\subsubsection{Longer Completion Time than Anticipated}


The average completion time was 30 minutes, exceeding the estimated duration of 15 minutes, which directly contradicted our design goal, \textbf{DG3}. One reason behind that could be the three rounds required to complete the task. Participants had to perform a total of seven clicks in various parts of the interface to create a single link. The visual representation of node-link diagrams could also contribute to the longer completion times.

\section{Re-designed and Final interface}
\label{sec:survey2}

The formative study produced encouraging results, as the crowd demonstrated the ability to create meaningful networks while exhibiting evidence of various types of causal illusions. However, we also identified two major design issues. Based on these findings, we revised the design of the protocol and interface for collecting causal networks from crowd workers.
%
 %
 We had three specific goals in mind. Firstly, we aimed to eliminate any potential selection bias that may have influenced participant responses.  Secondly, we sought to streamline the data collection process by reducing the number of micro-tasks required and the amount of completion time. Lastly, we decided to enforce a structured process for recruiting participants and performing post-hoc quality control. 
 

\begin{figure}
    \centering
    \includegraphics[width=0.9\textwidth]{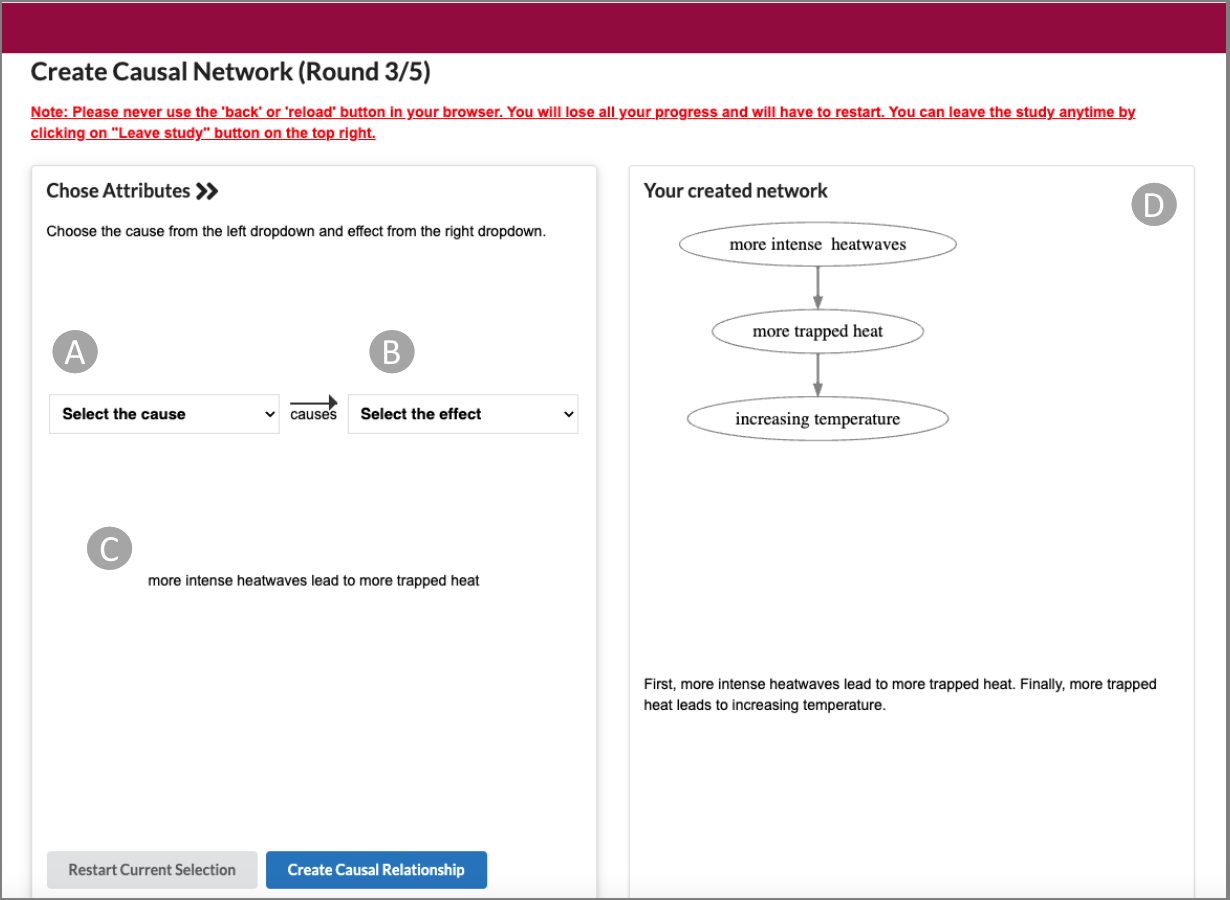}
    \caption{\textbf{Overview of the redesigned collection interface}. This example shows a causal network is being created. A shows the ``cause'' drop-down and B shows the ``effect'' drop-down. C shows the most recently created causal link as text. D contains the emerging causal network along with its textual narrative. }
    \Description{Figure \ref{fig:collection_interface_2} shows a screenshot of the collection interface. It shows a case where a crowd worker is in the process of making a causal network.}
    \label{fig:collection_interface_2}
\end{figure}

\subsection{Collection Interface Modules}\label{subsec:interface_modules}

The redesigned interface utilized a similar web-based technology as our initial interface. We excluded the attribute ``renewable energy'' from our input dataset, as it had a low number of votes in the formative study. The interface also has modules similar to the initial interface. We describe major changes to the modules below.


\subsubsection{Demographics Survey Module}
\label{demographics}


We decided to replace the previous questions gauging participants' self-assessed knowledge and awareness with a more standardized question set since crowdworkers may over or underestimate their knowledge about climate change~\cite{kittur2013future}. We decided to use the Six Americas Super Short Survey (SASSY)~\cite{sassy}. These questions classify individuals into six categories representing different levels of climate change awareness: Alarmed, Concerned, Cautious, Disengaged, Doubtful, and Dismissive.

Note that SASSY captures people's attitudes towards climate change, not their knowledge level. Our target audience is the \textit{general population}, thus neither knowledge nor attitude is an influential factor in the study. Nevertheless, we wanted to recruit people with diverse opinions and beliefs about climate change to validate our method and find a wide range of causal beliefs and illusions. With the lack of a standardized method for measuring knowledge about climate change, we believe SASSY is a reasonable proxy to determine the diversity of our participants' pool. 



\subsubsection{Causal Network Creation Module}
\label{creation}

We made four major changes to the Causal Network Creation Module (see Figure \ref{fig:collection_interface_2}). \textit{First}, instead of presenting the attributes using a node-link diagram, we provide two drop downs for users to select causes and effects (Figure~\ref{fig:collection_interface_2}A and B). This design reduces the possibilities of biases induced by the position of the nodes in a node-link diagram and scales the interface to any number of attributes. \textit{Secondly}, instead of randomly selecting a subset of attributes, we presented all attributes to the users in the drop-downs. This design choice again reduces any possibilities of selection bias. \textit{Thirdly}, since a user can see the whole set of attributes from the beginning, we only asked participants to complete a single session (creating five relations), eliminating the need for the additional two sessions. This design choice significantly reduces the completion time. \textit{Finally}, even though users do not create networks using node-link diagrams, we show the created diagrams to the users using GraphViz~\cite{ellson2002graphviz}, a well-known graph visualization library that has built-in rendering mechanisms (Figure~\ref{fig:collection_interface_2}D).  We also reduced the number of micro-tasks for optimizing the completion time. For example, trend selection was a separate micro-task in the initial interface, but now the trends are embedded with the attribute names. Users do not need to select trends separately, reducing two micro-tasks in total (one for selecting cause and one for effect).
We attached a video to demonstrate the overall workflow. Appendix B.1 describes the workflow in more detail.

\section{Final Study}
\label{sec:experiment2}


We recruited $72$ crowd workers from Amazon Mechanical Turk (AMT) and $60$ workers on Prolific (132 workers in total) to collect causal perceptions on climate change using our final interface. We excluded the work of 31 workers due to incomplete and fraudulent results, totaling to 101 valid responses. Successful crowd workers were compensated \$3.75 on AMT and \$3 on Prolific upon completion. The compensation amount on Prolific was determined automatically based on the estimated completion time, while on AMT we estimated it ourselves. The average completion time was around 12 minutes. The study design and protocol remained similar to our formative study, except for the following changes. We mention the detailed protocol in Appendix \ref{final-detailed-protocol}.

\subsection{Crowd Workers' Validation Process and Post-hoc Quality Control}

\begin{figure}
    \centering
    \includegraphics[width=0.9\textwidth]{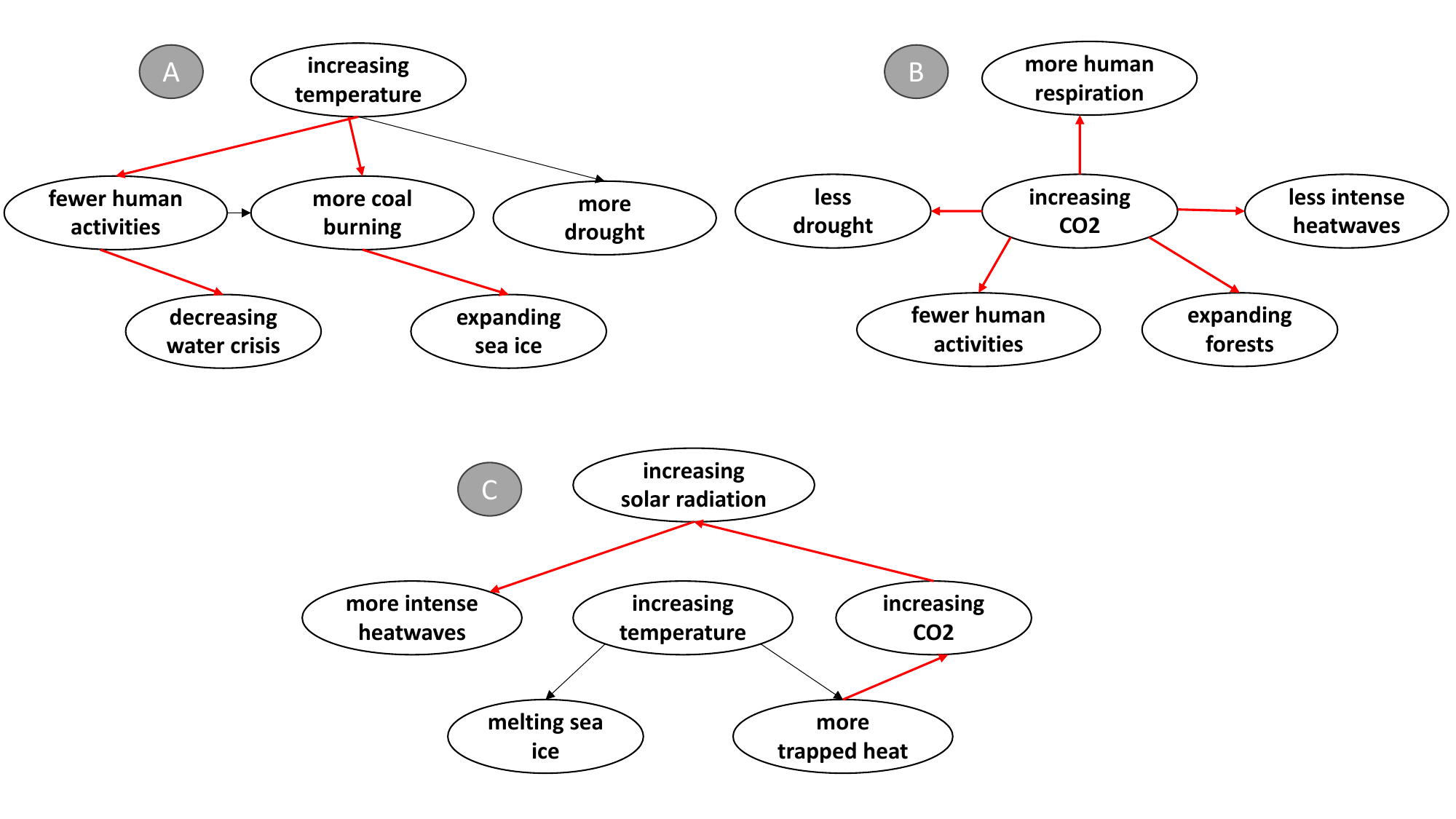}
    \caption{\textbf{Three examples of flagged data collected during the final study}. \textbf{A} and \textbf{B} are fraudulent data that were rejected, whereas \textbf{C} was accepted. Network \textbf{A} has 4 and Network \textbf{B} has 5 links that show patterns of inconsistencies and indications of random or careless selection (marked in \textcolor{red}{red}). For example, in Network \textbf{A}, ``increasing temperature $\rightarrow$ more coal burning'' is a spurious link, the scientific evidence supports the opposite relation. In Network \textbf{B}, ``increasing CO2 $\rightarrow$ less intense heatwaves'' is another example of such a spurious link, as increasing CO2 levels contribute to intensified heatwaves, not diminished ones. Links inside one network also mostly do not bear any relevance to each other in Networks \textbf{A} and \textbf{B}. On the other hand, Network \textbf{C}, has three scientifically invalid links (marked in \textcolor{red}{red}), but all links are fitting and express a coherent narrative.}
    \label{fig:fraudulent}
\end{figure}

 In the final study, in contrast to the formative study, we implemented a rigorous quality control process to ensure data integrity. First, we noticed from the formative study that while we designed the \textit{crediblilty score} to measure causal beliefs and illusions, it can also be used to filter out potential fraudulent or random causal relations. For example, an excessive number (3 or more out of 5) of non-credible links (credibility score = 0) could indicate randomly created relations or fraudulent behaviors. We flagged such relations and crowd workers in the final study. Then, the research team manually reviewed each case.
    During the review, we examined patterns of inconsistencies, repeated responses, and indications of random or careless selection. Based on the manual review, we identified and excluded workers who produced unreliable or fraudulent data. We provide two examples of such data in Figure~\ref{fig:fraudulent}-A and B. Note that the lack of credibility (a score of 0) does not necessarily mean fraudulent behavior. We accepted submission as long as the overall network exhibited a pattern of understanding and relevance, even if some links had a credibility score of 0 (Figure~\ref{fig:fraudulent}-C). We interpreted these results as beliefs stemming from potentially flawed understanding, which are of potential interest to us, and accepted the results. Thus, the combined networks had links with credibility scores spanning from 0 to 3.

\begin{figure}
    \centering
    \includegraphics[width=0.9\textwidth]{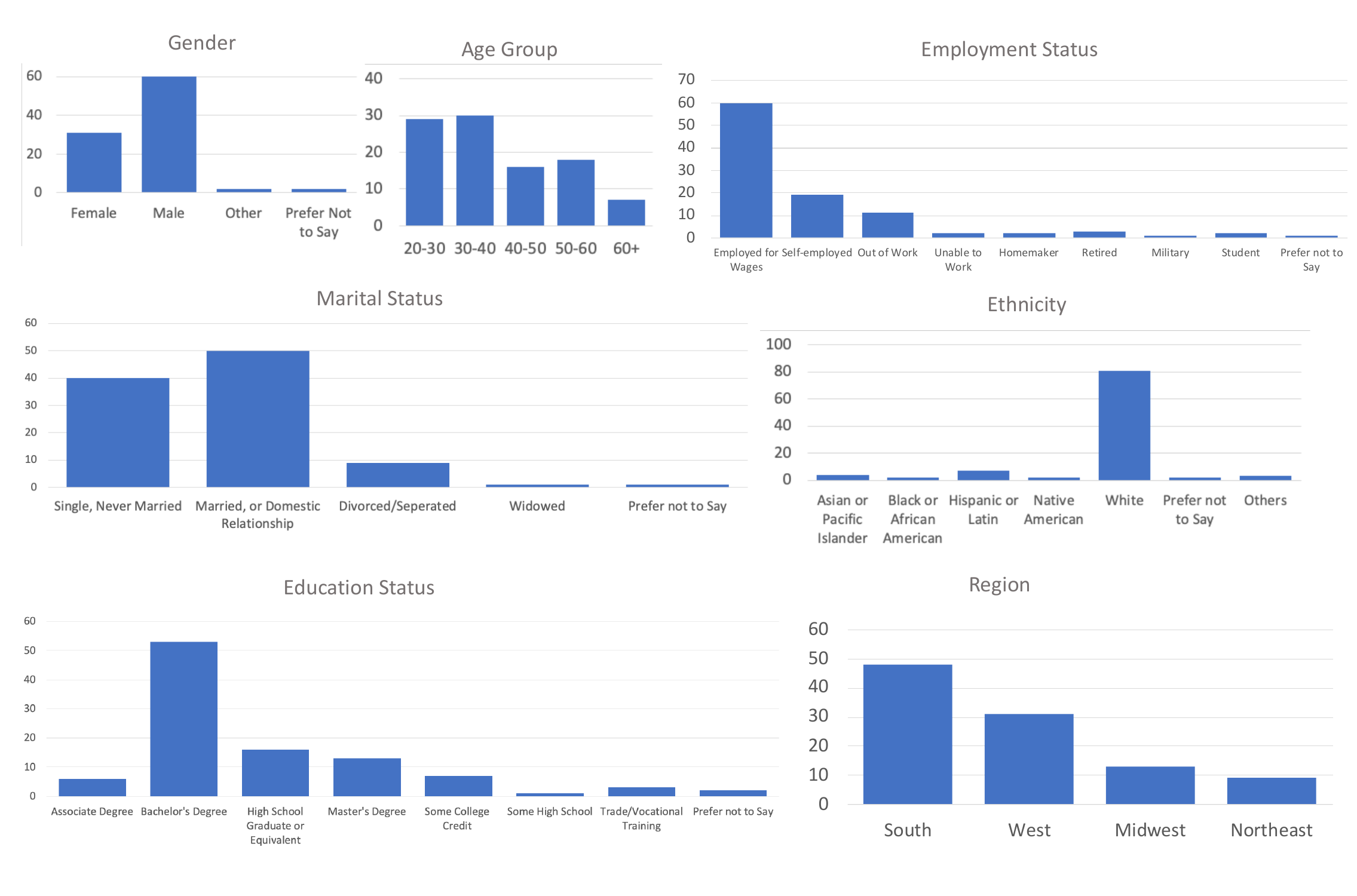}
    \caption{\textbf{Demographics of the crowd workers in the final study.} Y-axes represent counts for each category. 
    }
    \Description{Figure \ref{fig:demographics_2} shows a collection of bar charts, each containing statistics of various demographic variables. The variables in order from the top-left are: Gender, Age Group, Employment Status, Marital Status, Region, Ethnicity, and Education Status.}
    \label{fig:demographics-2}
\end{figure}

 We conducted the study in phases, 10 participants at a time, allowing us to calculate node exploration in the aggregated network and perform quality control at each phase. All 31 workers whose results were rejected due to incomplete or fraudulent data were from AMT. Thus, we decided to conduct the rest of the study on Prolific, as it is known for its more stringent participant vetting process and higher-quality data. We ensured all participants finished the study exactly once. Therefore, every participant from AMT and Prolific constructed their own network from the ground up without building upon one another's work. This led us to have 101 valid workers. We present various aspects of the crowd workers' demographics in Figure \ref{fig:demographics-2}. Geographically, we only collected results from the United States. 

\begin{figure}
    \centering
    \includegraphics[width=0.9\textwidth]{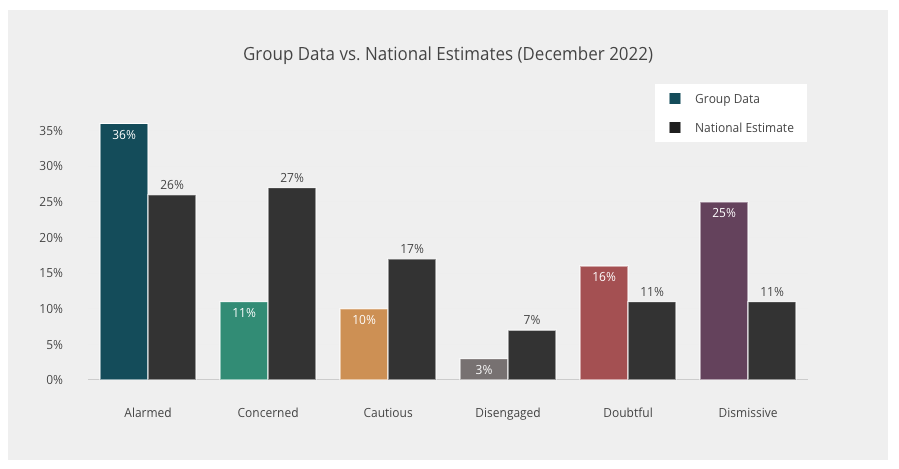}
    \caption{\textbf{Crowd workers' view on climate change, compared to the national estimates.} The black bars represent the national estimates and the colored bars represent our sample. The chart was generated using the group scoring tool provided by \cite{sassy}. }
   
    \label{fig:sassy-national-averages}
\end{figure}

\subsection{Crowd Workers' View on Climate Change}
We employed the ``\textit{Six Americas Super Short SurveY (SASSY)}'' Group Scoring Tool to segment our participants into different groups based on their responses to climate change~\cite{sassy}. We present the results in Figure \ref{fig:sassy-national-averages}. The majority of workers (36) were ``Alarmed'' and expressed high concern about climate change. There were also significant numbers of workers in the ``Dismissive'' group (25) who held dismissive or denying attitudes. Other categories included ``Concerned'' (11), ``Cautious'' (10), ``Doubtful'' (16), and ``Disengaged'' (3). The group scoring tool also allowed us to compare the groups to national averages (Figure \ref{fig:sassy-national-averages}). Compared to national averages from December 2022, we observed some variations in the distribution of attitudes toward climate change. Nevertheless, all groups were present in our participant pool, reflecting a range of perspectives and attitudes toward climate change.



\subsection{Stopping Criteria}

We conducted the study in phases (10 participants at a time). After each phase, we analyzed the combined network created by the crowd and compared it with the previous phases. We also checked how many nodes/attributes have been explored in the combined network. We stopped the study once the network was saturated, displaying minimal changes from previous phases. We present the final node exploration status in Appendix \ref{final-node-exploration}. Additionally, we considered the representation of the participants in terms of the SASSY groups. We wanted representatives from each group.

\section{Findings from the Final Study}
\label{sec:evaluation}
The analysis pipeline in the final study remained consistent with that of the formative study. We present the findings below.

\subsection{Aggregated Evaluation}\label{subsec:aggregated-2}

\begin{figure}
    \centering
    \includegraphics[width=0.85\textwidth]{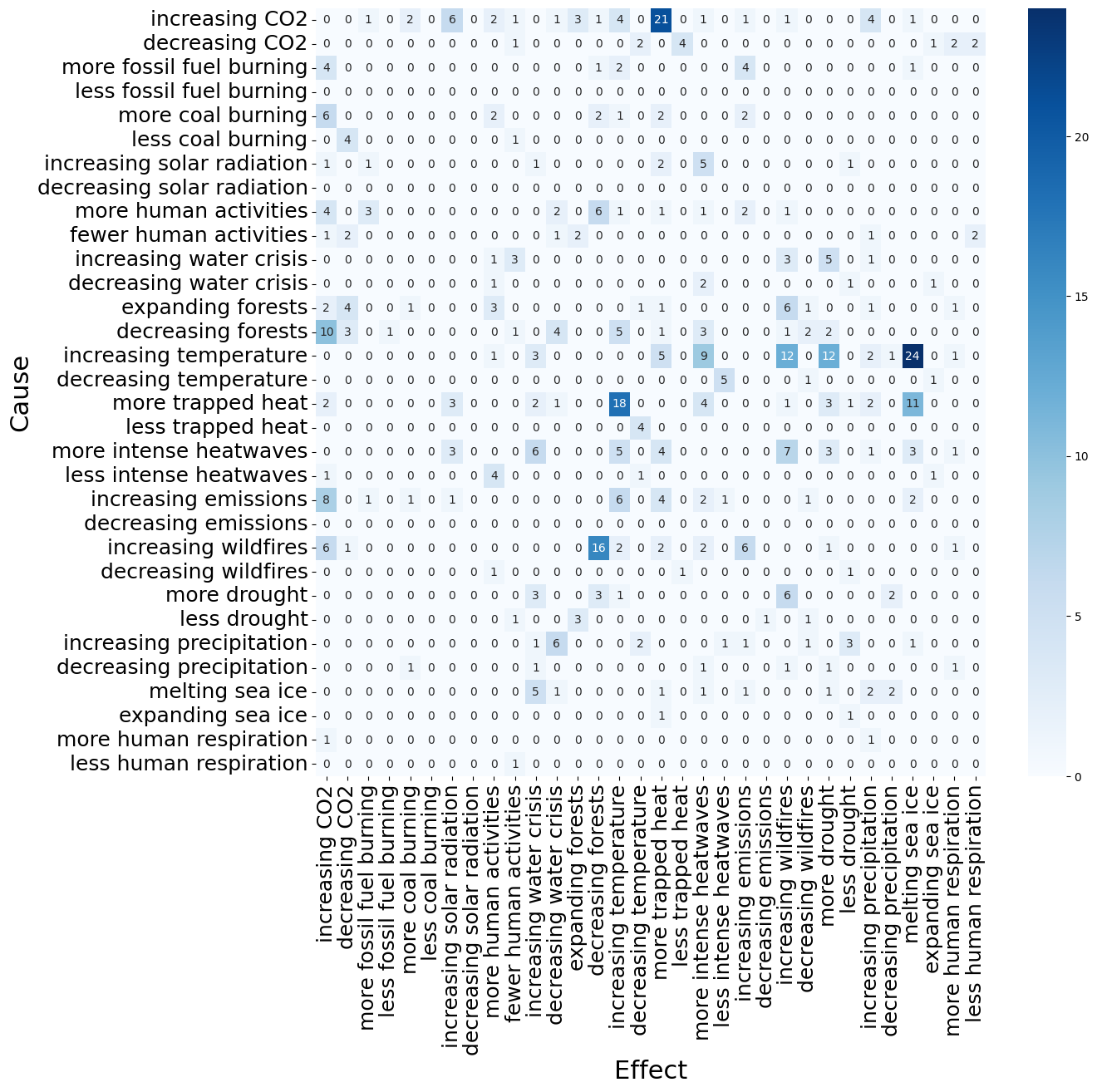}
    \caption{\textbf{The Adjacency Matrix Heatmap Representation of the 101 causal networks collected in the final study.} The cell values represent the total number of votes for that specific causal relation.}
    \Description{Figure \ref{fig:adjacency2} demonstrates an adjacency matrix heatmap representation, where both rows and columns denote causal attributes and each cell shows the total number of votes of specific causal relations with numbers and color intensity.}
    \label{fig:adjacency2}
\end{figure}
\subsubsection{Combined Network and Total Votes Per Link}
We present the adjacency matrix of the 101 combined networks in Figure \ref{fig:adjacency2}. The top-3 most voted links were, ``increasing temperature $\rightarrow$ melting sea ice'', ``increasing CO2 $\rightarrow$ more trapped heat,'' and ``more trapped heat $\rightarrow$ increasing temperature'' with 24, 21, and 18 votes, respectively. The combined network is sparse, and most relations have zero votes, similar to the formative study. The Pearson Correlation Coefficient $(r)$ between the total votes and the credibility scores for the causal links is 0.63 with $p<10^{-105}$. This indicates a stronger alignment between the crowd and expert consensus compared to the formative study. Although the level of consensus remains moderate, the stronger correlation suggests that the crowd was able to create networks that contain more scientifically accurate causal links in general.
\subsubsection{Average Network Credibility Scores (ANC) and Average Confidence Scores (AC)}

\begin{figure}
     \centering
     \subfloat[]{
        \includegraphics[width=0.45\textwidth]{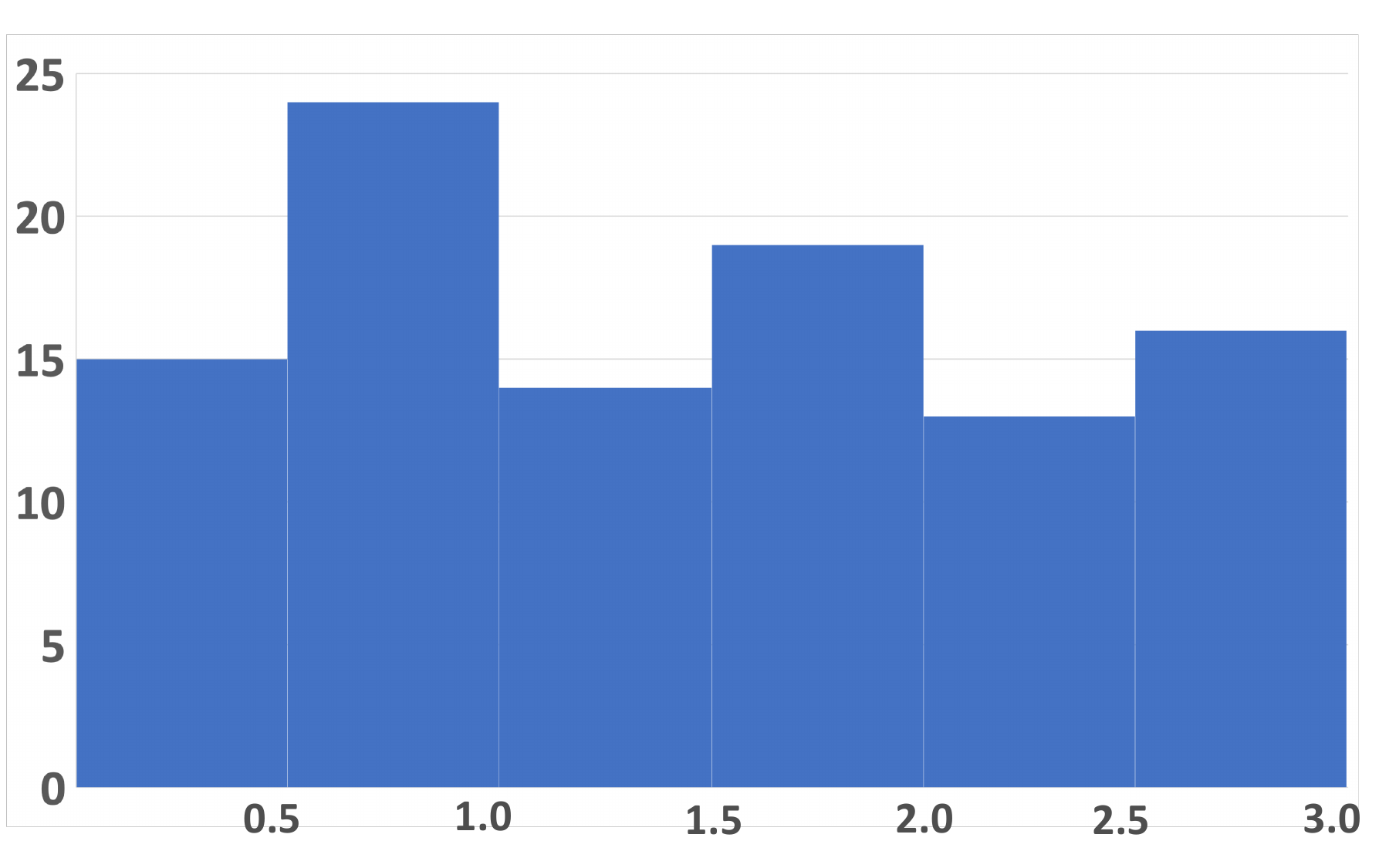}
        \label{fig:credibility_network_2}
     }
    \subfloat[]{
         \includegraphics[width=0.45\textwidth]{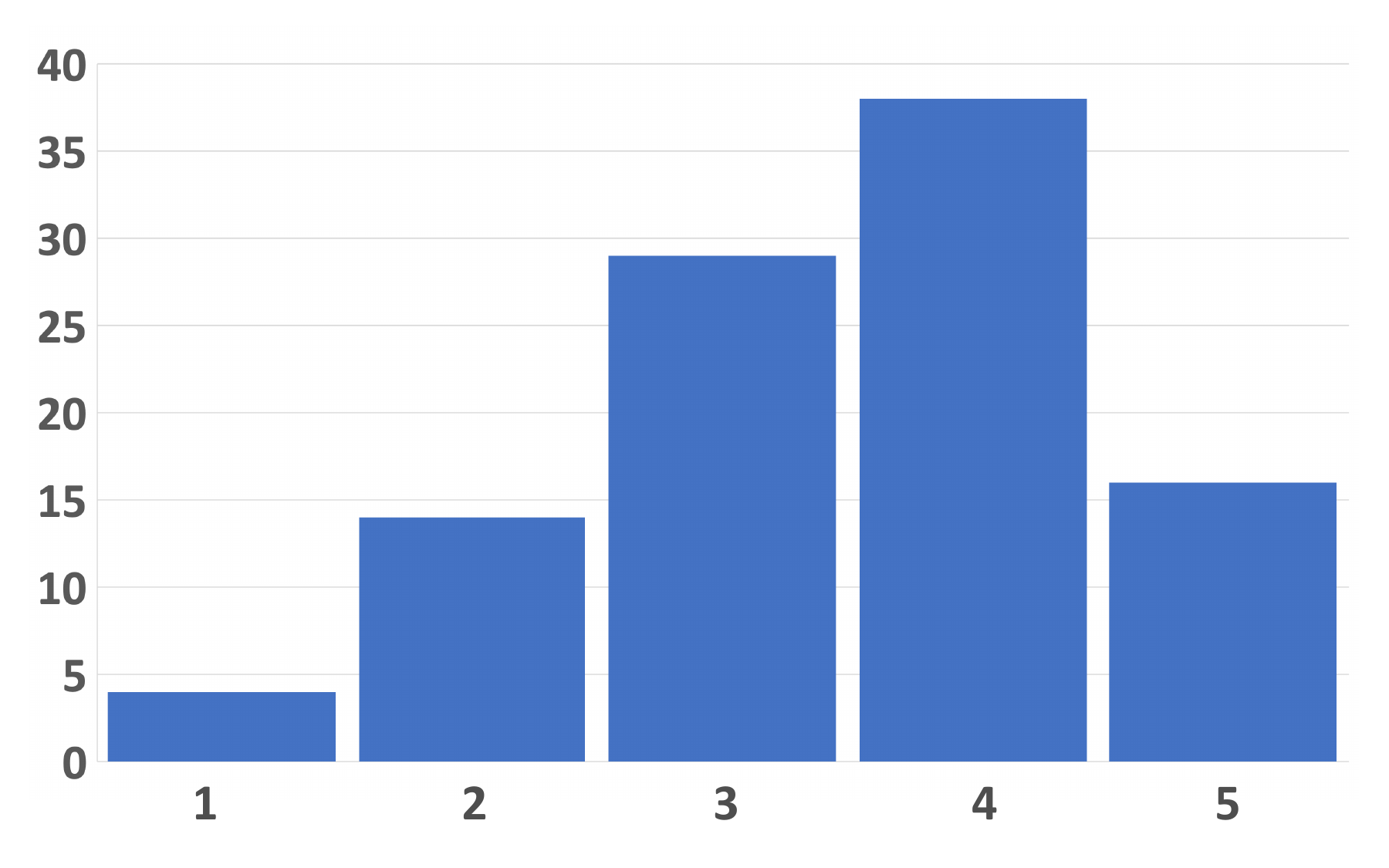}
         \label{fig:confidence_2}
    }
    \caption{\textbf{The actual average network credibility scores and the crowd's evaluations/confidence on the causal networks collected in the final study.} (a) Distribution of Average Network Credibility Scores (0= incorrect link, 3= correct link). (b) Distribution of the crowd's provided confidence scores (1= not confident at all, 5= completely confident).}
    \Description{Figure \ref{fig:evaluation2} contains two histograms of Average Network Credibility Scores and Confidence scores on the created causal networks. }
    \label{fig:evaluation2}
\end{figure}

Figure \ref{fig:credibility_network_2} shows the distribution of ANC scores, which reflects that the majority of the crowd workers created less credible networks $(ANC<2)$ in general. We present the confidence scores in Figure \ref{fig:confidence_2}, which reflects that in less than 18\% of the cases, the workers did not feel very confident (Confidence Score 1-2) in their own created network and nearly 54\% of the time, they felt positively confident (Confidence Score 4-5). 


\subsection{A Closer Look into the Most Popular Causal Relations} \label{subsec:closer-look-2}
Similar to the formative study, we mention the most noteworthy aspect of the most popular causal links below.
\begin{itemize}
    \item  Similar to the formative study, the most voted causal relations in the final study reflected widely acknowledged scientific facts.  The top relation (24 votes, average confidence (AC) = 3.5, credibility score (CS) = 3) was ``increasing temperature $\rightarrow$ melting sea ice''. Other top relations attributed the ``more trapped heat'' to ``increasing CO2'' (21 votes, 3.5 AC, 3 CS) and ``increasing temperature'' to ``more trapped heat (18 votes, 3.3 AC, 3 CS).
    \item There was evidence of recognizing underlying factors of climate change, such as linking ``more human activities'' to ``decreasing forests'' (6 votes, 3.7 AC, 2 CS) and ``increasing CO2'' (4 votes, 3.5 AC, 2 CS) and considering ``increasing CO2'' as an effect of ``decreasing forests'' (10 votes, 3.7 AC, 3 CS) and ``increasing emissions'' (8 votes, 3.5 AC, 3.5 CS) \cite{forestcarbon, carbondioxide}.
    \item There was also evidence of physical understanding; many marked that ``increasing wildfire'' leads to ``decreasing forests'' (16 votes, 3.5 AC, 3 CS) and ``increasing emissions'' (6 votes, 3.7 AC, 3 CS) and ``increasing CO2'' (6 votes, 3.5 AC, 2 CS), and, in turn, ``increasing wildfires'' is caused by ``increasing temperature'' (12 votes, 3.2 AC, 2 CS) and ``more intense heatwaves'' (7 votes, 4.14, 2 CS). 
    \item Among the causal links that received at least 10 votes, all were highly credible links (CS=3) except for 2 links: ``increasing temperature'' $\rightarrow$ ``increasing wildfires'' and ``more trapped heat'' $\rightarrow$ ``melting sea ice'' (CS=2). Interestingly, these links also received the lowest AC scores (3.2 and 2.8, respectively) compared to others, exhibiting a parallel trend to the formative study findings. While these links are partially correct, they lack crucial mediators. It is important to acknowledge that wildfires are primarily caused by dry weather resulting from \textit{drought}, which is a consequence of rising temperatures. Similarly, the consistent trapping of heat contributes to \textit{increasing temperatures}, which leads to the melting of sea ice~\cite{droughtwildfire, meltingseaice}.
    
\end{itemize}

\subsection{Causal Illusion Detection}\label{potential_causal_illusion}

\begin{figure}
    \centering
    \includegraphics[width=\textwidth]{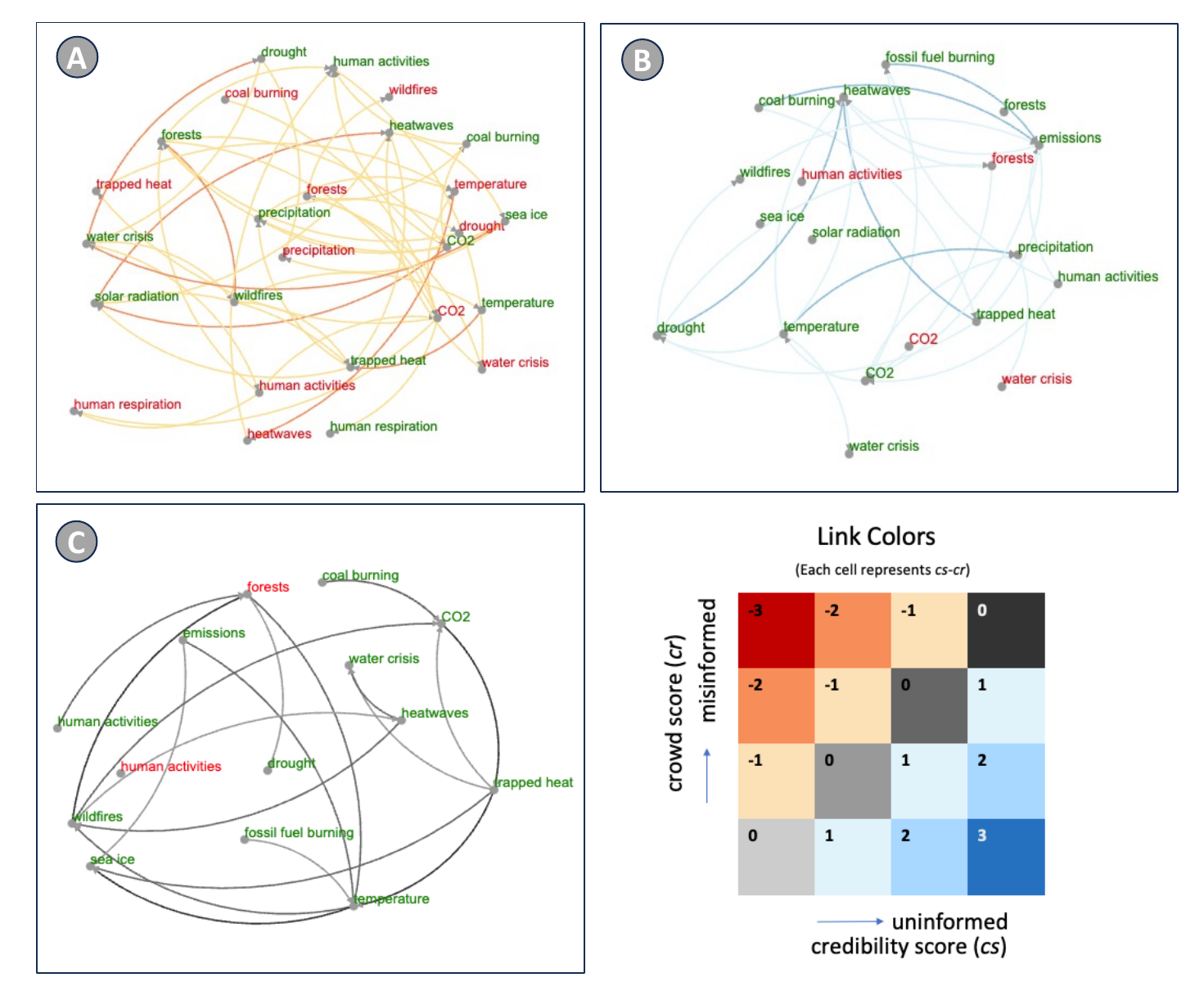}
    \caption{\textbf{The discrepancy networks generated from the combined crowd network and the ground truth network in the final study.} Each link color represents the discrepancy between the crowd and ground truth for that specific causal relation. The link colors denote the degree of discrepancy or illusion and the type (\textit{being misinformed} or \textit{being oblivious}). (A) shows the cases of potentially misinformed links, (B) shows the cases of potentially oblivious links, and (C) shows the cases where the crowd correctly predicted the credibility scores.
    }
    \Description{Figure \ref{fig:discrepancy-2} shows three different discrepancy networks and one legend table describing the link colors.}
    \label{fig:discrepancy-2}
\end{figure}
In Figure \ref{fig:discrepancy-2} we present the \textit{Discrepancy Network} along with various levels of discrepancies. 
%
%
Similar to the formative study, the ratio of \textit{misinformed} links consistently stays higher (47 out of 92 visible links). In contrast to the formative study, the results of the current study reveal a multitude of cases where the crowd demonstrated a significant degree of obliviousness, with 26 out of 92 visible links exhibiting this phenomenon. We present the statistics of various values of different discrepancy/illusion scores in Table \ref{tab:stats-discrepancy-2}. 

\begin{table}[t]
\begin{tabular}{cllll}
\toprule
\multicolumn{1}{l}{\textbf{Type}}& \textbf{Link Color}      &\textbf{Discrepancy Score (\textit{cs-cr})} & \textbf{Count (All)} & \textbf{Count (Visible)} \\ \midrule
\multirow{3}{*}{Misinformed} &\color{mydarkred}{Red}&-3                        & 0          & 0              \\
&\color{myorange}{Orange}& -2                        & 7          & 7              \\
&\color{myyellow}{Yellow}&-1                        & 40          & 40              \\ \midrule
\multirow{4}{*}{Correct}&\color{mydarkestgrey}{Grey(Darkest)}&0 (cs = 3)                & 4          & 4              \\
& \color{mydarkergrey}{Grey(Darker)}&0 (cs = 2)                & 9           & 9               \\
&\color{mygrey}{Grey}&0 (cs = 1)                & 6          & 6               \\
&\color{mylightgrey}{Grey(Light)}&0 (cs = 0)                & 75        & 0               \\ \midrule
\multirow{3}{*}{Oblivious}&\color{mylightestblue}{Blue(Very Light)}&1                         & 27          & 21               \\
&\color{mylightblue}{Blue(Light)}& 2                         & 9           & 5               \\
&\color{myblue}{Blue}&3                         & 4           & 0\\ \midrule
\multicolumn{1}{l}{}& &Total                     & 181         & 92                          \\ \bottomrule
\end{tabular}
\caption{\textbf{Statistics of various discrepancy/illusion scores in the final study.}}
\Description{Table \ref{tab:stats-discrepancy} contains the statistics of various discrepancy scores.}
\label{tab:stats-discrepancy-2}
\end{table}

\begin{figure}
    \centering
    \includegraphics[width=0.9\textwidth]{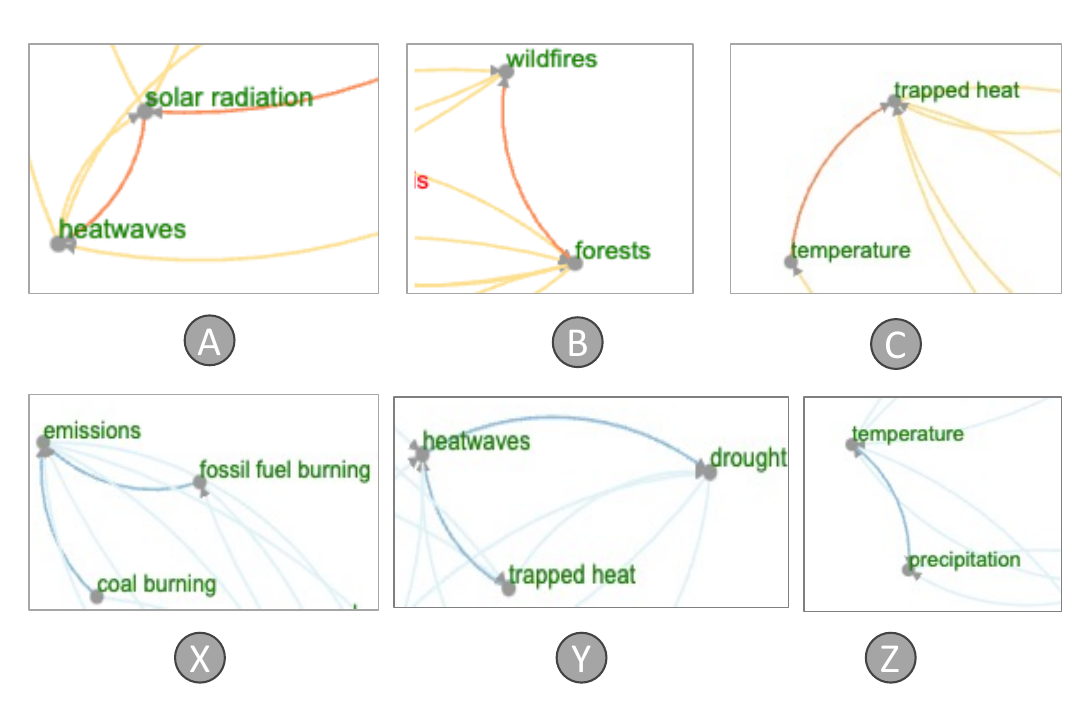}
    \caption{\textbf{\textit{Misinformed} and \textit{Oblivious} cases in discrepancy network.}}
    \Description{Figure \ref{fig:misconception-cases-2} shows seven screenshots of various interesting Misconception and Obliviousness cases.}
    \label{fig:misconception-cases-2}
\end{figure}

 In Figure \ref{fig:misconception-cases-2}, we present some noticeable cases within the discrepancy network. The crowd selected \textcolor{mygreen}{increasing solar radiation} as a cause for \textcolor{mygreen}{more intense heatwaves}, this misconception is consistent among the participants of both studies (Figure \ref{fig:misconception-cases-2}-A). The crowd also seemed to think that \textcolor{mygreen}{expanding forest} causes \textcolor{mygreen}{increasing wildfires} (Figure \ref{fig:misconception-cases-2}-B), whereas, the relationship between expanding forests and increased wildfires is more nuanced. Factors such as climate conditions, human activities, and forest management practices play significant roles in determining wildfire risk~\cite{wildfirecauses, droughtwildfire}.

Several intriguing instances of obliviousness were observed within the crowd as depicted in Figure \ref{fig:misconception-cases-2}-X, Y, Z. Notably, the crowd exhibited a preference for attributing emissions to \textcolor{mygreen}{increasing wildfires}, which is indeed a valid relationship. However, there was a comparative disregard for the influence of \textcolor{mygreen}{fossil fuel burning} and \textcolor{mygreen}{coal burning} on emissions. Additionally, a certain level of unawareness was evident regarding the interplay between variables such as drought, precipitation, and temperature-related factors like trapped heat and heatwaves. This highlights the need to prioritize the dissemination of accurate information concerning the impact of rising temperatures on drought conditions.


\begin{figure}
    \centering
    \includegraphics[width=0.9\textwidth]{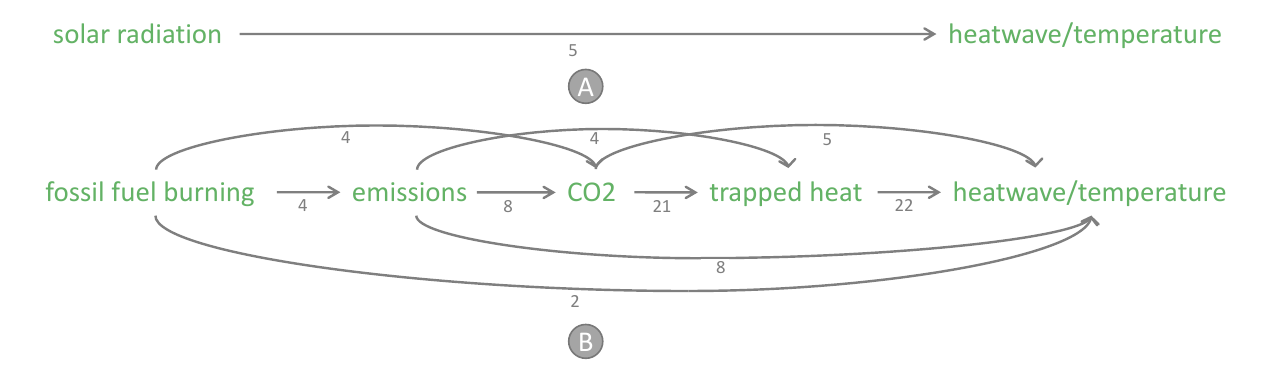}
    \caption{\textbf{Replication of Figure~\ref{fig:fourth} in the final study.} 
    }
    \label{fig:causal illusions}
\end{figure}
\subsubsection{Causal Illusion Quantification.}\label{subsubsec:illusion_quant_2}


Figure \ref{fig:causal illusions} illustrates the causal links associated with the two trial matrices described in Section \ref{subsec:formulating-trial-matrices} (Table \ref{tab:trial-matrix-5-6}). The strongest 4-hop path (\textcolor{mygreen}{more fossil fuel burning} $\rightarrow$ \textcolor{mygreen}{increasing emissions} $\rightarrow$ \textcolor{mygreen}{increasing CO2} $\rightarrow$ \textcolor{mygreen}{more trapped heat} $\rightarrow$ \textcolor{mygreen}{increased temperature}/\textcolor{mygreen}{more intense heat waves}) received 4 votes (based on the weakest link criterion) or 13.75 votes (based on the average link criterion). The optimal 3-hop path (\textcolor{mygreen}{more fossil fuel burning} $\rightarrow$ \textcolor{mygreen}{increasing CO2} $\rightarrow$ \textcolor{mygreen}{more trapped heat} $\rightarrow$ \textcolor{mygreen}{increased temperature}/\textcolor{mygreen}{more intense heat waves}) garnered 4 votes (weakest link criterion) or 15.67 votes (average link criterion). The best 2-hop path (\textcolor{mygreen}{more fossil fuel burning} $\rightarrow$ \textcolor{mygreen}{increasing CO2} $\rightarrow$ \textcolor{mygreen}{increased temperature}/\textcolor{mygreen}{more intense heat waves}) obtained 4 votes (weakest link criterion) or 4.5 votes (average link criterion). Lastly, the simplest 1-hop path (\textcolor{mygreen}{more fossil fuel burning} $\rightarrow$ \textcolor{mygreen}{increasing temperature/more intense heat waves}) received 2 votes. To assess the degree of illusion between the bogus cause and the true cause, we can examine the vote ratios. For the most accurate 4-hop path, this ratio is 5/4 = 1.25 for the weakest link criterion and 5/13.75 = 0.36 for the average criterion. Although the weakest link criterion suggests a mild degree of illusion, the number of votes is too low. Another encouraging finding is that a significant number of workers correctly identified the direct cause of \textcolor{mygreen}{increased temperature}/\textcolor{mygreen}{more intense heat waves}, which is \textcolor{mygreen}{more trapped heat} (21 votes).

\subsection{Interface Usability and Knowledge}
\begin{figure}
         \centering
         \includegraphics[width=\textwidth]{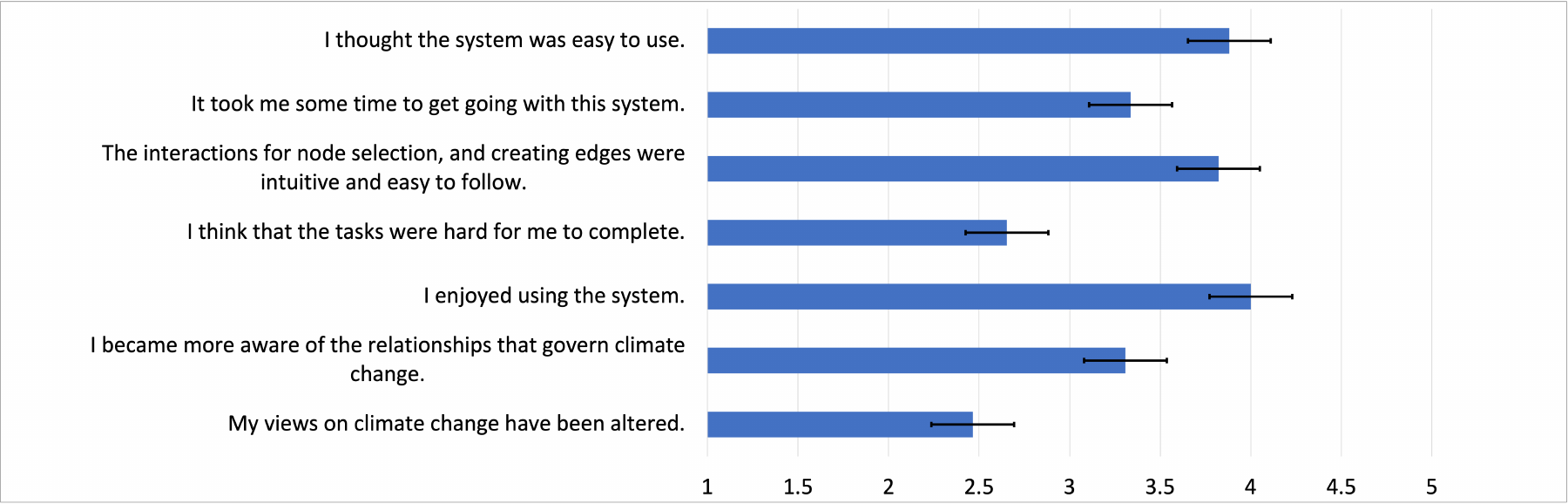}
         \caption{\textbf{Average ratings provided by the crowd workers of the final study on 7 subjective usability and knowledge-related statements.} Error bars represent standard errors. (1=Strongly Disagree, 5=Strongly Agree).}
        \Description{Figure \ref{fig:usability_box} contains a horizontal bar chart along with error bars expressing the average ratings of the crowd workers on the usability and knowledge-related statements.} \label{fig:usability_box}
\end{figure}

Figure \ref{fig:usability_box} presents crowd workers' agreement on the usability and learning statements. The results reflect an overall positive outlook toward the usability of our system. For the knowledge-related statements, the majority of workers felt neutral to positive towards the statement ``I became more aware of the relationships that govern climate change'' and disagreed with the statement ``My views on climate change have been altered''. We think this happened because they might not have felt comfortable agreeing to such a strong statement.

\section{Discussion, Limitations, and Future work}
\label{sec:discussion}
 We briefly summarize the study findings, design implications, and limitations of our work below. 





\subsection{Reflecting on the User Studies} 
The two studies show that Belief Miner is data-driven and agnostic to the method used to collect the data. Regardless of the collection mechanism, Belief Miner can extract causal illusions given a set of causal networks collected from the crowd and ground truth collected from experts. Thus, we were able to apply the same method in both studies, even though the underlying protocol and interfaces were different. 

We found a wide range of causal beliefs and illusions in the studies. Some findings from the studies align while others do not. For example, in both studies, participants with causal illusions assigned lower confidence scores to their networks. However, we could not replicate some results from the formative study in the final study. For example, we did not find the causal illusion reported in Figure~\ref{fig:fourth} from the formative study in the final study. Several factors could contribute to this phenomenon: the redesigned interface, the participants' pool, or the changes in the general knowledge about climate change within the timeframe of the studies (Fall 2021 to Spring 2023).


\subsection{Design Implications}

\subsubsection{Belief Miner as an Intervention against Causal Illusion.}
In our current system, we collect the causal beliefs from crowd workers through the interactive visual interface and detect the causal illusions later. Both of our studies show the phenomenon of people rating their own causal networks containing spurious causal links with low confidence scores once they see the complete picture. We see this as an indication of people correcting or educating themselves while they see the externalized version of their mental causal model. Motivated by this, we envision using Belief Miner as an intervention tool against causal illusion. Several crowdsourcing methods could be useful here. For example, seeing other people's beliefs often positively affect a crowdsourcing task~\cite{yen2021narratives+, kittur2011crowdforge}. In our case, we can expose the controversy or disagreement about a causal relation at the time of the experiment by looking at the networks already created by others. Another potential solution is enabling peer review~\cite{DBLP:conf/cscw/WhitingGGGGBMCR17}, allowing crowdworkers to provide feedback to each other. Other potential solutions include automatically extracting digestible scientific documents related to the relevant causal attributes. We believe these additions will facilitate informed decision-making and will promote scientific thinking which is identified as the best defense against causal illusions~\cite{matute2015illusions}. 

\subsubsection{From Causal Illusion to Misconception}
Belief Miner investigates the concept of causal illusion, which is related to people's inherent bias to draw connections between coincidental events. A closely related concept is misconception, which is the inaccurate or wrong interpretation of concepts~\cite{shi2021towardsemi}. The terms misconception and misinformation are often used interchangeably. While misconception generally comes from a lack of knowledge, misinformation is often deliberately created for deception and spread intentionally or unintentionally~\cite{wu2019misinformation}. Existing misconception and misinformation discovery methods mainly rely on natural language processing (NLP) and machine learning (ML). These methods fall under the broader category of content-based detection~\cite{mazid2022climatechangemyths,benamira2019semisupervised}, context-based detection~\cite{Kwon_Cha_2014}, propagation-based detection~\cite{kim2019homogeneity}, etc.  We believe our method could provide a realistic tool to measure misconceptions where content-based analysis is not feasible.

\subsubsection{What Kinds of Causal Illusions Appear Together and Who Falls Victim to Them?}
 Another interesting future direction is utilizing the demographic profile of the crowdworkers and the cases of potential causal illusions to identify groups of illusions along with the population groups and their geo-locations who fall victim to them. One way of doing this is to find all possible cases and cluster them based on their structures, i.e., the attributes, type of the link, and level of causal illusions (introduced in Section \ref{potential_causal_illusion_1}). Another potential way of finding such groups can be using frequent itemset mining algorithms such as Apriori \cite{agrawal1994fast}. In the case of the causal belief dataset, each causal network made by a particular worker consisting of a collection of causal relations and their casual illusion levels represents one transaction. These clusters/frequent itemsets can potentially bring out population groups susceptible to that specific groups of causal illusions. Eventually, these population groups can represent different schools of thought. Therefore, we think it is worth exploring this idea as an extension of our current methodology.

\subsubsection{Belief Miner for Causal ML}
Prior research has used causal crowdsourcing as a way to develop training datasets for machine learning~\cite{yen2021narratives+}. While our work focuses on behavioral analysis, we believe detecting causal illusions would be useful for machine learning too. If not detected, causal illusions can lead to incorrect decision-making and machine-intelligence.

\section{Conclusion}
\label{sec:conclusion}
We presented \textit{Belief Miner}, a methodology for collecting and evaluating the crowd's causal beliefs and discovering causal illusions. We developed two interactive interfaces to collect such beliefs, where people can create small causal networks to create a large causal network collectively. Two separate crowdsourcing studies show that all participants successfully created the small networks using our interactive interface, except for a few dropouts. Our evaluation methodology can find potential discrepancies and illusions in the causal relations created by the crowd. We hope our work will start the discussion on different methods of analyzing people's causal opinions and beliefs and what they may reveal about the people themselves.

\begin{acks}
    This research was partially supported by NSF grant IIS 1941613 and IIS 1527200.
\end{acks}

\bibliographystyle{ACM-Reference-Format}
\bibliography{references}


\begin{thebibliography}{84}


\ifx \showCODEN    \undefined \def \showCODEN     #1{\unskip}     \fi
\ifx \showDOI      \undefined \def \showDOI       #1{#1}\fi
\ifx \showISBNx    \undefined \def \showISBNx     #1{\unskip}     \fi
\ifx \showISBNxiii \undefined \def \showISBNxiii  #1{\unskip}     \fi
\ifx \showISSN     \undefined \def \showISSN      #1{\unskip}     \fi
\ifx \showLCCN     \undefined \def \showLCCN      #1{\unskip}     \fi
\ifx \shownote     \undefined \def \shownote      #1{#1}          \fi
\ifx \showarticletitle \undefined \def \showarticletitle #1{#1}   \fi
\ifx \showURL      \undefined \def \showURL       {\relax}        \fi
\providecommand\bibfield[2]{#2}
\providecommand\bibinfo[2]{#2}
\providecommand\natexlab[1]{#1}
\providecommand\showeprint[2][]{arXiv:#2}

\bibitem[Agley et~al\mbox{.}(2022)]%
        {agley2022quality}
\bibfield{author}{\bibinfo{person}{Jon Agley}, \bibinfo{person}{Yunyu Xiao},
  \bibinfo{person}{Rachael Nolan}, {and} \bibinfo{person}{Lilian
  Golzarri-Arroyo}.} \bibinfo{year}{2022}\natexlab{}.
\newblock \showarticletitle{Quality control questions on Amazon’s Mechanical
  Turk (MTurk): A randomized trial of impact on the USAUDIT, PHQ-9, and GAD-7}.
\newblock \bibinfo{journal}{\emph{Behavior research methods}}
  \bibinfo{volume}{54}, \bibinfo{number}{2} (\bibinfo{year}{2022}),
  \bibinfo{pages}{885--897}.
\newblock


\bibitem[Agrawal et~al\mbox{.}(1994)]%
        {agrawal1994fast}
\bibfield{author}{\bibinfo{person}{Rakesh Agrawal},
  \bibinfo{person}{Ramakrishnan Srikant}, {et~al\mbox{.}}}
  \bibinfo{year}{1994}\natexlab{}.
\newblock \showarticletitle{Fast algorithms for mining association rules}. In
  \bibinfo{booktitle}{\emph{Proc. 20th int. conf. very large data bases,
  VLDB}}, Vol.~\bibinfo{volume}{1215}. Santiago, Chile,
  \bibinfo{pages}{487--499}.
\newblock


\bibitem[Allan(1980)]%
        {allan1980note}
\bibfield{author}{\bibinfo{person}{Lorraine~G Allan}.}
  \bibinfo{year}{1980}\natexlab{}.
\newblock \showarticletitle{A note on measurement of contingency between two
  binary variables in judgment tasks}.
\newblock \bibinfo{journal}{\emph{Bulletin of the Psychonomic Society}}
  \bibinfo{volume}{15}, \bibinfo{number}{3} (\bibinfo{year}{1980}),
  \bibinfo{pages}{147--149}.
\newblock


\bibitem[Benamira et~al\mbox{.}(2019)]%
        {benamira2019semisupervised}
\bibfield{author}{\bibinfo{person}{Adrien Benamira}, \bibinfo{person}{Benjamin
  Devillers}, \bibinfo{person}{Etienne Lesot}, \bibinfo{person}{Ayush~K. Ray},
  \bibinfo{person}{Manal Saadi}, {and} \bibinfo{person}{Fragkiskos~D.
  Malliaros}.} \bibinfo{year}{2019}\natexlab{}.
\newblock \showarticletitle{Semi-Supervised Learning and Graph Neural Networks
  for Fake News Detection}. In \bibinfo{booktitle}{\emph{Proceedings of the
  2019 IEEE/ACM International Conference on Advances in Social Networks
  Analysis and Mining}} (Vancouver, British Columbia, Canada)
  \emph{(\bibinfo{series}{ASONAM '19})}. \bibinfo{publisher}{Association for
  Computing Machinery}, \bibinfo{address}{New York, NY, USA},
  \bibinfo{pages}{568–569}.
\newblock
\showISBNx{9781450368681}
\urldef\tempurl%
\url{https://doi.org/10.1145/3341161.3342958}
\showDOI{\tempurl}


\bibitem[Berenberg and Bagrow(2018)]%
        {berenberg2018efficient}
\bibfield{author}{\bibinfo{person}{Daniel Berenberg} {and}
  \bibinfo{person}{James~P. Bagrow}.} \bibinfo{year}{2018}\natexlab{}.
\newblock \showarticletitle{Efficient Crowd Exploration of Large Networks: The
  Case of Causal Attribution}.
\newblock \bibinfo{journal}{\emph{Proc. {ACM} Hum. Comput. Interact.}}
  \bibinfo{volume}{2}, \bibinfo{number}{{CSCW}} (\bibinfo{year}{2018}),
  \bibinfo{pages}{24:1--24:25}.
\newblock
\urldef\tempurl%
\url{https://doi.org/10.1145/3274293}
\showDOI{\tempurl}


\bibitem[Bernstein et~al\mbox{.}(2010)]%
        {bernstein2010soylent}
\bibfield{author}{\bibinfo{person}{Michael~S. Bernstein}, \bibinfo{person}{Greg
  Little}, \bibinfo{person}{Robert~C. Miller}, \bibinfo{person}{Bj{\"{o}}rn
  Hartmann}, \bibinfo{person}{Mark~S. Ackerman}, \bibinfo{person}{David~R.
  Karger}, \bibinfo{person}{David Crowell}, {and} \bibinfo{person}{Katrina
  Panovich}.} \bibinfo{year}{2010}\natexlab{}.
\newblock \showarticletitle{Soylent: a word processor with a crowd inside}. In
  \bibinfo{booktitle}{\emph{Proceedings of the 23rd Annual {ACM} Symposium on
  User Interface Software and Technology, New York, NY, USA, October 3-6,
  2010}}. \bibinfo{publisher}{{ACM}}, \bibinfo{address}{New York, NY, USA},
  \bibinfo{pages}{313--322}.
\newblock
\urldef\tempurl%
\url{https://doi.org/10.1145/1866029.1866078}
\showDOI{\tempurl}


\bibitem[Bigham et~al\mbox{.}(2015)]%
        {bigham2015human}
\bibfield{author}{\bibinfo{person}{Jeffrey~P Bigham},
  \bibinfo{person}{Michael~S Bernstein}, {and} \bibinfo{person}{Eytan Adar}.}
  \bibinfo{year}{2015}\natexlab{}.
\newblock \showarticletitle{Human-computer interaction and collective
  intelligence}.
\newblock \bibinfo{journal}{\emph{Handbook of collective intelligence}}
  \bibinfo{volume}{57} (\bibinfo{year}{2015}).
\newblock


\bibitem[Blanco et~al\mbox{.}(2012)]%
        {blanco2012mediating}
\bibfield{author}{\bibinfo{person}{Fernando Blanco}, \bibinfo{person}{Helena
  Matute}, {and} \bibinfo{person}{Miguel A.~Vadillo}.}
  \bibinfo{year}{2012}\natexlab{}.
\newblock \showarticletitle{Mediating role of activity level in the depressive
  realism effect}.
\newblock  (\bibinfo{year}{2012}).
\newblock


\bibitem[Blanco et~al\mbox{.}(2013)]%
        {blanco2013interactive}
\bibfield{author}{\bibinfo{person}{Fernando Blanco}, \bibinfo{person}{Helena
  Matute}, {and} \bibinfo{person}{Miguel~A Vadillo}.}
  \bibinfo{year}{2013}\natexlab{}.
\newblock \showarticletitle{Interactive effects of the probability of the cue
  and the probability of the outcome on the overestimation of null
  contingency}.
\newblock \bibinfo{journal}{\emph{Learning \& Behavior}} \bibinfo{volume}{41},
  \bibinfo{number}{4} (\bibinfo{year}{2013}), \bibinfo{pages}{333--340}.
\newblock


\bibitem[Blanzieri(2012)]%
        {blanzieri2012role}
\bibfield{author}{\bibinfo{person}{Enrico Blanzieri}.}
  \bibinfo{year}{2012}\natexlab{}.
\newblock \showarticletitle{The role of causal beliefs in technology-supported
  policy}.
\newblock \bibinfo{journal}{\emph{IFAC Proceedings Volumes}}
  \bibinfo{volume}{45}, \bibinfo{number}{10} (\bibinfo{year}{2012}),
  \bibinfo{pages}{171--176}.
\newblock


\bibitem[Bostock et~al\mbox{.}(2011)]%
        {bostock2011d3}
\bibfield{author}{\bibinfo{person}{Michael Bostock}, \bibinfo{person}{Vadim
  Ogievetsky}, {and} \bibinfo{person}{Jeffrey Heer}.}
  \bibinfo{year}{2011}\natexlab{}.
\newblock \showarticletitle{D$^3$ data-driven documents}.
\newblock \bibinfo{journal}{\emph{IEEE transactions on visualization and
  computer graphics}} \bibinfo{volume}{17}, \bibinfo{number}{12}
  (\bibinfo{year}{2011}), \bibinfo{pages}{2301--2309}.
\newblock


\bibitem[Callison-Burch(2009)]%
        {callison2009fast}
\bibfield{author}{\bibinfo{person}{Chris Callison-Burch}.}
  \bibinfo{year}{2009}\natexlab{}.
\newblock \showarticletitle{Fast, cheap, and creative: Evaluating translation
  quality using Amazon’s Mechanical Turk}. In
  \bibinfo{booktitle}{\emph{Proceedings of the 2009 conference on empirical
  methods in natural language processing}}. \bibinfo{publisher}{{ACL}},
  \bibinfo{pages}{286--295}.
\newblock
\urldef\tempurl%
\url{https://aclanthology.org/D09-1030/}
\showURL{%
\tempurl}


\bibitem[Cartwright(2011)]%
        {cartwright2011alternative}
\bibfield{author}{\bibinfo{person}{MM Cartwright}.}
  \bibinfo{year}{2011}\natexlab{}.
\newblock \showarticletitle{Alternative medicine \& the death of Steve Jobs}.
\newblock \bibinfo{journal}{\emph{Psychology Today. October}}
  \bibinfo{volume}{21} (\bibinfo{year}{2011}).
\newblock


\bibitem[Caselli and Inel(2018)]%
        {caselli-inel-2018-crowdsourcing}
\bibfield{author}{\bibinfo{person}{Tommaso Caselli} {and} \bibinfo{person}{Oana
  Inel}.} \bibinfo{year}{2018}\natexlab{}.
\newblock \showarticletitle{Crowdsourcing {S}tory{L}ines: Harnessing the Crowd
  for Causal Relation Annotation}. In \bibinfo{booktitle}{\emph{Proceedings of
  the Workshop Events and Stories in the News 2018}}.
  \bibinfo{publisher}{Association for Computational Linguistics},
  \bibinfo{address}{Santa Fe, New Mexico, U.S.A}, \bibinfo{pages}{44--54}.
\newblock
\urldef\tempurl%
\url{https://aclanthology.org/W18-4306}
\showURL{%
\tempurl}


\bibitem[Chang et~al\mbox{.}(2017)]%
        {DBLP:conf/chi/ChangAK17}
\bibfield{author}{\bibinfo{person}{Joseph~Chee Chang}, \bibinfo{person}{Saleema
  Amershi}, {and} \bibinfo{person}{Ece Kamar}.}
  \bibinfo{year}{2017}\natexlab{}.
\newblock \showarticletitle{Revolt: Collaborative Crowdsourcing for Labeling
  Machine Learning Datasets}. In \bibinfo{booktitle}{\emph{Proceedings of the
  2017 {CHI} Conference on Human Factors in Computing Systems, Denver, CO, USA,
  May 06-11, 2017}}. \bibinfo{publisher}{{ACM}}, \bibinfo{address}{New York,
  NY, USA}, \bibinfo{pages}{2334--2346}.
\newblock
\urldef\tempurl%
\url{https://doi.org/10.1145/3025453.3026044}
\showDOI{\tempurl}


\bibitem[Chiang et~al\mbox{.}(2021)]%
        {chiang2021freetime}
\bibfield{author}{\bibinfo{person}{Chia-En Chiang}, \bibinfo{person}{Yu-Chun
  Chen}, \bibinfo{person}{Fang-Yu Lin}, \bibinfo{person}{Felicia Feng},
  \bibinfo{person}{Hao-An Wu}, \bibinfo{person}{Hao-Ping Lee},
  \bibinfo{person}{Chang-Hsuan Yang}, {and} \bibinfo{person}{Yung-Ju Chang}.}
  \bibinfo{year}{2021}\natexlab{}.
\newblock \showarticletitle{“I Got Some Free Time”: Investigating
  Task-Execution and Task-Effort Metrics in Mobile Crowdsourcing Tasks}. In
  \bibinfo{booktitle}{\emph{Proceedings of the 2021 CHI Conference on Human
  Factors in Computing Systems}} (Yokohama, Japan) \emph{(\bibinfo{series}{CHI
  '21})}. \bibinfo{publisher}{Association for Computing Machinery},
  \bibinfo{address}{New York, NY, USA}, Article \bibinfo{articleno}{648},
  \bibinfo{numpages}{14}~pages.
\newblock
\showISBNx{9781450380966}
\urldef\tempurl%
\url{https://doi.org/10.1145/3411764.3445477}
\showDOI{\tempurl}


\bibitem[Choudhry et~al\mbox{.}(2021)]%
        {Choudhry2021OnceUA}
\bibfield{author}{\bibinfo{person}{Arjun Choudhry}, \bibinfo{person}{Mandar
  Sharma}, \bibinfo{person}{Pramod Chundury}, \bibinfo{person}{Thomas Kapler},
  \bibinfo{person}{Derek W.~S. Gray}, \bibinfo{person}{Naren Ramakrishnan},
  {and} \bibinfo{person}{Niklas Elmqvist}.} \bibinfo{year}{2021}\natexlab{}.
\newblock \showarticletitle{Once Upon A Time In Visualization: Understanding
  the Use of Textual Narratives for Causality}.
\newblock \bibinfo{journal}{\emph{IEEE Transactions on Visualization and
  Computer Graphics}}  \bibinfo{volume}{27} (\bibinfo{year}{2021}),
  \bibinfo{pages}{1332--1342}.
\newblock


\bibitem[Chryst et~al\mbox{.}({[n.\,d.]})]%
        {sassy}
\bibfield{author}{\bibinfo{person}{Breanne Chryst}, \bibinfo{person}{Jennifer
  Marlon}, \bibinfo{person}{Xinran Wang}, \bibinfo{person}{Sander van~der
  Linden}, \bibinfo{person}{Edward Maibach}, \bibinfo{person}{Connie
  Roser-Renouf}, {and} \bibinfo{person}{Anthony Leiserowitz}.}
  \bibinfo{year}{[n.\,d.]}\natexlab{}.
\newblock \bibinfo{title}{Six Americas Super Short Survey (SASSY!)}.
\newblock
  \bibinfo{howpublished}{\url{https://climatecommunication.yale.edu/visualizations-data/sassy/}}.
\newblock
\newblock
\shownote{(Accessed on 07/10/2023)}.


\bibitem[Chung et~al\mbox{.}(2019)]%
        {DBLP:journals/pacmhci/ChungSKHKL19}
\bibfield{author}{\bibinfo{person}{John Joon~Young Chung},
  \bibinfo{person}{Jean~Y. Song}, \bibinfo{person}{Sindhu Kutty},
  \bibinfo{person}{Sungsoo~(Ray) Hong}, \bibinfo{person}{Juho Kim}, {and}
  \bibinfo{person}{Walter~S. Lasecki}.} \bibinfo{year}{2019}\natexlab{}.
\newblock \showarticletitle{Efficient Elicitation Approaches to Estimate
  Collective Crowd Answers}.
\newblock \bibinfo{journal}{\emph{Proc. {ACM} Hum. Comput. Interact.}}
  \bibinfo{volume}{3}, \bibinfo{number}{{CSCW}} (\bibinfo{year}{2019}),
  \bibinfo{pages}{62:1--62:25}.
\newblock
\urldef\tempurl%
\url{https://doi.org/10.1145/3359164}
\showDOI{\tempurl}


\bibitem[Climate.gov({[n.\,d.]})]%
        {carbondioxide}
\bibfield{author}{\bibinfo{person}{Climate.gov}.}
  \bibinfo{year}{[n.\,d.]}\natexlab{}.
\newblock \bibinfo{title}{Climate Change: Atmospheric Carbon Dioxide}.
\newblock
  \bibinfo{howpublished}{\url{https://www.climate.gov/news-features/understanding-climate/climate-change-atmospheric-carbon-dioxide}}.
\newblock
\newblock
\shownote{(Accessed on 07/10/2022)}.


\bibitem[Devlin et~al\mbox{.}(2018)]%
        {devlin2018bert}
\bibfield{author}{\bibinfo{person}{Jacob Devlin}, \bibinfo{person}{Ming-Wei
  Chang}, \bibinfo{person}{Kenton Lee}, {and} \bibinfo{person}{Kristina
  Toutanova}.} \bibinfo{year}{2018}\natexlab{}.
\newblock \showarticletitle{Bert: Pre-training of deep bidirectional
  transformers for language understanding}.
\newblock \bibinfo{journal}{\emph{arXiv preprint arXiv:1810.04805}}
  (\bibinfo{year}{2018}).
\newblock


\bibitem[Dietvorst et~al\mbox{.}(2018)]%
        {dietvorst2018overcoming}
\bibfield{author}{\bibinfo{person}{Berkeley~J Dietvorst},
  \bibinfo{person}{Joseph~P Simmons}, {and} \bibinfo{person}{Cade Massey}.}
  \bibinfo{year}{2018}\natexlab{}.
\newblock \showarticletitle{Overcoming algorithm aversion: People will use
  imperfect algorithms if they can (even slightly) modify them}.
\newblock \bibinfo{journal}{\emph{Management Science}} \bibinfo{volume}{64},
  \bibinfo{number}{3} (\bibinfo{year}{2018}), \bibinfo{pages}{1155--1170}.
\newblock


\bibitem[Dow et~al\mbox{.}(2012)]%
        {dow2012shepherding}
\bibfield{author}{\bibinfo{person}{Steven Dow}, \bibinfo{person}{Anand~Pramod
  Kulkarni}, \bibinfo{person}{Scott~R. Klemmer}, {and}
  \bibinfo{person}{Bj{\"{o}}rn Hartmann}.} \bibinfo{year}{2012}\natexlab{}.
\newblock \showarticletitle{Shepherding the crowd yields better work}. In
  \bibinfo{booktitle}{\emph{{CSCW} '12 Computer Supported Cooperative Work,
  Seattle, WA, USA, February 11-15, 2012}}. \bibinfo{publisher}{{ACM}},
  \bibinfo{address}{New York, NY, USA}, \bibinfo{pages}{1013--1022}.
\newblock
\urldef\tempurl%
\url{https://doi.org/10.1145/2145204.2145355}
\showDOI{\tempurl}


\bibitem[Downs et~al\mbox{.}(2010)]%
        {downs2010your}
\bibfield{author}{\bibinfo{person}{Julie~S. Downs}, \bibinfo{person}{Mandy~B.
  Holbrook}, \bibinfo{person}{Steve Sheng}, {and} \bibinfo{person}{Lorrie~Faith
  Cranor}.} \bibinfo{year}{2010}\natexlab{}.
\newblock \showarticletitle{Are your participants gaming the system?: screening
  mechanical turk workers}. In \bibinfo{booktitle}{\emph{Proceedings of the
  28th International Conference on Human Factors in Computing Systems, {CHI}
  2010, Atlanta, Georgia, USA, April 10-15, 2010}}. \bibinfo{publisher}{{ACM}},
  \bibinfo{address}{New York, NY, USA}, \bibinfo{pages}{2399--2402}.
\newblock
\urldef\tempurl%
\url{https://doi.org/10.1145/1753326.1753688}
\showDOI{\tempurl}


\bibitem[Drought.gov(2022)]%
        {droughtwildfire}
\bibfield{author}{\bibinfo{person}{Drought.gov}.}
  \bibinfo{year}{2022}\natexlab{}.
\newblock \bibinfo{title}{Wildfire Management}.
\newblock
\newblock
\urldef\tempurl%
\url{https://www.drought.gov/sectors/wildfire-management}
\showURL{%
\tempurl}
\newblock
\shownote{(Accessed: 2023-07-10)}.


\bibitem[Ellson et~al\mbox{.}(2002)]%
        {ellson2002graphviz}
\bibfield{author}{\bibinfo{person}{John Ellson}, \bibinfo{person}{Emden
  Gansner}, \bibinfo{person}{Lefteris Koutsofios}, \bibinfo{person}{Stephen~C
  North}, {and} \bibinfo{person}{Gordon Woodhull}.}
  \bibinfo{year}{2002}\natexlab{}.
\newblock \showarticletitle{Graphviz—open source graph drawing tools}. In
  \bibinfo{booktitle}{\emph{Graph Drawing: 9th International Symposium, GD 2001
  Vienna, Austria, September 23--26, 2001 Revised Papers 9}}. Springer,
  \bibinfo{pages}{483--484}.
\newblock


\bibitem[Fleming et~al\mbox{.}(2021)]%
        {fleming2021causalmisconceptions}
\bibfield{author}{\bibinfo{person}{Whitney Fleming}, \bibinfo{person}{Adam~L.
  Hayes}, \bibinfo{person}{Katherine~M. Crosman}, {and} \bibinfo{person}{Ann
  Bostrom}.} \bibinfo{year}{2021}\natexlab{}.
\newblock \showarticletitle{Indiscriminate, Irrelevant, and Sometimes Wrong:
  Causal Misconceptions about Climate Change}.
\newblock \bibinfo{journal}{\emph{Risk Analysis}} \bibinfo{volume}{41},
  \bibinfo{number}{1} (\bibinfo{year}{2021}), \bibinfo{pages}{157--178}.
\newblock
\urldef\tempurl%
\url{https://doi.org/10.1111/risa.13587}
\showDOI{\tempurl}
\showeprint{https://onlinelibrary.wiley.com/doi/pdf/10.1111/risa.13587}


\bibitem[Forests(2022)]%
        {forestcarbon}
\bibfield{author}{\bibinfo{person}{American Forests}.}
  \bibinfo{year}{2022}\natexlab{}.
\newblock \bibinfo{title}{Forests as Carbon Sinks}.
\newblock
\newblock
\urldef\tempurl%
\url{https://www.americanforests.org/article/forests-as-carbon-sinks/}
\showURL{%
\tempurl}
\newblock
\shownote{(Accessed: 2022-02-04)}.


\bibitem[Freckelton(2012)]%
        {freckelton2012death}
\bibfield{author}{\bibinfo{person}{Ian Freckelton}.}
  \bibinfo{year}{2012}\natexlab{}.
\newblock \showarticletitle{Death by homeopathy: issues for civil, criminal and
  coronial law and for health service policy.}
\newblock \bibinfo{journal}{\emph{Journal of Law and medicine}}
  \bibinfo{volume}{19}, \bibinfo{number}{3} (\bibinfo{year}{2012}),
  \bibinfo{pages}{454--478}.
\newblock


\bibitem[Gadiraju et~al\mbox{.}(2015)]%
        {gadiraju2015malicious}
\bibfield{author}{\bibinfo{person}{Ujwal Gadiraju}, \bibinfo{person}{Ricardo
  Kawase}, \bibinfo{person}{Stefan Dietze}, {and} \bibinfo{person}{Gianluca
  Demartini}.} \bibinfo{year}{2015}\natexlab{}.
\newblock \showarticletitle{Understanding Malicious Behavior in Crowdsourcing
  Platforms: The Case of Online Surveys} \emph{(\bibinfo{series}{CHI '15})}.
  \bibinfo{publisher}{Association for Computing Machinery},
  \bibinfo{address}{New York, NY, USA}, \bibinfo{pages}{1631–1640}.
\newblock
\showISBNx{9781450331456}
\urldef\tempurl%
\url{https://doi.org/10.1145/2702123.2702443}
\showDOI{\tempurl}


\bibitem[Ghai and Mueller(2023)]%
        {DBLP:journals/tvcg/GhaiM23}
\bibfield{author}{\bibinfo{person}{Bhavya Ghai} {and} \bibinfo{person}{Klaus
  Mueller}.} \bibinfo{year}{2023}\natexlab{}.
\newblock \showarticletitle{{D-BIAS:} {A} Causality-Based Human-in-the-Loop
  System for Tackling Algorithmic Bias}.
\newblock \bibinfo{journal}{\emph{{{IEEE} Transactions on Visualization and
  Computer Graphics}}} \bibinfo{volume}{29}, \bibinfo{number}{1}
  (\bibinfo{year}{2023}), \bibinfo{pages}{473--482}.
\newblock
\urldef\tempurl%
\url{https://doi.org/10.1109/TVCG.2022.3209484}
\showDOI{\tempurl}


\bibitem[Global Climate~Change(2021)]%
        {nasa_sun}
\bibfield{author}{\bibinfo{person}{NASA Global Climate~Change}.}
  \bibinfo{year}{2021}\natexlab{}.
\newblock \bibinfo{title}{Is the Sun causing global warming?}
\newblock
\newblock
\urldef\tempurl%
\url{https://climate.nasa.gov/faq/14/is-the-sun-causing-global-warming/}
\showURL{%
\tempurl}
\newblock
\shownote{Accessed: 2022-06-04}.


\bibitem[Health and Council(2015)]%
        {national2015nhmrc}
\bibfield{author}{\bibinfo{person}{National Health} {and}
  \bibinfo{person}{Medical~Research Council}.} \bibinfo{year}{2015}\natexlab{}.
\newblock \bibinfo{booktitle}{\emph{NHMRC Information Paper: Evidence on the
  effectiveness of homeopathy for treating health conditions}}.
\newblock \bibinfo{publisher}{National Health and Medical Research Council}.
\newblock


\bibitem[Hoque and Mueller(2022)]%
        {hoque2021outcome}
\bibfield{author}{\bibinfo{person}{Md.~Naimul Hoque} {and}
  \bibinfo{person}{Klaus Mueller}.} \bibinfo{year}{2022}\natexlab{}.
\newblock \showarticletitle{Outcome-Explorer: {A} Causality Guided Interactive
  Visual Interface for Interpretable Algorithmic Decision Making}.
\newblock \bibinfo{journal}{\emph{{IEEE} Trans. Vis. Comput. Graph.}}
  \bibinfo{volume}{28}, \bibinfo{number}{12} (\bibinfo{year}{2022}),
  \bibinfo{pages}{4728--4740}.
\newblock
\urldef\tempurl%
\url{https://doi.org/10.1109/TVCG.2021.3102051}
\showDOI{\tempurl}


\bibitem[Horton and Chilton(2010)]%
        {horton2010labor}
\bibfield{author}{\bibinfo{person}{John~Joseph Horton} {and}
  \bibinfo{person}{Lydia~B Chilton}.} \bibinfo{year}{2010}\natexlab{}.
\newblock \showarticletitle{The labor economics of paid crowdsourcing}. In
  \bibinfo{booktitle}{\emph{Proceedings of the 11th ACM conference on
  Electronic commerce}}. \bibinfo{pages}{209--218}.
\newblock


\bibitem[Ipeirotis et~al\mbox{.}(2010)]%
        {ipeirotis2010quality}
\bibfield{author}{\bibinfo{person}{Panagiotis~G. Ipeirotis},
  \bibinfo{person}{Foster~J. Provost}, {and} \bibinfo{person}{Jing Wang}.}
  \bibinfo{year}{2010}\natexlab{}.
\newblock \showarticletitle{Quality management on Amazon Mechanical Turk}. In
  \bibinfo{booktitle}{\emph{Proceedings of the {ACM} {SIGKDD} Workshop on Human
  Computation, {HCOMP} '10, Washington DC, USA, July 25, 2010}}.
  \bibinfo{publisher}{{ACM}}, \bibinfo{address}{New York, NY, USA},
  \bibinfo{pages}{64--67}.
\newblock
\urldef\tempurl%
\url{https://doi.org/10.1145/1837885.1837906}
\showDOI{\tempurl}


\bibitem[Kairam and Heer(2016)]%
        {DBLP:conf/cscw/KairamH16}
\bibfield{author}{\bibinfo{person}{Sanjay Kairam} {and}
  \bibinfo{person}{Jeffrey Heer}.} \bibinfo{year}{2016}\natexlab{}.
\newblock \showarticletitle{Parting Crowds: Characterizing Divergent
  Interpretations in Crowdsourced Annotation Tasks}. In
  \bibinfo{booktitle}{\emph{Proceedings of the 19th {ACM} Conference on
  Computer-Supported Cooperative Work {\&} Social Computing, {CSCW} 2016, San
  Francisco, CA, USA, February 27 - March 2, 2016}}.
  \bibinfo{publisher}{{ACM}}, \bibinfo{address}{New York, NY, USA},
  \bibinfo{pages}{1635--1646}.
\newblock
\urldef\tempurl%
\url{https://doi.org/10.1145/2818048.2820016}
\showDOI{\tempurl}


\bibitem[Kim et~al\mbox{.}(2019)]%
        {kim2019homogeneity}
\bibfield{author}{\bibinfo{person}{Jooyeon Kim}, \bibinfo{person}{Dongkwan
  Kim}, {and} \bibinfo{person}{Alice Oh}.} \bibinfo{year}{2019}\natexlab{}.
\newblock \showarticletitle{Homogeneity-Based Transmissive Process to Model
  True and False News in Social Networks}. In
  \bibinfo{booktitle}{\emph{Proceedings of the Twelfth ACM International
  Conference on Web Search and Data Mining}} (Melbourne VIC, Australia)
  \emph{(\bibinfo{series}{WSDM '19})}. \bibinfo{publisher}{Association for
  Computing Machinery}, \bibinfo{address}{New York, NY, USA},
  \bibinfo{pages}{348–356}.
\newblock
\showISBNx{9781450359405}
\urldef\tempurl%
\url{https://doi.org/10.1145/3289600.3291009}
\showDOI{\tempurl}


\bibitem[Kim et~al\mbox{.}(2018)]%
        {kim2018hitorwait}
\bibfield{author}{\bibinfo{person}{Yongsung Kim}, \bibinfo{person}{Darren
  Gergle}, {and} \bibinfo{person}{Haoqi Zhang}.}
  \bibinfo{year}{2018}\natexlab{}.
\newblock \showarticletitle{Hit-or-Wait: Coordinating Opportunistic Low-Effort
  Contributions to Achieve Global Outcomes in On-the-Go Crowdsourcing}. In
  \bibinfo{booktitle}{\emph{Proceedings of the 2018 CHI Conference on Human
  Factors in Computing Systems}} (Montreal QC, Canada)
  \emph{(\bibinfo{series}{CHI '18})}. \bibinfo{publisher}{Association for
  Computing Machinery}, \bibinfo{address}{New York, NY, USA},
  \bibinfo{pages}{1–12}.
\newblock
\showISBNx{9781450356206}
\urldef\tempurl%
\url{https://doi.org/10.1145/3173574.3173670}
\showDOI{\tempurl}


\bibitem[Kittur et~al\mbox{.}(2008)]%
        {kittur2008crowdsourcing}
\bibfield{author}{\bibinfo{person}{Aniket Kittur}, \bibinfo{person}{Ed~H. Chi},
  {and} \bibinfo{person}{Bongwon Suh}.} \bibinfo{year}{2008}\natexlab{}.
\newblock \showarticletitle{Crowdsourcing user studies with Mechanical Turk}.
  In \bibinfo{booktitle}{\emph{Proceedings of the 2008 Conference on Human
  Factors in Computing Systems, {CHI} 2008, 2008, Florence, Italy, April 5-10,
  2008}}. \bibinfo{publisher}{{ACM}}, \bibinfo{address}{New York, NY, USA},
  \bibinfo{pages}{453--456}.
\newblock
\urldef\tempurl%
\url{https://doi.org/10.1145/1357054.1357127}
\showDOI{\tempurl}


\bibitem[Kittur et~al\mbox{.}(2012)]%
        {kittur2012crowdweaver}
\bibfield{author}{\bibinfo{person}{Aniket Kittur}, \bibinfo{person}{Susheel
  Khamkar}, \bibinfo{person}{Paul Andr{\'{e}}}, {and}
  \bibinfo{person}{Robert~E. Kraut}.} \bibinfo{year}{2012}\natexlab{}.
\newblock \showarticletitle{CrowdWeaver: visually managing complex crowd work}.
  In \bibinfo{booktitle}{\emph{{CSCW} '12 Computer Supported Cooperative Work,
  Seattle, WA, USA, February 11-15, 2012}}. \bibinfo{publisher}{{ACM}},
  \bibinfo{address}{New York, NY, USA}, \bibinfo{pages}{1033--1036}.
\newblock
\urldef\tempurl%
\url{https://doi.org/10.1145/2145204.2145357}
\showDOI{\tempurl}


\bibitem[Kittur et~al\mbox{.}(2013)]%
        {kittur2013future}
\bibfield{author}{\bibinfo{person}{Aniket Kittur}, \bibinfo{person}{Jeffrey~V.
  Nickerson}, \bibinfo{person}{Michael Bernstein}, \bibinfo{person}{Elizabeth
  Gerber}, \bibinfo{person}{Aaron Shaw}, \bibinfo{person}{John Zimmerman},
  \bibinfo{person}{Matt Lease}, {and} \bibinfo{person}{John Horton}.}
  \bibinfo{year}{2013}\natexlab{}.
\newblock \showarticletitle{The Future of Crowd Work}
  \emph{(\bibinfo{series}{CSCW '13})}. \bibinfo{publisher}{Association for
  Computing Machinery}, \bibinfo{address}{New York, NY, USA},
  \bibinfo{pages}{1301–1318}.
\newblock
\showISBNx{9781450313315}
\urldef\tempurl%
\url{https://doi.org/10.1145/2441776.2441923}
\showDOI{\tempurl}


\bibitem[Kittur et~al\mbox{.}(2011)]%
        {kittur2011crowdforge}
\bibfield{author}{\bibinfo{person}{Aniket Kittur}, \bibinfo{person}{Boris
  Smus}, \bibinfo{person}{Susheel Khamkar}, {and} \bibinfo{person}{Robert~E.
  Kraut}.} \bibinfo{year}{2011}\natexlab{}.
\newblock \showarticletitle{CrowdForge: crowdsourcing complex work}. In
  \bibinfo{booktitle}{\emph{Proceedings of the 24th Annual {ACM} Symposium on
  User Interface Software and Technology, Santa Barbara, CA, USA, October
  16-19, 2011}}. \bibinfo{publisher}{{ACM}}, \bibinfo{address}{New York, NY,
  USA}, \bibinfo{pages}{43--52}.
\newblock
\urldef\tempurl%
\url{https://doi.org/10.1145/2047196.2047202}
\showDOI{\tempurl}


\bibitem[Krishna et~al\mbox{.}(2016)]%
        {krishna2016embracingerrors}
\bibfield{author}{\bibinfo{person}{Ranjay~A. Krishna}, \bibinfo{person}{Kenji
  Hata}, \bibinfo{person}{Stephanie Chen}, \bibinfo{person}{Joshua Kravitz},
  \bibinfo{person}{David~A. Shamma}, \bibinfo{person}{Li Fei-Fei}, {and}
  \bibinfo{person}{Michael~S. Bernstein}.} \bibinfo{year}{2016}\natexlab{}.
\newblock \showarticletitle{Embracing Error to Enable Rapid Crowdsourcing}. In
  \bibinfo{booktitle}{\emph{Proceedings of the 2016 CHI Conference on Human
  Factors in Computing Systems}} (San Jose, California, USA)
  \emph{(\bibinfo{series}{CHI '16})}. \bibinfo{publisher}{Association for
  Computing Machinery}, \bibinfo{address}{New York, NY, USA},
  \bibinfo{pages}{3167–3179}.
\newblock
\showISBNx{9781450333627}
\urldef\tempurl%
\url{https://doi.org/10.1145/2858036.2858115}
\showDOI{\tempurl}


\bibitem[Kusner et~al\mbox{.}(2017)]%
        {kusner2017counterfactual}
\bibfield{author}{\bibinfo{person}{Matt~J. Kusner}, \bibinfo{person}{Joshua~R.
  Loftus}, \bibinfo{person}{Chris Russell}, {and} \bibinfo{person}{Ricardo
  Silva}.} \bibinfo{year}{2017}\natexlab{}.
\newblock \showarticletitle{Counterfactual Fairness}. In
  \bibinfo{booktitle}{\emph{Advances in Neural Information Processing Systems
  30: Annual Conference on Neural Information Processing Systems 2017, December
  4-9, 2017, Long Beach, CA, {USA}}}. \bibinfo{pages}{4066--4076}.
\newblock
\urldef\tempurl%
\url{https://proceedings.neurips.cc/paper/2017/hash/a486cd07e4ac3d270571622f4f316ec5-Abstract.html}
\showURL{%
\tempurl}


\bibitem[Kwon and Cha(2014)]%
        {Kwon_Cha_2014}
\bibfield{author}{\bibinfo{person}{Sejeong Kwon} {and}
  \bibinfo{person}{Meeyoung Cha}.} \bibinfo{year}{2014}\natexlab{}.
\newblock \showarticletitle{Modeling Bursty Temporal Pattern of Rumors}.
\newblock \bibinfo{journal}{\emph{Proceedings of the International AAAI
  Conference on Web and Social Media}} \bibinfo{volume}{8}, \bibinfo{number}{1}
  (\bibinfo{date}{May} \bibinfo{year}{2014}), \bibinfo{pages}{650--651}.
\newblock
\urldef\tempurl%
\url{https://ojs.aaai.org/index.php/ICWSM/article/view/14494}
\showURL{%
\tempurl}


\bibitem[Lasecki et~al\mbox{.}(2015)]%
        {lasecki2015sequence}
\bibfield{author}{\bibinfo{person}{Walter~S. Lasecki},
  \bibinfo{person}{Jeffrey~M. Rzeszotarski}, \bibinfo{person}{Adam Marcus},
  {and} \bibinfo{person}{Jeffrey~P. Bigham}.} \bibinfo{year}{2015}\natexlab{}.
\newblock \showarticletitle{The Effects of Sequence and Delay on Crowd Work}.
  In \bibinfo{booktitle}{\emph{Proceedings of the 33rd Annual ACM Conference on
  Human Factors in Computing Systems}} (Seoul, Republic of Korea)
  \emph{(\bibinfo{series}{CHI '15})}. \bibinfo{publisher}{Association for
  Computing Machinery}, \bibinfo{address}{New York, NY, USA},
  \bibinfo{pages}{1375–1378}.
\newblock
\showISBNx{9781450331456}
\urldef\tempurl%
\url{https://doi.org/10.1145/2702123.2702594}
\showDOI{\tempurl}


\bibitem[Leiserowitz et~al\mbox{.}(2019)]%
        {leiserowitz2019climate}
\bibfield{author}{\bibinfo{person}{Anthony Leiserowitz},
  \bibinfo{person}{Edward~W Maibach}, \bibinfo{person}{Seth Rosenthal},
  \bibinfo{person}{John Kotcher}, \bibinfo{person}{Parrish Bergquist},
  \bibinfo{person}{Matthew Ballew}, \bibinfo{person}{Matthew Goldberg}, {and}
  \bibinfo{person}{Abel Gustafson}.} \bibinfo{year}{2019}\natexlab{}.
\newblock \showarticletitle{Climate change in the American mind: April 2019}.
\newblock \bibinfo{journal}{\emph{Yale University and George Mason University.
  New Haven, CT: Yale Program on Climate Change Communication}}
  (\bibinfo{year}{2019}).
\newblock


\bibitem[Life({[n.\,d.]})]%
        {meltingseaice}
\bibfield{author}{\bibinfo{person}{World~Wild Life}.}
  \bibinfo{year}{[n.\,d.]}\natexlab{}.
\newblock \bibinfo{title}{Why are glaciers and sea ice melting?}
\newblock
  \bibinfo{howpublished}{\url{https://www.worldwildlife.org/pages/why-are-glaciers-and-sea-ice-melting}}.
\newblock
\newblock
\shownote{(Accessed on 07/10/2023)}.


\bibitem[Liu et~al\mbox{.}(2018)]%
        {liu2018conceptscape}
\bibfield{author}{\bibinfo{person}{Ching Liu}, \bibinfo{person}{Juho Kim},
  {and} \bibinfo{person}{Hao-Chuan Wang}.} \bibinfo{year}{2018}\natexlab{}.
\newblock \showarticletitle{ConceptScape: Collaborative Concept Mapping for
  Video Learning}. In \bibinfo{booktitle}{\emph{Proceedings of the 2018 CHI
  Conference on Human Factors in Computing Systems}} (Montreal QC, Canada)
  \emph{(\bibinfo{series}{CHI '18})}. \bibinfo{publisher}{Association for
  Computing Machinery}, \bibinfo{address}{New York, NY, USA},
  \bibinfo{pages}{1–12}.
\newblock
\showISBNx{9781450356206}
\urldef\tempurl%
\url{https://doi.org/10.1145/3173574.3173961}
\showDOI{\tempurl}


\bibitem[Liu et~al\mbox{.}(2019)]%
        {liu2019roberta}
\bibfield{author}{\bibinfo{person}{Yinhan Liu}, \bibinfo{person}{Myle Ott},
  \bibinfo{person}{Naman Goyal}, \bibinfo{person}{Jingfei Du},
  \bibinfo{person}{Mandar Joshi}, \bibinfo{person}{Danqi Chen},
  \bibinfo{person}{Omer Levy}, \bibinfo{person}{Mike Lewis},
  \bibinfo{person}{Luke Zettlemoyer}, {and} \bibinfo{person}{Veselin
  Stoyanov}.} \bibinfo{year}{2019}\natexlab{}.
\newblock \showarticletitle{Roberta: A robustly optimized bert pretraining
  approach}.
\newblock \bibinfo{journal}{\emph{arXiv preprint arXiv:1907.11692}}
  (\bibinfo{year}{2019}).
\newblock


\bibitem[MacKinnon et~al\mbox{.}(2007)]%
        {mackinnon2007mediation}
\bibfield{author}{\bibinfo{person}{David~P MacKinnon},
  \bibinfo{person}{Amanda~J Fairchild}, {and} \bibinfo{person}{Matthew~S
  Fritz}.} \bibinfo{year}{2007}\natexlab{}.
\newblock \showarticletitle{Mediation analysis}.
\newblock \bibinfo{journal}{\emph{Annual review of psychology}}
  \bibinfo{volume}{58} (\bibinfo{year}{2007}), \bibinfo{pages}{593}.
\newblock


\bibitem[Mahyar et~al\mbox{.}(2018)]%
        {mahyar2018communitycrit}
\bibfield{author}{\bibinfo{person}{Narges Mahyar}, \bibinfo{person}{Michael~R.
  James}, \bibinfo{person}{Michelle~M. Ng}, \bibinfo{person}{Reginald~A. Wu},
  {and} \bibinfo{person}{Steven~P. Dow}.} \bibinfo{year}{2018}\natexlab{}.
\newblock \showarticletitle{CommunityCrit: Inviting the Public to Improve and
  Evaluate Urban Design Ideas through Micro-Activities}. In
  \bibinfo{booktitle}{\emph{Proceedings of the 2018 CHI Conference on Human
  Factors in Computing Systems}} (Montreal QC, Canada)
  \emph{(\bibinfo{series}{CHI '18})}. \bibinfo{publisher}{Association for
  Computing Machinery}, \bibinfo{address}{New York, NY, USA},
  \bibinfo{pages}{1–14}.
\newblock
\showISBNx{9781450356206}
\urldef\tempurl%
\url{https://doi.org/10.1145/3173574.3173769}
\showDOI{\tempurl}


\bibitem[Matute et~al\mbox{.}(2015)]%
        {matute2015illusions}
\bibfield{author}{\bibinfo{person}{Helena Matute}, \bibinfo{person}{Fernando
  Blanco}, \bibinfo{person}{Ion Yarritu}, \bibinfo{person}{Marcos
  D{\'\i}az-Lago}, \bibinfo{person}{Miguel~A Vadillo}, {and}
  \bibinfo{person}{Itxaso Barberia}.} \bibinfo{year}{2015}\natexlab{}.
\newblock \showarticletitle{Illusions of causality: how they bias our everyday
  thinking and how they could be reduced}.
\newblock \bibinfo{journal}{\emph{Frontiers in psychology}}
  \bibinfo{volume}{6} (\bibinfo{year}{2015}), \bibinfo{pages}{888}.
\newblock


\bibitem[Mazid and Zarnaz(2022)]%
        {mazid2022climatechangemyths}
\bibfield{author}{\bibinfo{person}{Md~Abdullah~Al Mazid} {and}
  \bibinfo{person}{Zaima Zarnaz}.} \bibinfo{year}{2022}\natexlab{}.
\newblock \showarticletitle{Climate Change Myths Detection Using Dynamically
  Weighted Ensemble Based Stance Classifier}. In
  \bibinfo{booktitle}{\emph{Proceedings of the 2nd International Conference on
  Computing Advancements}} (Dhaka, Bangladesh) \emph{(\bibinfo{series}{ICCA
  '22})}. \bibinfo{publisher}{Association for Computing Machinery},
  \bibinfo{address}{New York, NY, USA}, \bibinfo{pages}{277–283}.
\newblock
\showISBNx{9781450397346}
\urldef\tempurl%
\url{https://doi.org/10.1145/3542954.3542995}
\showDOI{\tempurl}


\bibitem[Mikolov et~al\mbox{.}(2013)]%
        {mikolov2013efficient}
\bibfield{author}{\bibinfo{person}{Tomas Mikolov}, \bibinfo{person}{Kai Chen},
  \bibinfo{person}{Greg Corrado}, {and} \bibinfo{person}{Jeffrey Dean}.}
  \bibinfo{year}{2013}\natexlab{}.
\newblock \showarticletitle{Efficient estimation of word representations in
  vector space}.
\newblock \bibinfo{journal}{\emph{arXiv preprint arXiv:1301.3781}}
  (\bibinfo{year}{2013}).
\newblock


\bibitem[Noronha et~al\mbox{.}(2011)]%
        {noronha2011platemate}
\bibfield{author}{\bibinfo{person}{Jon Noronha}, \bibinfo{person}{Eric Hysen},
  \bibinfo{person}{Haoqi Zhang}, {and} \bibinfo{person}{Krzysztof~Z. Gajos}.}
  \bibinfo{year}{2011}\natexlab{}.
\newblock \showarticletitle{Platemate: crowdsourcing nutritional analysis from
  food photographs}. In \bibinfo{booktitle}{\emph{Proceedings of the 24th
  Annual {ACM} Symposium on User Interface Software and Technology, Santa
  Barbara, CA, USA, October 16-19, 2011}}. \bibinfo{publisher}{{ACM}},
  \bibinfo{address}{New York, NY, USA}, \bibinfo{pages}{1--12}.
\newblock
\urldef\tempurl%
\url{https://doi.org/10.1145/2047196.2047198}
\showDOI{\tempurl}


\bibitem[Nps.gov(2022)]%
        {wildfirecauses}
\bibfield{author}{\bibinfo{person}{Nps.gov}.} \bibinfo{year}{2022}\natexlab{}.
\newblock \bibinfo{title}{Wildfire Causes and Evaluations}.
\newblock
\newblock
\urldef\tempurl%
\url{https://www.nps.gov/articles/wildfire-causes-and-evaluation.}
\showURL{%
\tempurl}
\newblock
\shownote{(Accessed: 2023-07-10)}.


\bibitem[Organization({[n.\,d.]})]%
        {heatwaves}
\bibfield{author}{\bibinfo{person}{World~Meteorological Organization}.}
  \bibinfo{year}{[n.\,d.]}\natexlab{}.
\newblock \bibinfo{title}{Heatwaves}.
\newblock
  \bibinfo{howpublished}{\url{https://www.britannica.com/science/heat-wave-meteorology}}.
\newblock
\newblock
\shownote{(Accessed on 07/10/2022)}.


\bibitem[Panch et~al\mbox{.}(2019)]%
        {Panch2019-yv}
\bibfield{author}{\bibinfo{person}{Trishan Panch}, \bibinfo{person}{Heather
  Mattie}, {and} \bibinfo{person}{Rifat Atun}.}
  \bibinfo{year}{2019}\natexlab{}.
\newblock \showarticletitle{Artificial intelligence and algorithmic bias:
  implications for health systems}.
\newblock \bibinfo{journal}{\emph{J. Glob. Health}} \bibinfo{volume}{9},
  \bibinfo{number}{2} (\bibinfo{date}{Dec.} \bibinfo{year}{2019}),
  \bibinfo{pages}{010318}.
\newblock


\bibitem[Pearl and Mackenzie(2018)]%
        {pearl2018book}
\bibfield{author}{\bibinfo{person}{Judea Pearl} {and} \bibinfo{person}{Dana
  Mackenzie}.} \bibinfo{year}{2018}\natexlab{}.
\newblock \bibinfo{booktitle}{\emph{The book of why: the new science of cause
  and effect}}.
\newblock \bibinfo{publisher}{Basic books}.
\newblock


\bibitem[Pennington et~al\mbox{.}(2014)]%
        {pennington2014glove}
\bibfield{author}{\bibinfo{person}{Jeffrey Pennington},
  \bibinfo{person}{Richard Socher}, {and} \bibinfo{person}{Christopher~D
  Manning}.} \bibinfo{year}{2014}\natexlab{}.
\newblock \showarticletitle{Glove: Global vectors for word representation}. In
  \bibinfo{booktitle}{\emph{Proceedings of the 2014 conference on empirical
  methods in natural language processing (EMNLP)}}.
  \bibinfo{pages}{1532--1543}.
\newblock


\bibitem[Qiu et~al\mbox{.}(2020)]%
        {qiu2020conversationalmicrotask}
\bibfield{author}{\bibinfo{person}{Sihang Qiu}, \bibinfo{person}{Ujwal
  Gadiraju}, {and} \bibinfo{person}{Alessandro Bozzon}.}
  \bibinfo{year}{2020}\natexlab{}.
\newblock \showarticletitle{Improving Worker Engagement Through Conversational
  Microtask Crowdsourcing}. In \bibinfo{booktitle}{\emph{Proceedings of the
  2020 CHI Conference on Human Factors in Computing Systems}} (Honolulu, HI,
  USA) \emph{(\bibinfo{series}{CHI '20})}. \bibinfo{publisher}{Association for
  Computing Machinery}, \bibinfo{address}{New York, NY, USA},
  \bibinfo{pages}{1–12}.
\newblock
\showISBNx{9781450367080}
\urldef\tempurl%
\url{https://doi.org/10.1145/3313831.3376403}
\showDOI{\tempurl}


\bibitem[Rastogi et~al\mbox{.}(2022)]%
        {10.1145/3512930}
\bibfield{author}{\bibinfo{person}{Charvi Rastogi}, \bibinfo{person}{Yunfeng
  Zhang}, \bibinfo{person}{Dennis Wei}, \bibinfo{person}{Kush~R. Varshney},
  \bibinfo{person}{Amit Dhurandhar}, {and} \bibinfo{person}{Richard Tomsett}.}
  \bibinfo{year}{2022}\natexlab{}.
\newblock \showarticletitle{Deciding Fast and Slow: The Role of Cognitive
  Biases in AI-Assisted Decision-Making}.
\newblock \bibinfo{journal}{\emph{Proc. ACM Hum.-Comput. Interact.}}
  \bibinfo{volume}{6}, \bibinfo{number}{CSCW1}, Article \bibinfo{articleno}{83}
  (\bibinfo{date}{apr} \bibinfo{year}{2022}), \bibinfo{numpages}{22}~pages.
\newblock
\urldef\tempurl%
\url{https://doi.org/10.1145/3512930}
\showDOI{\tempurl}


\bibitem[Robert and Romero(2015)]%
        {robert2015crowddiversity}
\bibfield{author}{\bibinfo{person}{Lionel Robert} {and}
  \bibinfo{person}{Daniel~M. Romero}.} \bibinfo{year}{2015}\natexlab{}.
\newblock \showarticletitle{Crowd Size, Diversity and Performance}. In
  \bibinfo{booktitle}{\emph{Proceedings of the 33rd Annual ACM Conference on
  Human Factors in Computing Systems}} (Seoul, Republic of Korea)
  \emph{(\bibinfo{series}{CHI '15})}. \bibinfo{publisher}{Association for
  Computing Machinery}, \bibinfo{address}{New York, NY, USA},
  \bibinfo{pages}{1379–1382}.
\newblock
\showISBNx{9781450331456}
\urldef\tempurl%
\url{https://doi.org/10.1145/2702123.2702469}
\showDOI{\tempurl}


\bibitem[Rzeszotarski and Kittur(2012)]%
        {rzeszotarski2012crowdscape}
\bibfield{author}{\bibinfo{person}{Jeffrey~M. Rzeszotarski} {and}
  \bibinfo{person}{Aniket Kittur}.} \bibinfo{year}{2012}\natexlab{}.
\newblock \showarticletitle{CrowdScape: interactively visualizing user behavior
  and output}. In \bibinfo{booktitle}{\emph{The 25th Annual {ACM} Symposium on
  User Interface Software and Technology, {UIST} '12, Cambridge, MA, USA,
  October 7-10, 2012}}. \bibinfo{publisher}{{ACM}}, \bibinfo{address}{New York,
  NY, USA}, \bibinfo{pages}{55--62}.
\newblock
\urldef\tempurl%
\url{https://doi.org/10.1145/2380116.2380125}
\showDOI{\tempurl}


\bibitem[Saha et~al\mbox{.}(2019)]%
        {saha2019project}
\bibfield{author}{\bibinfo{person}{Manaswi Saha}, \bibinfo{person}{Michael
  Saugstad}, \bibinfo{person}{Hanuma~Teja Maddali}, \bibinfo{person}{Aileen
  Zeng}, \bibinfo{person}{Ryan Holland}, \bibinfo{person}{Steven Bower},
  \bibinfo{person}{Aditya Dash}, \bibinfo{person}{Sage Chen},
  \bibinfo{person}{Anthony Li}, \bibinfo{person}{Kotaro Hara}, {and}
  \bibinfo{person}{Jon Froehlich}.} \bibinfo{year}{2019}\natexlab{}.
\newblock \showarticletitle{Project Sidewalk: {A} Web-based Crowdsourcing Tool
  for Collecting Sidewalk Accessibility Data At Scale}. In
  \bibinfo{booktitle}{\emph{Proceedings of the 2019 {CHI} Conference on Human
  Factors in Computing Systems, {CHI} 2019, Glasgow, Scotland, UK, May 04-09,
  2019}}. \bibinfo{publisher}{{ACM}}, \bibinfo{address}{New York, NY, USA},
  \bibinfo{pages}{62}.
\newblock
\urldef\tempurl%
\url{https://doi.org/10.1145/3290605.3300292}
\showDOI{\tempurl}


\bibitem[Schwitzgebel(2011)]%
        {schwitzgebel2011belief}
\bibfield{author}{\bibinfo{person}{Eric Schwitzgebel}.}
  \bibinfo{year}{2011}\natexlab{}.
\newblock \showarticletitle{Belief}.
\newblock In \bibinfo{booktitle}{\emph{The Routledge Companion to
  Epistemology}}. \bibinfo{publisher}{Routledge}, \bibinfo{pages}{40--50}.
\newblock


\bibitem[Shi et~al\mbox{.}(2021)]%
        {shi2021towardsemi}
\bibfield{author}{\bibinfo{person}{Yang Shi}, \bibinfo{person}{Krupal Shah},
  \bibinfo{person}{Wengran Wang}, \bibinfo{person}{Samiha Marwan},
  \bibinfo{person}{Poorvaja Penmetsa}, {and} \bibinfo{person}{Thomas Price}.}
  \bibinfo{year}{2021}\natexlab{}.
\newblock \showarticletitle{Toward Semi-Automatic Misconception Discovery Using
  Code Embeddings}. In \bibinfo{booktitle}{\emph{LAK21: 11th International
  Learning Analytics and Knowledge Conference}} (Irvine, CA, USA)
  \emph{(\bibinfo{series}{LAK21})}. \bibinfo{publisher}{Association for
  Computing Machinery}, \bibinfo{address}{New York, NY, USA},
  \bibinfo{pages}{606–612}.
\newblock
\showISBNx{9781450389358}
\urldef\tempurl%
\url{https://doi.org/10.1145/3448139.3448205}
\showDOI{\tempurl}


\bibitem[Singh and Ernst(2008)]%
        {singh2008trick}
\bibfield{author}{\bibinfo{person}{Simon Singh} {and} \bibinfo{person}{Edzard
  Ernst}.} \bibinfo{year}{2008}\natexlab{}.
\newblock \bibinfo{booktitle}{\emph{Trick or treatment: The undeniable facts
  about alternative medicine}}.
\newblock \bibinfo{publisher}{WW Norton \& Company}.
\newblock


\bibitem[Today(2022)]%
        {climatechangenotbelief}
\bibfield{author}{\bibinfo{person}{Psychology Today}.}
  \bibinfo{year}{2022}\natexlab{}.
\newblock \bibinfo{title}{Why Don't People Believe in Climate Change?}
\newblock
  \bibinfo{howpublished}{\url{https://www.psychologytoday.com/us/blog/psych-unseen/202204/why-dont-people-believe-in-climate-change}}.
\newblock
\newblock
\shownote{(Accessed on 07/10/2022)}.


\bibitem[usability.gov({[n.\,d.]})]%
        {sus}
\bibfield{author}{\bibinfo{person}{usability.gov}.}
  \bibinfo{year}{[n.\,d.]}\natexlab{}.
\newblock \bibinfo{title}{System Usability Scale (SUS)}.
\newblock
  \bibinfo{howpublished}{\url{https://www.usability.gov/how-to-and-tools/methods/system-usability-scale.html}}.
\newblock
\newblock
\shownote{(Accessed on 07/10/2022)}.


\bibitem[Vadillo et~al\mbox{.}(2013)]%
        {vadillo2013fighting}
\bibfield{author}{\bibinfo{person}{Miguel~A Vadillo}, \bibinfo{person}{Helena
  Matute}, {and} \bibinfo{person}{Fernando Blanco}.}
  \bibinfo{year}{2013}\natexlab{}.
\newblock \showarticletitle{Fighting the illusion of control: How to make use
  of cue competition and alternative explanations}.
\newblock \bibinfo{journal}{\emph{Universitas Psychologica}}
  \bibinfo{volume}{12}, \bibinfo{number}{1} (\bibinfo{year}{2013}),
  \bibinfo{pages}{261--270}.
\newblock


\bibitem[Vadillo et~al\mbox{.}(2011)]%
        {vadillo2011contrasting}
\bibfield{author}{\bibinfo{person}{Miguel~A Vadillo}, \bibinfo{person}{Serban~C
  Musca}, \bibinfo{person}{Fernando Blanco}, {and} \bibinfo{person}{Helena
  Matute}.} \bibinfo{year}{2011}\natexlab{}.
\newblock \showarticletitle{Contrasting cue-density effects in causal and
  prediction judgments}.
\newblock \bibinfo{journal}{\emph{Psychonomic Bulletin \& Review}}
  \bibinfo{volume}{18}, \bibinfo{number}{1} (\bibinfo{year}{2011}),
  \bibinfo{pages}{110--115}.
\newblock


\bibitem[Walsh and Sloman(2004)]%
        {walsh2004revising}
\bibfield{author}{\bibinfo{person}{Clare~R Walsh} {and}
  \bibinfo{person}{Steven~A Sloman}.} \bibinfo{year}{2004}\natexlab{}.
\newblock \showarticletitle{Revising causal beliefs}. In
  \bibinfo{booktitle}{\emph{Proceedings of the Annual Meeting of the Cognitive
  Science Society}}, Vol.~\bibinfo{volume}{26}.
\newblock


\bibitem[Wang and Mueller(2015)]%
        {wang2015visual}
\bibfield{author}{\bibinfo{person}{Jun Wang} {and} \bibinfo{person}{Klaus
  Mueller}.} \bibinfo{year}{2015}\natexlab{}.
\newblock \showarticletitle{The visual causality analyst: An interactive
  interface for causal reasoning}.
\newblock \bibinfo{journal}{\emph{IEEE transactions on visualization and
  computer graphics}} \bibinfo{volume}{22}, \bibinfo{number}{1}
  (\bibinfo{year}{2015}), \bibinfo{pages}{230--239}.
\newblock


\bibitem[Whiting et~al\mbox{.}(2017)]%
        {DBLP:conf/cscw/WhitingGGGGBMCR17}
\bibfield{author}{\bibinfo{person}{Mark~E. Whiting}, \bibinfo{person}{Dilrukshi
  Gamage}, \bibinfo{person}{Snehalkumar (Neil)~S. Gaikwad},
  \bibinfo{person}{Aaron Gilbee}, \bibinfo{person}{Shirish Goyal},
  \bibinfo{person}{Alipta Ballav}, \bibinfo{person}{Dinesh Majeti},
  \bibinfo{person}{Nalin Chhibber}, \bibinfo{person}{Angela Richmond{-}Fuller},
  \bibinfo{person}{Freddie Vargus}, \bibinfo{person}{Tejas~Seshadri Sarma},
  \bibinfo{person}{Varshine Chandrakanthan}, \bibinfo{person}{Te{\'{o}}genes
  Moura}, \bibinfo{person}{Mohamed~Hashim Salih}, \bibinfo{person}{Gabriel
  Bayomi~Tinoco Kalejaiye}, \bibinfo{person}{Adam Ginzberg},
  \bibinfo{person}{Catherine~A. Mullings}, \bibinfo{person}{Yoni Dayan},
  \bibinfo{person}{Kristy Milland}, \bibinfo{person}{Henrique Orefice},
  \bibinfo{person}{Jeff Regino}, \bibinfo{person}{Sayna Parsi},
  \bibinfo{person}{Kunz Mainali}, \bibinfo{person}{Vibhor Sehgal},
  \bibinfo{person}{Sekandar Matin}, \bibinfo{person}{Akshansh Sinha},
  \bibinfo{person}{Rajan Vaish}, {and} \bibinfo{person}{Michael~S. Bernstein}.}
  \bibinfo{year}{2017}\natexlab{}.
\newblock \showarticletitle{Crowd Guilds: Worker-led Reputation and Feedback on
  Crowdsourcing Platforms}. In \bibinfo{booktitle}{\emph{Proceedings of the
  2017 {ACM} Conference on Computer Supported Cooperative Work and Social
  Computing, {CSCW} 2017, Portland, OR, USA, February 25 - March 1, 2017}}.
  \bibinfo{publisher}{{ACM}}, \bibinfo{address}{New York, NY, USA},
  \bibinfo{pages}{1902--1913}.
\newblock
\urldef\tempurl%
\url{https://doi.org/10.1145/2998181.2998234}
\showDOI{\tempurl}


\bibitem[Wu et~al\mbox{.}(2019)]%
        {wu2019misinformation}
\bibfield{author}{\bibinfo{person}{Liang Wu}, \bibinfo{person}{Fred
  Morstatter}, \bibinfo{person}{Kathleen~M. Carley}, {and}
  \bibinfo{person}{Huan Liu}.} \bibinfo{year}{2019}\natexlab{}.
\newblock \showarticletitle{Misinformation in Social Media: Definition,
  Manipulation, and Detection}.
\newblock \bibinfo{journal}{\emph{SIGKDD Explor. Newsl.}} \bibinfo{volume}{21},
  \bibinfo{number}{2} (\bibinfo{date}{nov} \bibinfo{year}{2019}),
  \bibinfo{pages}{80–90}.
\newblock
\showISSN{1931-0145}
\urldef\tempurl%
\url{https://doi.org/10.1145/3373464.3373475}
\showDOI{\tempurl}


\bibitem[Yapo and Weiss(2018)]%
        {yapo2018ethical}
\bibfield{author}{\bibinfo{person}{Adrienne Yapo} {and} \bibinfo{person}{Joseph
  Weiss}.} \bibinfo{year}{2018}\natexlab{}.
\newblock \showarticletitle{Ethical implications of bias in machine learning}.
\newblock  (\bibinfo{year}{2018}).
\newblock


\bibitem[Yarritu et~al\mbox{.}(2015)]%
        {yarritu2015dark}
\bibfield{author}{\bibinfo{person}{Ion Yarritu}, \bibinfo{person}{Helena
  Matute}, {and} \bibinfo{person}{David Luque}.}
  \bibinfo{year}{2015}\natexlab{}.
\newblock \showarticletitle{The dark side of cognitive illusions: When an
  illusory belief interferes with the acquisition of evidence-based knowledge}.
\newblock \bibinfo{journal}{\emph{British Journal of Psychology}}
  \bibinfo{volume}{106}, \bibinfo{number}{4} (\bibinfo{year}{2015}),
  \bibinfo{pages}{597--608}.
\newblock


\bibitem[Yarritu et~al\mbox{.}(2014)]%
        {yarritu2014illusion}
\bibfield{author}{\bibinfo{person}{Ion Yarritu}, \bibinfo{person}{Helena
  Matute}, {and} \bibinfo{person}{Miguel~A Vadillo}.}
  \bibinfo{year}{2014}\natexlab{}.
\newblock \showarticletitle{Illusion of control: the role of personal
  involvement.}
\newblock \bibinfo{journal}{\emph{Experimental psychology}}
  \bibinfo{volume}{61}, \bibinfo{number}{1} (\bibinfo{year}{2014}),
  \bibinfo{pages}{38}.
\newblock


\bibitem[Yen et~al\mbox{.}(2021)]%
        {yen2021narratives+}
\bibfield{author}{\bibinfo{person}{Chi{-}Hsien~(Eric) Yen},
  \bibinfo{person}{Haocong Cheng}, \bibinfo{person}{Yu{-}Chun~(Grace) Yen},
  \bibinfo{person}{Brian~P. Bailey}, {and} \bibinfo{person}{Yun Huang}.}
  \bibinfo{year}{2021}\natexlab{}.
\newblock \showarticletitle{Narratives + Diagrams: An Integrated Approach for
  Externalizing and Sharing People's Causal Beliefs}.
\newblock \bibinfo{journal}{\emph{Proc. {ACM} Hum. Comput. Interact.}}
  \bibinfo{volume}{5}, \bibinfo{number}{{CSCW2}} (\bibinfo{year}{2021}),
  \bibinfo{pages}{444:1--444:27}.
\newblock
\urldef\tempurl%
\url{https://doi.org/10.1145/3479588}
\showDOI{\tempurl}


\bibitem[Yen et~al\mbox{.}(2023)]%
        {yen2023crowdidea}
\bibfield{author}{\bibinfo{person}{Chi-Hsien Yen}, \bibinfo{person}{Haocong
  Cheng}, \bibinfo{person}{Yilin Xia}, {and} \bibinfo{person}{Yun Huang}.}
  \bibinfo{year}{2023}\natexlab{}.
\newblock \showarticletitle{CrowdIDEA: Blending Crowd Intelligence and Data
  Analytics to Empower Causal Reasoning}. In
  \bibinfo{booktitle}{\emph{Proceedings of the 2023 CHI Conference on Human
  Factors in Computing Systems}} (Hamburg, Germany) \emph{(\bibinfo{series}{CHI
  '23})}. \bibinfo{publisher}{Association for Computing Machinery},
  \bibinfo{address}{New York, NY, USA}, Article \bibinfo{articleno}{463},
  \bibinfo{numpages}{17}~pages.
\newblock
\showISBNx{9781450394215}
\urldef\tempurl%
\url{https://doi.org/10.1145/3544548.3581021}
\showDOI{\tempurl}


\bibitem[Yin et~al\mbox{.}(2018)]%
        {yin2018runningout}
\bibfield{author}{\bibinfo{person}{Ming Yin}, \bibinfo{person}{Siddharth Suri},
  {and} \bibinfo{person}{Mary~L. Gray}.} \bibinfo{year}{2018}\natexlab{}.
\newblock \showarticletitle{Running Out of Time: The Impact and Value of
  Flexibility in On-Demand Crowdwork}. In \bibinfo{booktitle}{\emph{Proceedings
  of the 2018 CHI Conference on Human Factors in Computing Systems}} (Montreal
  QC, Canada) \emph{(\bibinfo{series}{CHI '18})}.
  \bibinfo{publisher}{Association for Computing Machinery},
  \bibinfo{address}{New York, NY, USA}, \bibinfo{pages}{1–11}.
\newblock
\showISBNx{9781450356206}
\urldef\tempurl%
\url{https://doi.org/10.1145/3173574.3174004}
\showDOI{\tempurl}


\end{thebibliography}
\appendix
\appendix

\section{Formative Study}
\subsection{Crowd Workers' Expertise Level Regarding Climate Change}\label{formative-concern}
\begin{figure}[h]
    \centering
    \includegraphics[width=0.75\textwidth]{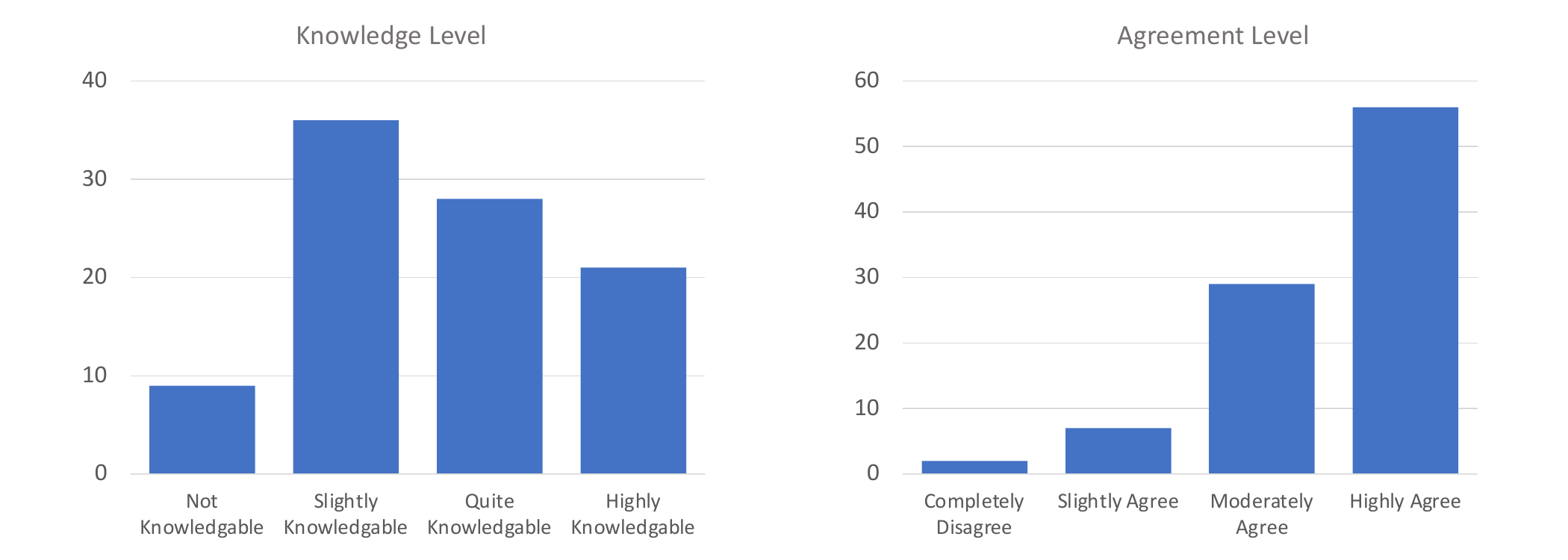}
    \caption{\textbf{Self-reported knowledge and agreement level of the crowd workers regarding climate change.} Y-axes represent counts for each category. The higher agreement level denotes a higher inclination to believe in the existence of climate change.}
    \Description{Figure \ref{fig:expertise} shows a collection of bar charts, each containing statistics of knowledge level and agreement level regarding climate change.}
    \label{fig:expertise}
\end{figure}

We present the self-reported knowledge and agreement level of the crowd workers regarding various climate change-related attributes and statements in Figure \ref{fig:expertise}.  Examples of climate change-related attributes are greenhouse gases, deforestation, and the melting of ice. One example of climate change-related statements is: ``Climate change is happening right now''. We present the attributes and statements in the supplemental material containing survey questions. We aggregate a particular worker's selected levels and bin them into different knowledge and agreement levels.

In terms of knowledge level, only a small portion (<10\%) of the crowd workers did not deem themselves knowledgeable about the climate change-related attributes. On the other hand, the majority of the crowd workers seemed to highly agree with climate change-related statements, which indicates that they can be categorized as climate change believers.  



\subsection{Noteworthy and Interesting Causal Illusion Cases}
\label{formative-misconception-cases}
\begin{figure}
    \centering
    \includegraphics[width=0.9\textwidth]{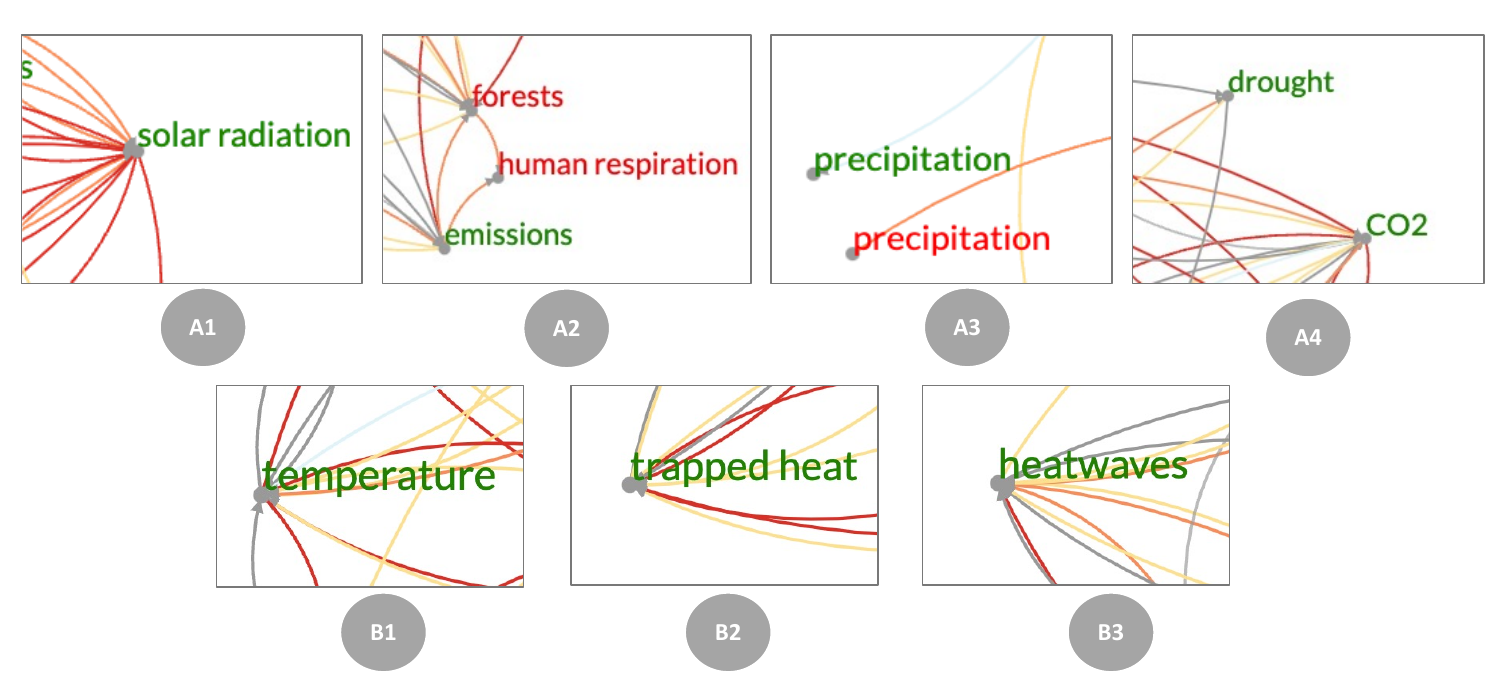}
    \caption{\textbf{\textit{Misinformed} and \textit{Oblivious} cases in discrepancy network.}}
    \Description{Figure \ref{fig:misconception-cases} shows seven screenshots of various interesting Misconception and Obliviousness cases.}
    \label{fig:misconception-cases}
\end{figure}

In Figure \ref{fig:misconception-cases}, we present some noticeable cases within the discrepancy network. A large number of high levels of \textit{misinformed} links (with red and orange links) are related to \textcolor{mygreen}{increasing solar radiation}. We previously identified this as a prevalent misconception among climate-change deniers (Figure \ref{fig:misconception-cases}-A1). \textcolor{red}{Less human respiration} is an attribute that always appears with links with $cs=0$ in the ground truth data, but the crowd linked it with \textcolor{red}{decreasing forests} and \textcolor{mygreen}{increasing emissions} as effects of both (Figure \ref{fig:misconception-cases}-A2). Another interesting finding is that \textcolor{mygreen}{increasing precipitation} has more significant or visible links than \textcolor{red}{decreasing precipitation}, meaning more people have voted for them. This can be rationalized by the higher ratio of crowd workers from the states of California and Texas (Figure \ref{fig:misconception-cases}-A3). Here it is also interesting that the \textcolor{red}{decreasing precipitation} is linked with being misinformed (orange link) while \textcolor{mygreen}{increasing precipitation} is linked with being oblivious (lightest blue link).

Some attributes that present some level of visible alignment of the crowd's opinion with the ground truth are \textcolor{mygreen}{more drought} and \textcolor{mygreen}{increasing CO2} (a good mixture of red, orange, grey, and blue colored links)(Figure \ref{fig:misconception-cases}-A4).
Other various levels of being misinformed are linked with temperature-related attributes such as \textcolor{mygreen}{increasing temperature}, \textcolor{mygreen}{more intense heatwaves}, and \textcolor{mygreen}{more trapped heat}. We think this happened because of a certain ambiguity related to the physical relationships among these three attributes. This validates further refinement of the causal attributes for the next phases of our experiments (Figure \ref{fig:misconception-cases}-B1, B2, B3).

\section{Final Study}

\subsection{Detailed Experimental Protocol}
\label{final-detailed-protocol}
We mention the sequential workflow of the crowd workers below:
 \begin{enumerate}
      \item \textbf{Read the instructions and perform and pass the test.} These tasks were performed in the \textit{Instructions and Overview Module} (Section~\ref{instruction}). Each crowd worker reads an overview of the whole interface as a step-by-step guide, along with the explanation and purpose of each step. They have the opportunity to go back to previous pages or restart anytime. They can refer to the pictures of the interface in different phases of their upcoming workflow that are provided in the module. Next, they encounter a simple instance of building a causal relation between two attributes. If they can pass this test, they proceed to the next step. If they fail this test, they can always restart this module and retry this test to improve their understanding.
     \item \textbf{Complete the demographics survey.} This task was performed in the \textit{Demographics and Climate Change Awareness Survey Module} (Section~\ref{demographics}). Each crowd worker answers 8 questions regarding their demographics, ad 4 questions on climate change awareness. The demographics questions are about their ethnicity, gender, marital status, geographical location (state and county), education, employment status, and age group. In each demographic question, they have an option not to provide their information. We provide the demographics questions in the supplemental materials.
     \item\textbf{Create a causal network.} This task was performed in the \textit{Causal Network Creation Module} (Section~\ref{creation}). Each crowd worker will create five different causal links to build a small causal network. They need to perform two micro-tasks to create each causal link.
\begin{itemize}
    \item Choose the ``cause'' (and its respective trend) from the cause drop-down. 
    \item Choose the ``effect'' (and its respective trend) from the effect drop-down.
\end{itemize} 
    The crowd workers are free to choose the attributes and their trends based on their personal perceptions. For example, they may choose the attribute ``CO2'' with an ``increasing'' or ``decreasing'' trend. Performing these two microtasks will automatically create a causal relationship. Say, for the chosen two trended attributes, \textit{increasing emissions} as the ``cause'' and \textit{increasing CO2} as the ``effect'', the created causal relation would be \textit{increasing emissions} leads to \textit{increasing CO2}. 

After the first link, the workers are asked to select a new node that has not been selected before and an already selected node in order to create a new causal relation, adding to the emerging small network. To ensure that the worker will evolve the graph from an already selected node, the cause (or effect) drop-down menu (whatever the user chooses first) will only show already selected nodes. Let’s say, the user has chosen the cause first, where the drop-down had listed only already selected nodes, already part of the evolving graph. In this case, the effect drop-down menu will now show all variables. This will make sure that only a single graph of connected nodes is created.
     \item\textbf{Alter causal network. }This task is performed in the \textit{Causal Network Alteration Module} (Section~\ref{alternation}). Each crowd worker can alter their created causal network from the previous step by choosing the link they want to change by left-clicking on it and
selecting from the available options for alteration. The options are: \textit{change direction} and \textit{do not modify anything}. 
     \item\textbf{Evaluate causal network.} This task is performed in the \textit{Causal Network Interpretation and Evaluation Module} (Section~\ref{interpretation}). Each crowd worker can read their causal network and view it in a node-link diagram or Directed Acyclic Graph (DAG) format. On a scale of 1 to 5, they also provide their confidence level to the created networks.
     
     \item \textbf{Evaluate the interface}. After creating a causal network, each crowd worker is asked to evaluate the interface based on seven usability and learning statements on a 5-point Likert Scale using the \textit{Interface Evaluation Module} (Section~\ref{evaluation_interface}). 
     
     
     \item \textbf{Verification and compensation}. Each crowd worker is provided with a unique code at the end of their participation (on AMT) or redirected to a specific link (on Prolific). We used these mechanisms to validate the results, discard any incomplete data, and compensate the crowd worker. 
 \end{enumerate}
 
\subsection{Final Aggregated Network Node Exploration Status}
\label{final-node-exploration}
\begin{figure}
    \centering
    \includegraphics[width=0.8\textwidth]{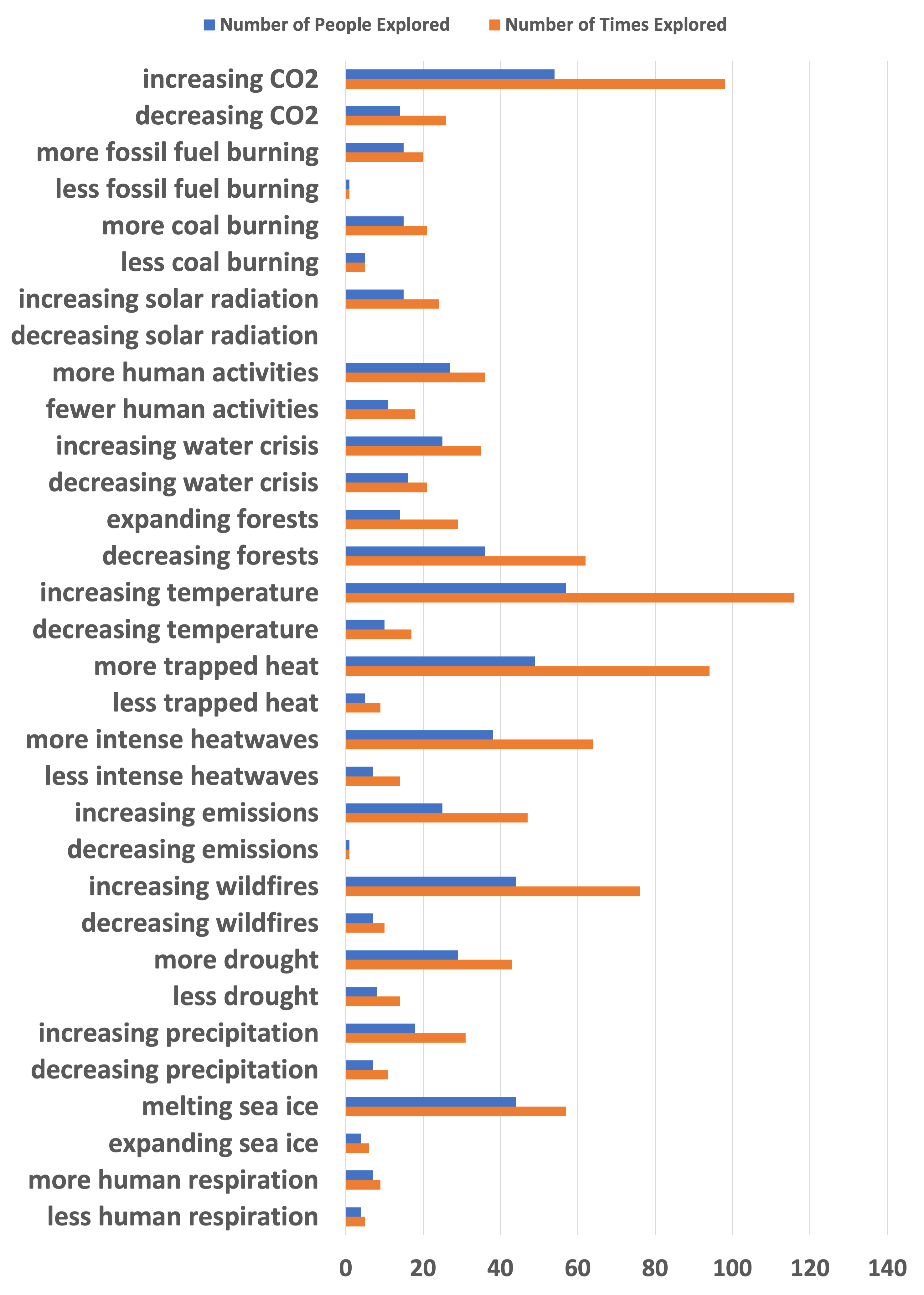}
    \caption{\textbf{The Exploration Status of all nodes denoting the causal attributes in the final aggregated network.} The orange bars denote the number of times the node has been explored either as cause or effect in any causal link created by the crowd. The blue bars denote the number of people who visited and explored that specific causal attribute/node either as cause or effect.}
    \Description{Figure \ref{fig:node-exploration} shows seven screenshots of various interesting Misconception and Obliviousness cases.}
    \label{fig:node-exploration}
\end{figure}

We provide the final node exploration status of the aggregated network in Figure \ref{fig:node-exploration}. The crowd explored all attributes to some extent, except for ``decreasing solar radiation'' which did not appear in any crowd worker's network. 

\end{document}